\documentclass[10pt,twocolumn]{article}

\usepackage[utf8]{inputenc}
\usepackage[T1]{fontenc}
\usepackage{amsmath,amssymb}
\usepackage{graphicx}
\usepackage{hyperref}
\usepackage{xcolor}
\usepackage{listings}
\usepackage{geometry}
\usepackage{booktabs}
\usepackage{enumitem}
\usepackage{cite}
\usepackage{mdframed}
\usepackage{titlesec}
\usepackage{parskip}
\usepackage{caption}
\usepackage{tikz}
\usepackage{soul}
\usetikzlibrary{shapes.geometric, arrows.meta, positioning}
\titlespacing*{\section}{0pt}{8pt}{4pt}
\titlespacing*{\subsection}{0pt}{4pt}{3pt}
\titlespacing*{\subsubsection}{0pt}{3pt}{3pt}
\titlespacing*{\paragraph}{0pt}{2pt}{1pt}
\definecolor{codeblue}{rgb}{0.13,0.29,0.53}
\definecolor{codegray}{rgb}{0.5,0.5,0.5}
\definecolor{codegreen}{rgb}{0.0,0.45,0.0}
\definecolor{codeback}{rgb}{0.97,0.97,0.97}
\definecolor{highlightblue}{rgb}{0.88,0.93,0.98}
\definecolor{boxborder}{rgb}{0.6,0.6,0.8}
\definecolor{synthback}{rgb}{0.93,0.97,0.93}

\lstdefinestyle{python}{
    backgroundcolor=\color{codeback},
    commentstyle=\color{codegreen}\itshape,
    keywordstyle=\color{codeblue}\bfseries,
    basicstyle=\ttfamily\small,
    breaklines=true,
    keepspaces=true,
    numbers=none,
    showstringspaces=false,
    language=Python,
    frame=single,
    framesep=2pt,
    rulecolor=\color{codegray},
    xleftmargin=4pt,
    xrightmargin=4pt,
    aboveskip=3pt,
    belowskip=3pt,
}

\lstdefinestyle{synth}{
    backgroundcolor=\color{synthback},
    commentstyle=\color{codegreen}\itshape,
    keywordstyle=\color{codeblue}\bfseries,
    basicstyle=\ttfamily\small,
    breaklines=true,
    keepspaces=true,
    numbers=none,
    showstringspaces=false,
    language=Python,
    frame=single,
    framesep=2pt,
    rulecolor=\color{codegreen!60},
    xleftmargin=4pt,
    xrightmargin=4pt,
    aboveskip=3pt,
    belowskip=3pt,
}

\newcommand{\ket}[1]{|#1\rangle}
\newcommand{\bigO}[1]{\mathcal{O}(#1)}

\hypersetup{colorlinks=true, linkcolor=codeblue,
            citecolor=codegreen, urlcolor=codeblue}

\newcommand{\loadfigs}[6]{%
  \vspace{4pt}
  \noindent
  \begin{minipage}[c]{0.47\linewidth}
    \centering
    \includegraphics[width=\linewidth]{figures/#1}
  \end{minipage}
  \hfill
  \begin{minipage}[c]{0.47\linewidth}
    \centering
    \includegraphics[width=#5\linewidth]{figures/#2}
  \end{minipage}
  \captionof{figure}{\textit{Left:} #3\quad\textit{Right:} PyEncode circuit.}
  \label{fig:#6}
}

\newcommand{\loadfigsv}[6]{%
  \vspace{4pt}
  \noindent
  \begin{center}
    \includegraphics[width=#4\linewidth]{figures/#1}\\[4pt]
    \includegraphics[width=#5\linewidth]{figures/#2}
  \end{center}
  \captionof{figure}{\textit{Top:} #3\quad\textit{Bottom:} PyEncode circuit.}
  \label{fig:#6}
}

\title{\textbf{PyEncode: An Open-Source Library for Structured Quantum State Preparation}}
\author{Krishnan Suresh\\
\small University of Wisconsin--Madison\\
\small \texttt{ksuresh@wisc.edu}
\and
Sanjay Suresh\\
\small University of Wisconsin--Madison\\
\small \texttt{ssuresh27@wisc.edu}}
\date{}

\begin{document}
\maketitle
\thispagestyle{empty}

\begin{abstract}
Quantum algorithms require encoding classical vectors as quantum states,
a step known as amplitude encoding. General-purpose routines produce
circuits with $\bigO{2^m}$ gates for vectors of length $N = 2^m$, \textcolor{black}{for an $m$-qubit register}.
However, vectors arising in scientific and engineering applications
often exhibit mathematical structure that admits far more efficient
encoding. Theoretical work over the last decade has established efficient circuits for several structured vector classes, but without open-source
implementations.

We present \textbf{PyEncode}, an open-source Python library that
implements this body of theory in a unified framework. It covers
ten exact pattern families: \emph{sparse, step, square, Walsh,
Fourier, geometric, Hamming, staircase, Dicke}, and \emph{polynomial}.
A function \texttt{encode} maps each pattern to a verified Qiskit
circuit, with no vector materialization and no approximation;
for example, \texttt{encode(SPARSE([(19, 1.0)]), N=64)} 
encodes the vector $\mathbf{e}_{19}$  of length $N = 64$. Sparse, step,
Walsh, Hamming, and staircase patterns require $\bigO{m}$ gates;
square and Fourier patterns require $\bigO{m^2}$; Dicke states
$|D^m_k\rangle$ require $\bigO{k(m-k)}$, \textcolor{black}{ that denotes uniform superpositions over indices of Hamming weight $k$}; degree-$d$ polynomials
require $\bigO{m^{d+1}}$. A companion \texttt{predict\_gates}
function estimates transpiled gate counts without synthesis, \textcolor{black}{and a
reverse-lookup utility \texttt{match\_vector} identifies which family
best fits a given numerical vector.}
Three composition primitives are supported: \texttt{SUM} for
weighted superpositions, \texttt{PARTITION} for ancilla-free
composition of disjoint-support patterns, and \texttt{TENSOR}
for separable states over disjoint subregisters. For amplitude
vectors outside these exact families, PyEncode also provides
a matrix product state (MPS) loader, \texttt{encode\_mps} for approximate vector encoding. The library is available at
\url{https://github.com/UW-ERSL/PyEncode}.
\end{abstract}

\section{Motivation}
\label{sec:motivation}

\textcolor{black}{Quantum computers operate on data encoded in the amplitudes of a
quantum state. Before any quantum algorithm~\cite{harrow2009,gilyen2019quantum,martyn2021grand}  can process a classical
vector, whether it is a force vector, a probability distribution, or a set of
Hamiltonian coefficients, that vector must first be loaded into a
quantum register, a step called \emph{amplitude encoding} or state
preparation. This loading step is often the hidden cost of quantum
algorithms: for a vector of length $N$, preparing a general quantum
state requires a number of gates that grows linearly with $N$, and
therefore exponentially with the number of qubits. For many
algorithms this cost is large enough to cancel the very speedup the
algorithm was meant to provide. PyEncode is a Python library that
sidesteps this cost for the large class of vectors that are not
arbitrary but instead follow a recognizable mathematical pattern.
A user declares the pattern --- a sinusoid, a step, an exponential
decay, a sparse set of entries --- and PyEncode returns a compact,
verified quantum circuit that prepares it, with a gate
count that grows only polynomially in the number of qubits $m$. This
paper documents the library, its supported patterns, and its use.}

\textcolor{black}{Throughout, $N = 2^m$ denotes the vector length on an $m$-qubit
register; $s$ is the number of nonzero amplitudes (sparsity), $d$ the
polynomial degree, and $k$ the Hamming weight for Dicke states.
Per-pattern parameters (the interval bounds $k_s$, $k_e$, the ratio
$r$, the amplitude $c$, and so on) are defined with each constructor
in Section~\ref{sec:patterns}.}

Amplitude encoding, also known as state
preparation~\cite{weigold2021}, is a well-known bottleneck~\cite{aaronson2015}: \textcolor{black}{for  encoding a vector of length $N = 2^m$}, general state preparation requires
$\bigO{2^m}$ gates, erasing any algorithmic speedup. The same bottleneck appears across quantum chemistry, where fault-tolerant
phase estimation requires the PREP oracle to amplitude-encode the
coefficient vector of the Hamiltonian~\cite{childs2018,babbush2018}, and
in quantum Monte Carlo, where amplitude estimation achieves a quadratic
speedup only if the probability distribution can be loaded
efficiently~\cite{montanaro2015,Herbert2022}.

However, what is common to all of these settings is that the vector to be encoded
is rarely arbitrary. In computational mechanics, the force often has a known mathematical form: a
sinusoidal mode, a uniform pressure, a point force. For lattice
Hamiltonians in quantum chemistry and condensed matter, such as the
Fermi--Hubbard and Heisenberg models, translational invariance
collapses the Pauli coefficient vector to a small number of distinct
values.  In quantum finance, the discretized probability distribution is piecewise constant.

\begin{mdframed}[backgroundcolor=highlightblue,
                 linecolor=codeblue, linewidth=0.8pt,
                 innertopmargin=4pt, innerbottommargin=4pt,
                 innerleftmargin=6pt, innerrightmargin=6pt]
\small
Theoretical work over the last decade has established efficient circuits
for such structured vector classes, but without open-source
implementations. PyEncode fills this gap: a single
function \texttt{encode} maps a structured declaration directly to a verified Qiskit circuit, with no vector materialization and no approximation. In addition, the \texttt{SUM} constructor
enables exact weighted superpositions of pattern states (exact on the
post-selected ancilla-$|0\rangle$ subspace, with analytically
determined success probability); the \texttt{PARTITION}
constructor handles disjoint-support compositions ancilla-free with
success probability one; and the \texttt{TENSOR} constructor
composes patterns over disjoint subregisters for separable
multi-dimensional vectors.  PyEncode also provides
a matrix product state loader,
\texttt{encode\_mps} for approximate vector encoding.
\end{mdframed}
\vspace{4pt}

For example, consider a vector of length $N = 2^6 = 64$
with a single nonzero entry at index $19$, i.e.,\ $\mathbf{f} = \mathbf{e}_{19}$.
Qiskit's \texttt{StatePreparation} produces \textbf{97 gates}.
In PyEncode, \texttt{encode(SPARSE([(19, 1.0)]), N=64)} yields a circuit with \textbf{3 gates}. \textcolor{black}{Throughout this paper, gate counts
are reported after \emph{transpilation} --- the compilation of a
circuit into hardware-native gates followed by optimization --- to the
standard two-element basis $\{\textsc{cx}, U\}$, consisting of a
two-qubit entangling gate ($\textsc{cx}$) and a general single-qubit
rotation ($U$). We use Qiskit~2.3.1 at \texttt{optimization\_level=3},
its most aggressive optimization setting to reduce gate count, and decompose Qiskit's
\texttt{StatePreparation} with \texttt{reps=3}.}

\section{Prior Work}
\label{sec:priorwork}

\textbf{General-purpose state preparation}\\
The problem of preparing an arbitrary $N = 2^m$-dimensional quantum
state has been studied extensively.
Shende, Bullock, and Markov~\cite{shende2006} established the
$\mathcal{O}(2^m)$ lower bound on gate count and gave an explicit
constructive procedure.
Araujo et al.~\cite{araujo2021divide} reduce circuit depth to
$\mathcal{O}(\log^2 N)$ via a divide-and-conquer strategy, at the cost
of $\mathcal{O}(N)$ ancilla qubits.
Gui et al.~\cite{gui2024spacetime} propose a deterministic
$\mathcal{O}(\log N)$-depth protocol with optimal spacetime allocation
$\mathcal{O}(N)$.
All of these methods treat the input vector as an opaque array and
do not exploit any structure in the amplitudes.

\textbf{Approximate state preparation}\\
Several methods achieve sub-exponential complexity by allowing a
controlled approximation error.
Welch et al.~\cite{welch2014} established the connection between Walsh
functions and diagonal unitary circuits, showing that truncated
Walsh--Fourier series yield approximately minimal-depth circuits.
O'Brien and S\"{u}nderhauf~\cite{obrien2025} achieve efficient approximate
state preparation for piecewise-defined functions, whose amplitudes are
well approximated by piecewise polynomials, via quantum singular value
transformation (QSVT).
Marin-Sanchez et al.~\cite{marinsanchez2023} reduce the
Grover--Rudolph circuit complexity (see discussion below) from
$\mathcal{O}(2^n)$ to $\mathcal{O}(2^{k_0(\epsilon)})$ for smooth,
real-valued functions.
Zylberman and Debbasch~\cite{zylberman2024walsh} introduce the Walsh
Series Loader (WSL), achieving circuit depth
$\mathcal{O}(1/\sqrt{\epsilon})$ independent of the number of qubits.
Xie and Ben-Ami~\cite{xie2025gaussian} target the discretized Gaussian
specifically, exploiting a separable product state as an intermediate
step before a final QFT.
Finally, \textbf{matrix product state (MPS)} approximations exploit bounded
entanglement to load smooth, differentiable functions in linear
depth~\cite{holmes2020,melnikov2023quantum}, building on the
sequential generation framework of Sch\"on et al.~\cite{schon2005}
and the decomposition of
Ran~\cite{ran2020}; \emph{PyEncode implements an MPS loader as described in Section~\ref{sec:mps}}.

\textbf{Structure-exploiting state preparation}\\
A significant body of work exploits specific structural properties of the
target vector to reduce gate complexity well below the general
$\mathcal{O}(2^m)$ bound. PyEncode implements thirteen such
constructions from the literature (ten pattern families and three
composition rules), each under a named constructor with
the original complexity guarantees preserved. These constructions are:
\begin{itemize}[leftmargin=*, itemsep=2pt, topsep=2pt]
  \item \emph{Sparse}: Gleinig and Hoefler~\cite{gleinig2021} gave
        an exact $\mathcal{O}(sm)$-gate algorithm for $s$-sparse
        states based on pairwise merging over the nonzero index set.
  \item \emph{Step}: Shukla and Vedula~\cite{shukla2024} derived
        closed-form circuits for \emph{interval uniform
        superpositions}, establishing $\mathcal{O}(m)$ cost for the
        prefix case $[0, k_e)$ via multi-controlled operations on the
        binary representation of $k_e$.
  \item \emph{Walsh}: Welch et al.~\cite{welch2014} established
        the correspondence between Walsh series and diagonal operator
        compilation, introducing the $R_y(\theta) + H^{\otimes m}$
        two-level construction; PyEncode implements it in $m{+}1$
        gates, with a generalized form supporting asymmetric levels.
  \item \emph{Staircase}: Hackbusch's hierarchical-matrix
        arithmetic~\cite{hackbusch1999} is
        implemented in PyEncode via cascaded controlled-$R_y$
        rotations at $\mathcal{O}(m)$ cost.
  \item \emph{Square}: The general \emph{square}
        interval $[k_s, k_e)$ is obtained by composing \emph{step}~\cite{shukla2024} with
        Draper's adder~\cite{draper2000}. PyEncode's general interval $[k_s, k_e)$ construction
        is $\mathcal{O}(m^2)$ in general, $\mathcal{O}(m)$ for
        power-of-2-aligned blocks.
  \item \emph{Geometric}: Grover and
        Rudolph~\cite{grover2002creating} showed that whenever the
        cumulative amplitudes $\sum_{i=0}^{k} f_i^2$ can be computed
        efficiently, the state can be prepared in $\mathcal{O}(m)$
        depth via cumulative-integral-determined controlled $R_y$
        rotations; the exponential-decay specialization
        $f_i = c\,r^i$ factorizes across qubits, yielding a depth-1
        product state with zero entangling gates. 
  \item \emph{Fourier}: Finite-term Fourier
         is a standard amplitude-encoding
        primitive~\cite{nielsenchuang2010}, at $\mathcal{O}(m^2)$.  Gonzalez-Conde et
        al.~\cite{gonzalezconde2024} and Moosa et al.~\cite{moosa2024}
        independently sharpened by loading DFT
        coefficients as a sparse state and applying the inverse QFT
        at cost independent of the number of modes; this is the
         implementation in PyEncode.
  \item \emph{Polynomial}: Walsh framework combined with a
        sparse anchor load~\cite{welch2014, gonzalezconde2024}
        yields an exact, unit-success pipeline for degree-$d$
        amplitude vectors at $\mathcal{O}(m^{d+1})$ cost.
\item \emph{Dicke}: B\"artschi and
        Eidenbenz~\cite{baertschi2019dicke} gave a deterministic
        split-cyclic-shift cascade for Dicke states
        $|D^m_k\rangle$ at $\mathcal{O}(k(m-k))$ two-qubit cost; PyEncode adds a $k > m/2$ symmetry
        optimization that synthesizes the lighter
        $|D^m_{m-k}\rangle$ and appends $X^{\otimes m}$,  for high-weight targets.
  \item \emph{Hamming}: The identity
        $r^{\mathrm{wt}(i)} = \prod_j r^{\,b_j(i)}$ factorizes
        Hamming-weight-indexed product states
        $f_i = c\,r^{\mathrm{wt}(i)}$ into identical single-qubit
        rotations on every qubit, yielding a depth-1 product-state
        circuit~\cite{cruz2019efficient}.
  \item \emph{Partition}: Bentley and Saxe~\cite{bentley1980}
        introduced dyadic decomposition of half-open intervals in their decomposable-searching framework;
        PyEncode uses this $\mathcal{O}(\log N)$ as the
        basis for ancilla-free composition of disjoint-support
        patterns at $\mathcal{O}(Lm)$ cost.
    \item \emph{Sum}: The linear-combination-of-unitaries technique
        of Childs and Wiebe~\cite{childs2018} and Babbush et
        al.~\cite{babbush2018} enables weighted superpositions of
        block-encoded unitaries with ancilla-based post-selection;
        PyEncode applies this directly to pattern circuits, with
        analytically determined success probability and automatic
        disjoint-support detection that collapses to \emph{partition}
        whenever applicable.
    \item \emph{Tensor}: When the target factorizes as
        $\bigotimes_j |f^{(j)}\rangle$,
        each component is prepared independently on its own register;
        the composition runs in parallel, with total gate cost
        $\sum_j C_j$ and depth $\max_j D_j$~\cite{cruz2019efficient}.
\end{itemize}

\section{PyEncode Library}
\label{sec:patterns}
While the constructions above are individually well-studied, their
specifications have remained scattered across the literature without
open-source implementations. PyEncode assembles them into a single,
immediately deployable Python library in which every pattern is
made available through a typed constructor, returns a Qiskit circuit with
pre-computed gate-count and success-probability metadata. Table~\ref{tab:patterns} summarizes the ten supported patterns and three compositions.

PyEncode exposes two primary entry points, together with a reverse-lookup
utility \texttt{match\_vector} (Section~\ref{sec:match}) and a separate MPS
entry point (Section~\ref{sec:mps}). The first synthesizes a
Qiskit circuit for each pattern:
\begin{mdframed}[backgroundcolor=highlightblue,
                 linecolor=codeblue, linewidth=0.8pt,
                 innertopmargin=2pt, innerbottommargin=2pt,
                 innerleftmargin=6pt, innerrightmargin=6pt]
\small
\texttt{encode(pattern, N, validate=False, tol=1e-6)}
\end{mdframed}
It returns a tuple \texttt{(circuit, info)} containing the synthesized
Qiskit circuit and an \texttt{EncodingInfo} dataclass recording the
recognized pattern, gate counts, circuit depth, success probability  (see
Section~\ref{sec:returnValue}). By default, no classical vector is ever materialized during synthesis. However, if \texttt{validate=True}, the analytic amplitude vector is checked against
 the prepared state.
All ten exact patterns and the three composition rules accept
real or complex amplitudes.  The complex code-paths activate only when at least one parameter carries a
non-zero imaginary part, and add at most $\mathcal{O}(m)$ phase
gates that the transpiler typically absorbs into adjacent
single-qubit rotations at \texttt{optimization\_level=3}. \textcolor{black}{Since every pattern costs at least $\mathcal{O}(m)$ gates,
this additive $\mathcal{O}(m)$ phase layer leaves the asymptotic
gate complexity of every pattern unchanged; it holds for both
real and complex amplitudes.}

The second entry point returns the gate count and depth \emph{without} building
the circuit. This is useful for cost estimation in
optimization loops and resource planning at problem sizes where synthesis would be expensive:
\begin{mdframed}[backgroundcolor=highlightblue,
                 linecolor=codeblue, linewidth=0.8pt,
                 innertopmargin=2pt, innerbottommargin=2pt,
                 innerleftmargin=6pt, innerrightmargin=6pt]
\small
\texttt{predict\_gates(pattern, N)}
\end{mdframed}
It returns a dictionary reporting the predicted transpiled single- and
two-qubit gate counts, circuit depth, asymptotic complexity, and an
\texttt{exact} flag marking whether the prediction is guaranteed exact
(several patterns admit closed-form counts) or an empirically fitted
upper bound (see Section~\ref{sec:predict}).

PyEncode library is illustrated next through several examples. All examples below import:
\begin{lstlisting}[style=python]
from pyencode import (encode, SPARSE, STEP, SQUARE, FOURIER, WALSH, GEOMETRIC, HAMMING, STAIRCASE, DICKE, POLYNOMIAL, SUM, PARTITION, TENSOR)
\end{lstlisting}

\begin{table*}
\centering
\caption{Recognized patterns and compositions in PyEncode. $^\dagger$\texttt{SQUARE} uses a Draper QFT-based constant adder~\cite{draper2000}: $\mathcal{O}(m)$ for $k_s=0$ or power-of-2-aligned blocks; $\mathcal{O}(m^2)$ in general. $^\ddagger$\texttt{GEOMETRIC} with \texttt{k\_s=0} prepares $c\,r^{\,i}$ as a depth-1 product state at $\mathcal{O}(m)$; with an arbitrary offset $k_s>0$, the nonzero window $[k_s,N)$ decomposes into dyadic sub-blocks and the cost becomes $\mathcal{O}(m^2)$.}
\label{tab:patterns}
\begin{tabular}{lllll}
\toprule
\textbf{Name} & \textbf{Constructor} &
\textbf{Form} & $\mathcal{O}(\cdot)$ & \textbf{Source} \\
\midrule
\multicolumn{5}{l}{\textit{Patterns}} \\
\midrule
Sparse         & \texttt{SPARSE([(x,a),...])}    & $\sum_j \alpha_j |x_j\rangle$              & $\mathcal{O}(sm)$          & \cite{gleinig2021} \\
Step           & \texttt{STEP(k\_e,c)}           & $c\,\mathbf{1}[i < k_e]$                  & $\mathcal{O}(m)$           & \cite{shukla2024} \\
Square         & \texttt{SQUARE(k\_s,k\_e,c)}    & $c\,\mathbf{1}[k_s\leq i<k_e]$            & $\mathcal{O}(m^2)^\dagger$ & \cite{shukla2024,draper2000} \\
Walsh          & \texttt{WALSH(k,c0,c1)}         & $c_0/c_1$ on $b_k(i)=0/1$            & $\mathcal{O}(m)$           & \cite{welch2014} \\
Geometric      & \texttt{GEOMETRIC(r,k\_s)}     & $c\,r^{\,i-k_s}$ on $[k_s,N)$ & $\mathcal{O}(m^2)^\ddagger$ & \cite{grover2002creating,xie2025gaussian,bentley1980}\\
Hamming        & \texttt{HAMMING(r,c)}           & $c\,r^{\,\mathrm{wt}(i)}$                 & $\mathcal{O}(m)$           & \cite{baertschi2019dicke,nielsenchuang2010} \\
Staircase      & \texttt{STAIRCASE(r,c)}         & $c\,r^{\,k}$ on $i=2^k{-}1$               & $\mathcal{O}(m)$           & \cite{hackbusch1999} \\
Dicke          & \texttt{DICKE(k,c)}             & $c\,\mathbf{1}[\mathrm{wt}(i){=}k]$       & $\mathcal{O}(k(m{-}k))$    & \cite{baertschi2019dicke} \\
Polynomial     & \texttt{POLYNOMIAL(coeffs)}     & $\sum_{j=0}^{d} c_j\,(i/(N{-}1))^j$       & $\mathcal{O}(m^{d+1})$     & \cite{welch2014,gonzalezconde2024} \\
Fourier        & \texttt{FOURIER(modes=[...])}   & $\sum_t A_t\sin(2\pi n_t i/N+\varphi_t)$  & $\mathcal{O}(m^2)$         & \cite{nielsenchuang2010} \\
\midrule
\multicolumn{5}{l}{\textit{Compositions}} \\
\midrule
Sum            & \texttt{SUM([(w,pattern),\ldots])}        & $\sum_j w_j|f^{(j)}\rangle$                 & $\mathcal{O}(\sum_i C_i)$ & \cite{childs2018,babbush2018} \\
Partition      & \texttt{PARTITION([pattern,\ldots])}      & $\sum_j |f^{(j)}\rangle$, disjoint support  & $\mathcal{O}(L\cdot m)$   & \cite{bentley1980,gleinig2021} \\
Tensor         & \texttt{TENSOR([(pattern,$N_i$),\ldots])} & $\bigotimes_j |f^{(j)}\rangle$              & $\mathcal{O}(\sum_i C_i)$ & \cite{nielsenchuang2010} \\
\bottomrule
\end{tabular}
\end{table*}

\subsection{Sparse}
This pattern represents a superposition of $s$ basis states at
explicitly declared indices with (possibly distinct) real or complex
amplitudes $\alpha_j \in \mathbb{C}$, as defined by Gleinig and
Hoefler~\cite{gleinig2021}:
\begin{equation}
  |\psi\rangle = \frac{1}{\mathcal{N}}
    \sum_{j=1}^{s} \alpha_j\,|x_j\rangle,
  \quad S = \{x_1, \ldots, x_s\} \subseteq \{0,\ldots,N{-}1\}
\end{equation}
where $\mathcal{N}$ is the normalization constant.
Constructor: \texttt{SPARSE([(x1, alpha1), (x2, alpha2), ...])}.
Circuit complexity: $\mathcal{O}(s\,m)$~\cite{gleinig2021}.
The Gleinig--Hoefler algorithm builds the state recursively by
partitioning $S$ along qubit boundaries; at each level a single
controlled-$R_y$ rotation selects the relative weight between the two
halves.

\textbf{Example.}

The basis vector $\mathbf{e}_{19}$ on $N=64$ ($s=1)$:
\begin{lstlisting}[style=python]
circuit, info = encode(SPARSE([(19, 1.0)]), N=64)
# info.complexity  -> "O(s*m)"
# info.gate_count  -> 3   (Hamming weight of 19)
\end{lstlisting}
Figure~\ref{fig:ex_sparse_single} shows the vector and circuit.

\loadfigs{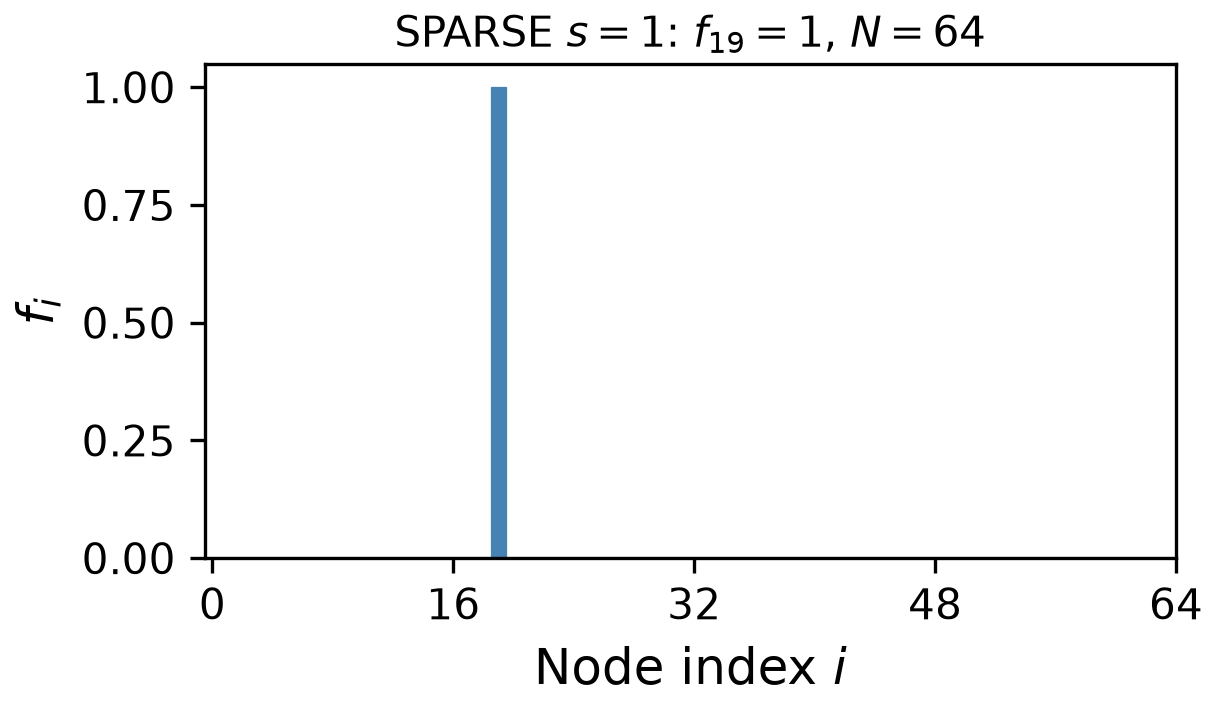}{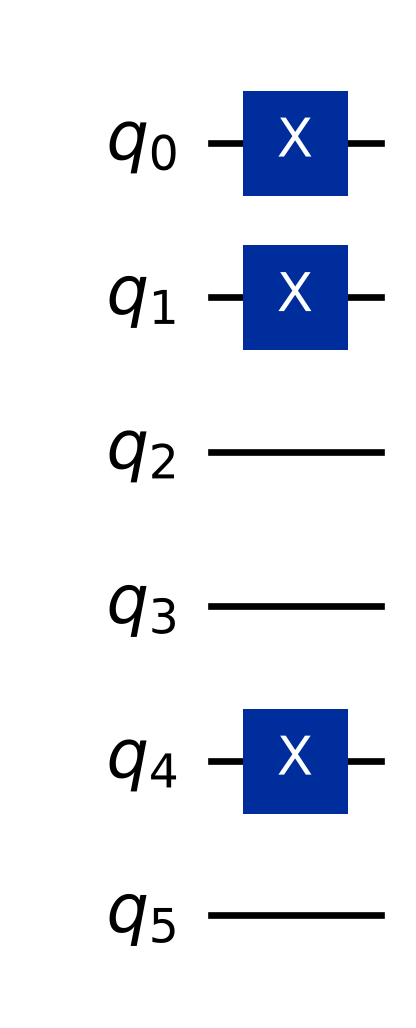}
  {Sparse ($s=1$): $\alpha_{19}=1$, $N=64$}
  {unused}
  {0.35}{ex_sparse_single}

\textbf{Multi-entry example.}

The Gleinig--Hoefler pairwise merging step accepts arbitrary real or
complex amplitudes; for complex inputs a single phase gate per merge
strips the relative phase before the standard rotation, and any
residual phase on a single basis state is recorded as
\texttt{qc.global\_phase} (observable when the circuit is used as a
controlled sub-block in \texttt{SUM}).  When every $\alpha_j$ is real
the phase layer collapses to identity and the original signed-real
circuit is recovered gate-for-gate.
\begin{lstlisting}[style=python]
circuit, info = encode(SPARSE([(1, 3.0), (6, -4.0)]), N=8)
# info.complexity  -> "O(s*m)"   (s=2)
# info.gate_count  -> 5
\end{lstlisting}
Figure~\ref{fig:ex_sparse_multi} shows the input amplitudes. The encoder
automatically normalizes, so the prepared state is
$\bigl(3|1\rangle - 4|6\rangle\bigr)/5$.

\loadfigs{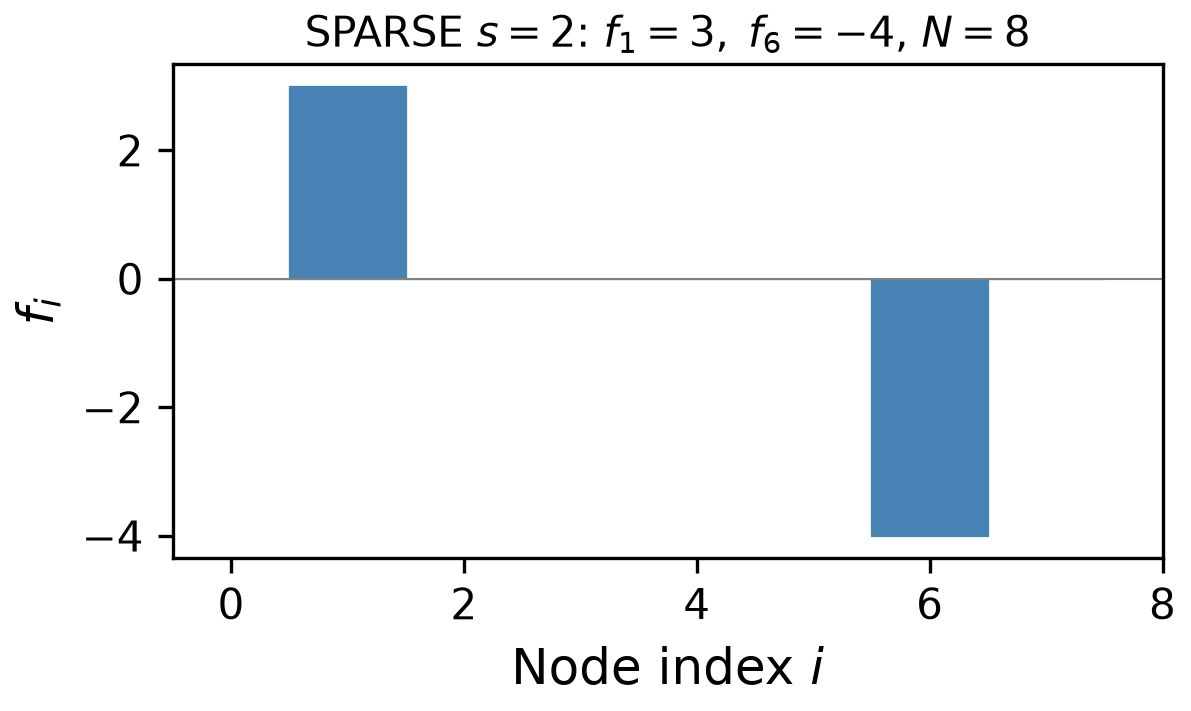}{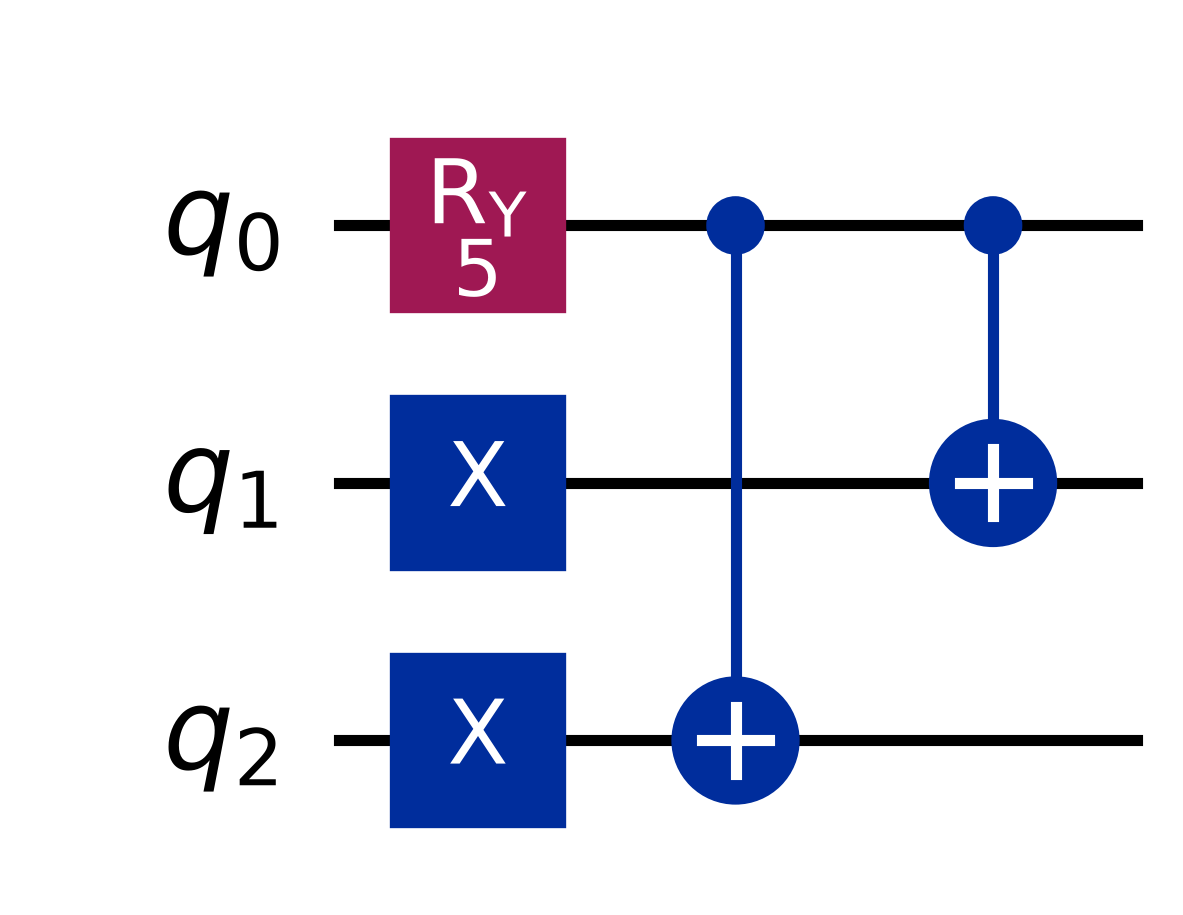}
  {Sparse ($s=2$): $\alpha_1=3,\;\alpha_6=-4$, $N=8$}
  {unused}
  {0.8}{ex_sparse_multi}

\subsection{Step}
Constant amplitude $c$ on the prefix $[0, k_e)$, zero otherwise:
\begin{equation}
  f_i = c\,\mathbf{1}[i < k_e]
\end{equation}
Constructor: \texttt{STEP(k\_e, c)}.
Circuit complexity: $\mathcal{O}(m)$.
This is the interval uniform superposition studied by Shukla and
Vedula~\cite{shukla2024}. Their construction exploits the binary
representation of $k_e$ to decompose the multi-controlled preparation
into $\mathcal{O}(m)$ elementary gates. The special case $k_e = N$ covers the entire range, producing the
uniform superposition $H^{\otimes m}\ket{0}$ in $m$ gates.

\textbf{Example.}
\begin{lstlisting}[style=python]
circuit, info = encode(STEP(k_e=4, c=1.0), N=8)
# info.complexity  -> "O(m)"
# info.gate_count  -> 2
\end{lstlisting}
Figure~\ref{fig:ex_step} shows the step vector and its $\mathcal{O}(m)$ circuit.

\loadfigs{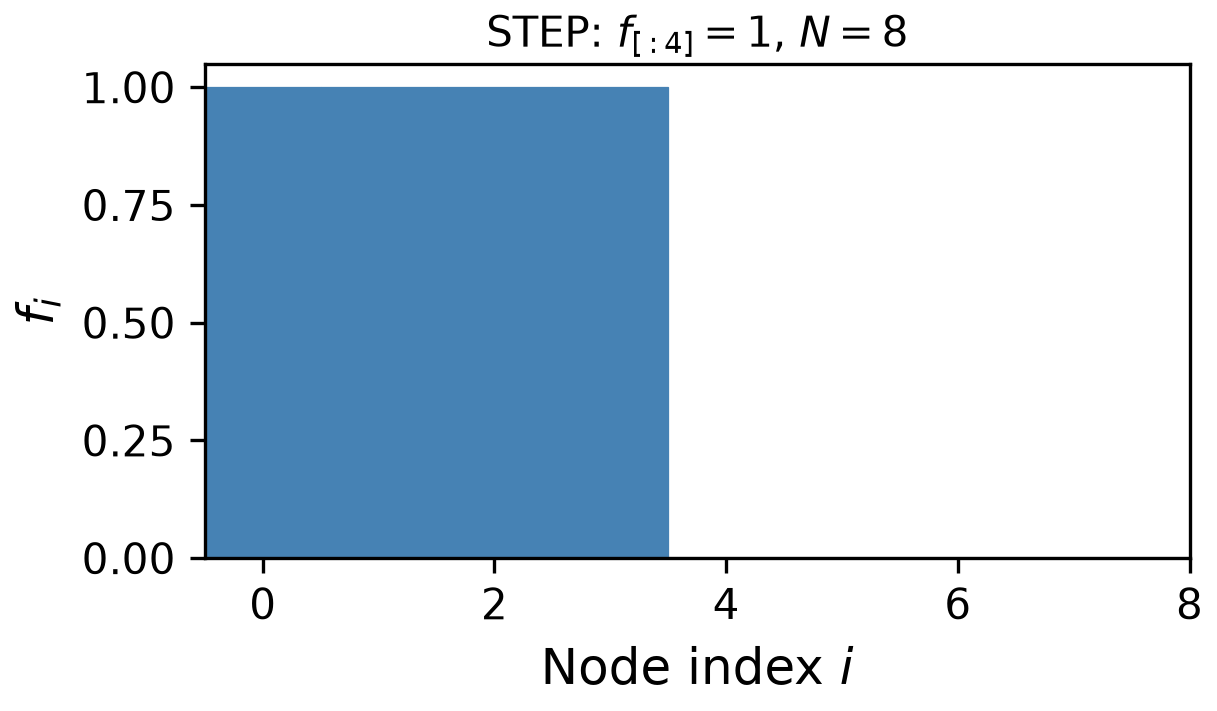}{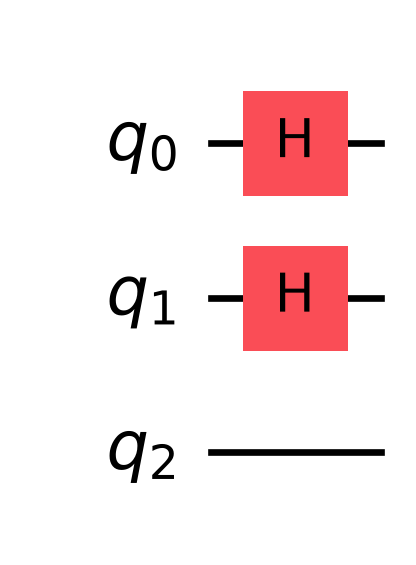}
  {Step: $f_{[:4]}=1$, $N=8$}
  {unused}
  {0.4}{ex_step}

\subsection{Square}
Constant amplitude $c$ on a general interval $[k_s, k_e)$, zero
otherwise:
\begin{equation}
  f_i = c\,\mathbf{1}[k_s \leq i < k_e]
\end{equation}
Constructor: \texttt{SQUARE(k\_s, k\_e, c)}. The interval $[k_s, k_e)$ is a shift of the prefix interval $[0, w)$
by the classical constant $k_s$, where $w = k_e - k_s$.
The circuit proceeds in two ancilla-free stages:
\begin{enumerate}[leftmargin=*, label=\arabic*., itemsep=0pt, topsep=2pt]
  \item \textbf{STEP}$(w)$: prepare the uniform superposition over
        $[0,w)$ using the Shukla--Vedula step circuit~\cite{shukla2024}
        at $\mathcal{O}(m)$ cost.
  \item \textbf{ADD}$(k_s)$: shift the register by $k_s$ in place
        using the Draper QFT-based constant adder~\cite{draper2000}:
        apply QFT, accumulate $\mathcal{O}(m)$ phase rotations,
        then apply QFT$^\dagger$.
\end{enumerate}
The result is the uniform superposition
$ \frac{1}{\sqrt{w}}\sum_{i=k_s}^{k_e-1}|i\rangle$,
prepared exactly with no ancilla qubits and no post-selection.

Alternative constructions using SUM or PARTITION are possible, but the Draper-adder
composition is both deterministic and the cheapest for a single
interval.

The Draper adder dominates at $\mathcal{O}(m^2)$ due to the QFT pair,
giving $\mathcal{O}(m^2)$ total gates in general.
Two special cases admit $\mathcal{O}(m)$ circuits:
\begin{itemize}[leftmargin=*, itemsep=0pt, topsep=2pt]
  \item $k_s = 0$: no shift is needed; reduces exactly to
        \texttt{STEP}$(k_e)$.
  \item Power-of-2-aligned block ($w = 2^p$, $k_s$ a multiple of $w$):
        the shift requires only $X$ gates on the upper qubits and
        $H^{\otimes p}$ on the lower qubits.
\end{itemize}

\textbf{Example.}
\begin{lstlisting}[style=python]
circuit, info = encode(SQUARE(k_s=2, k_e=6, c=1.0), N=8)
# info.complexity  -> "O(m^2)"
# info.gate_count  -> 7   (STEP(4) + Draper adder(2), m=3)
\end{lstlisting}
Figure~\ref{fig:ex_square} shows the interval vector and circuit.

\loadfigs{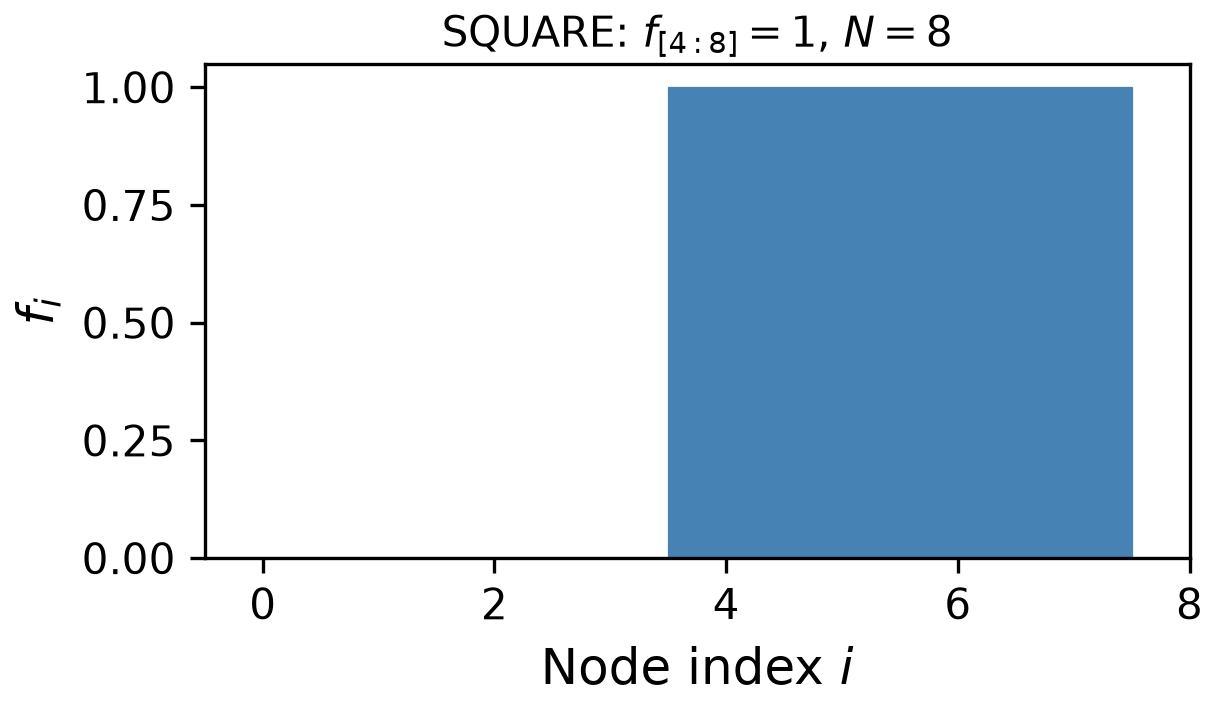}{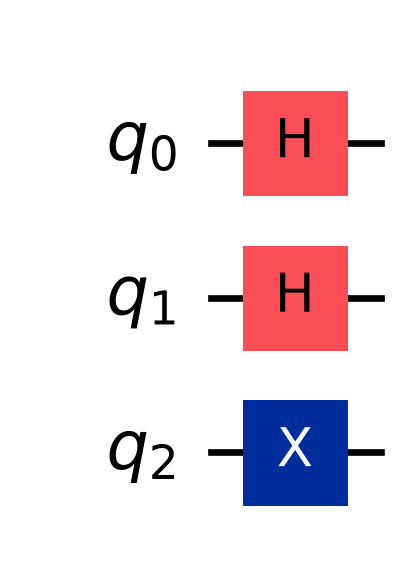}
  {Square: $f_{[2:6]}=1$, $N=8$}
  {unused}
  {0.4}{ex_square}

\subsection{Fourier}
A superposition of $T$ sinusoidal modes, each parameterized by
frequency $n_t$, amplitude $A_t$, and phase $\varphi_t$:
\begin{equation}
  f_i = \sum_{t=1}^{T} A_t \sin\!\left(\frac{2\pi n_t i}{N} + \varphi_t\right)
\end{equation}
Constructor: \texttt{FOURIER(modes=[(n1, A1, phi1), ...])}.
Circuit complexity: $\mathcal{O}(m^2)$ via the inverse Quantum Fourier
Transform~\cite{gonzalezconde2024,moosa2024}.
The DFT of each mode contributes exactly two nonzero coefficients
(a complex conjugate pair at $\pm n_t$), so the full vector has $2T$
nonzero DFT coefficients. PyEncode prepares this sparse Fourier state
using the \texttt{SPARSE} synthesizer and applies the inverse QFT,
yielding an $\mathcal{O}(m^2)$ circuit dominated by the QFT for
$T \ll m$; the \texttt{SPARSE} sub-circuit contributes
$\mathcal{O}(Tm)$ gates, which is subdominant when
$T = \mathcal{O}(1)$.
The single-mode case ($T=1$) subsumes sine ($\varphi=0$) and cosine
($\varphi=\pi/2$) as special cases of the same circuit.

\textbf{Example.}

A pure sine wave:
\begin{lstlisting}[style=python]
circuit, info = encode(FOURIER(modes=[(1, 1.0, 0)]), N=16)
# info.complexity  -> "O(m^2)"
\end{lstlisting}
Figure~\ref{fig:ex_fourier} shows the sinusoidal vector and inverse-QFT circuit.

\loadfigsv{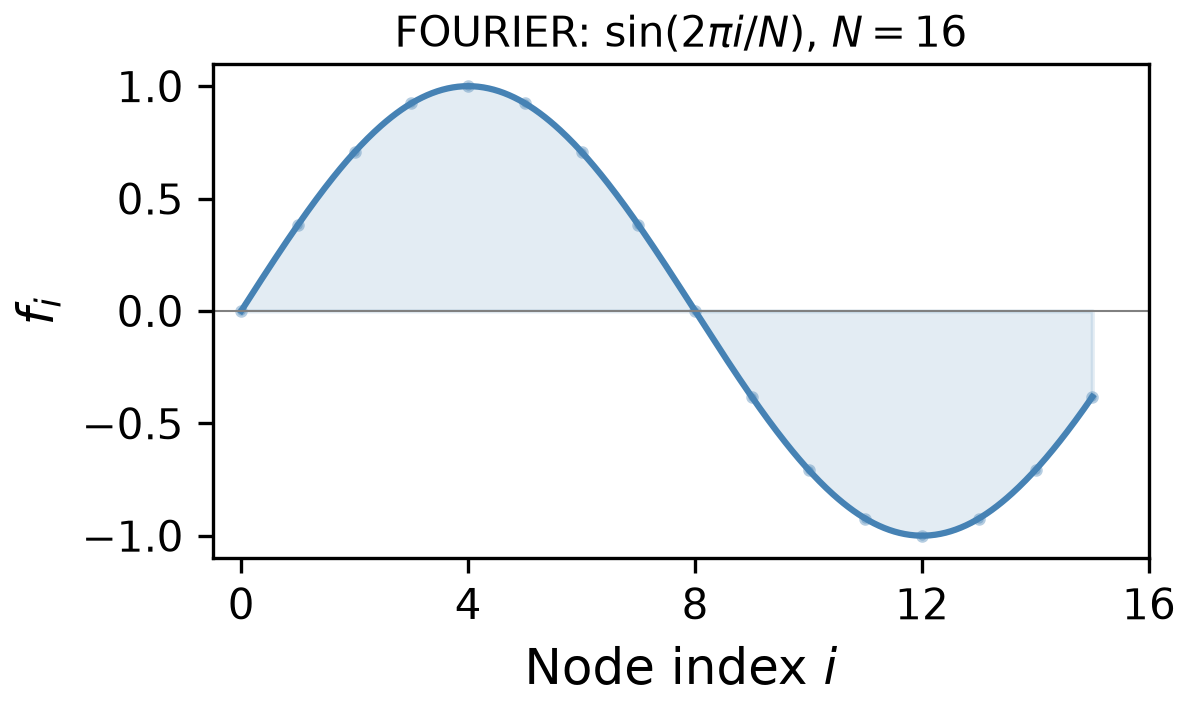}{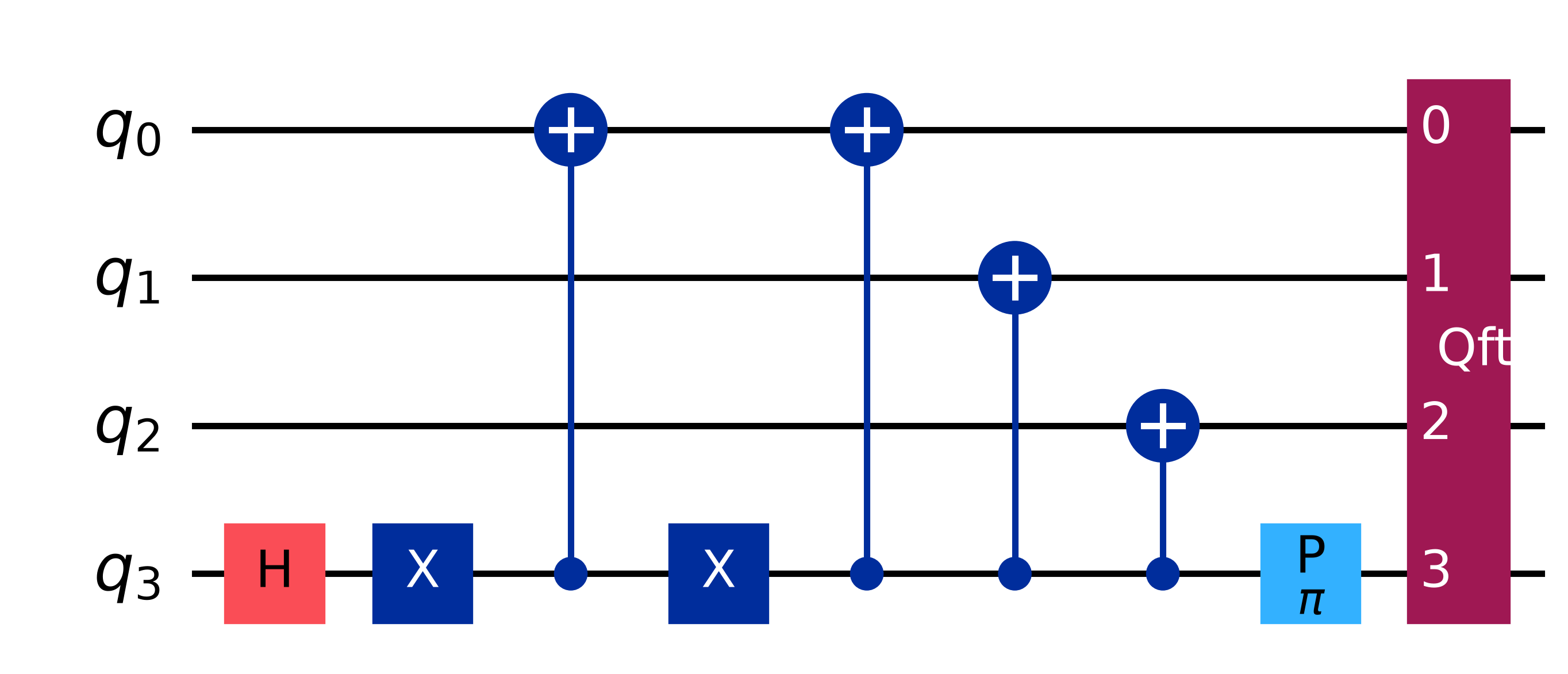}
  {Fourier ($T=1$): $\sin(2\pi i/N)$, $N=16$}
  {0.5}
  {0.7}{ex_fourier}

\subsection{Walsh}
Walsh functions are the natural ``binary cousins'' of sines and cosines.
They capture a two-level state determined by a single bit of the index:
\begin{equation}
  f_i = \begin{cases} c_0 & b_k(i) = 0 \\ c_1 & b_k(i) = 1, \end{cases}
  \qquad i \in \{0, 1, \ldots, N-1\},
\end{equation}
where $b_k(i)$ denotes bit $k$ of $i$ (LSB convention, $0 \le k < m$).
The amplitude is constant on blocks of $2^k$ consecutive indices,
alternating between $c_0$ and $c_1$ as $i$ crosses each $2^k$-boundary.

Constructor: \texttt{WALSH(k, c0, c1)}.
\textcolor{black}{Circuit complexity: $\mathcal{O}(m)$ ($m+1$ gates as emitted; the
final $R_y$--$H$ pair on qubit~$k$ fuses to a single $U$ under
\texttt{optimization\_level=3}, giving the $m$ post-transpile
count reported in Table~\ref{tab:gatecounts}).}
The circuit is a single-qubit rotation on qubit $k$ followed by
$H^{\otimes m}$~\cite{welch2014}.
The rotation angle is
\begin{equation}
  \theta = 2\,\mathrm{atan2}(c_0 - c_1,\; c_0 + c_1),
\end{equation}
chosen so that $R_y(\theta)|0\rangle = \cos(\theta/2)|0\rangle + \sin(\theta/2)|1\rangle$
distributes amplitude in the ratio $c_0 : c_1$ after the Hadamard layer.
When $c_1 = -c_0$ (the default), $\theta = \pi$ and $R_y(\pi) = X$,
recovering the standard Walsh function (signed uniform superposition).
For complex $c_0, c_1$ the same construction applies with one
additional phase gate on qubit $k$ to encode the relative phase
between the two levels; the cost remains $m+1$ rotations plus at most
one extra phase gate.

\textbf{Example.}

The standard form ($c_1=-c_0$) is a signed uniform superposition;
the generalized form encodes two distinct levels without ancilla.
Both use the same $m+1$-gate circuit~\cite{welch2014}:
\begin{lstlisting}[style=python]
# Two distinct levels, no ancilla
circuit, info = encode(WALSH(k=2, c0=1.0, c1=4.0), N=8)
# f = [1,1,1,1,4,4,4,4] / ||f||
# info.gate_count  -> 4   (still m+1)
\end{lstlisting}
Figure~\ref{fig:ex_walsh} shows the two-level state and $m+1$-gate circuit.
\loadfigs{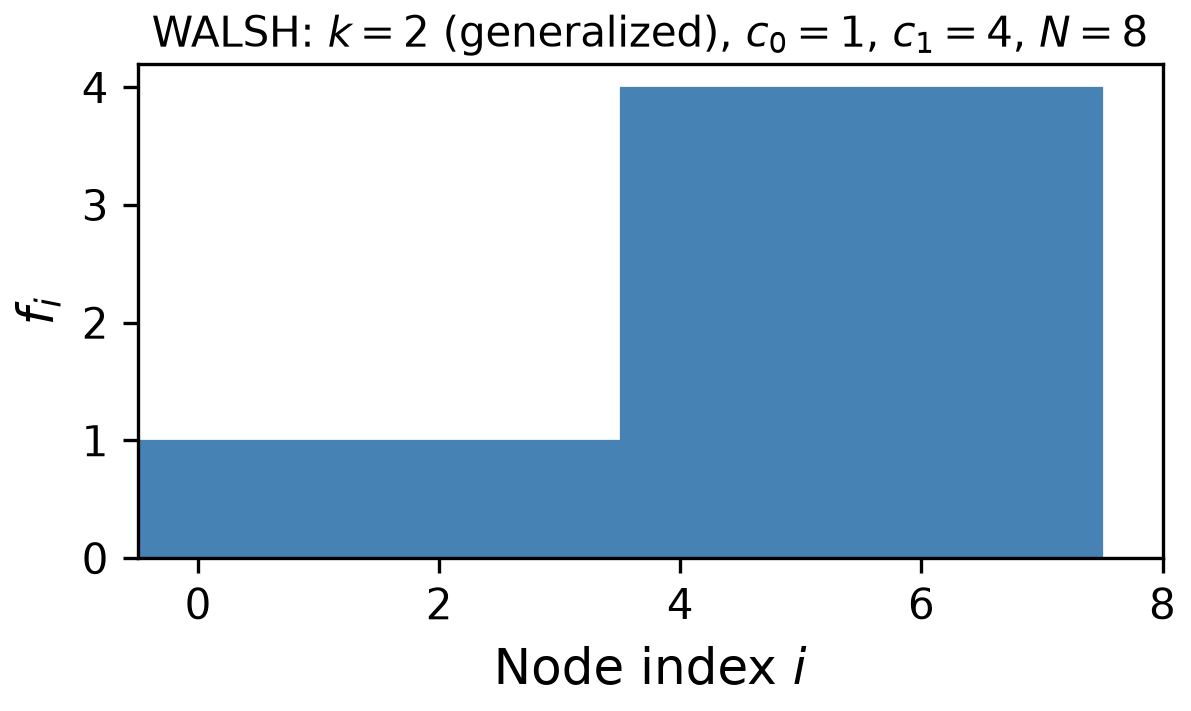}{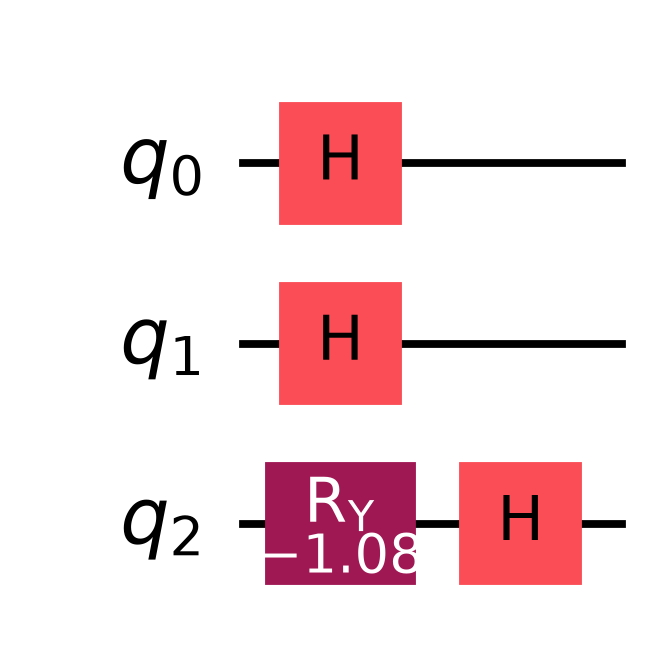}
  {Walsh $k=2$ (generalized): $c_0=1$, $c_1=4$, $N=8$}
  {unused}
  {0.6}{ex_walsh}

\subsection{Geometric}
\label{sec:geometric}
An exponential decay (or growth) sequence, optionally offset so that
the nonzero window begins at an arbitrary index $k_s$:
\begin{equation}
  f_i = \begin{cases} c\,r^{\,i-k_s} & i \ge k_s \\ 0 & i < k_s, \end{cases}
  \qquad r \in \mathbb{C}\setminus\{0\},\; r\neq 1.
\end{equation}
Constructor: \texttt{GEOMETRIC(r, k\_s=0, c=1)}.

\textbf{Example $k_s = 0$.}

For $k_s=0$ the cost is $\mathcal{O}(m)$ --- exactly $m$
single-qubit gates, zero two-qubit gates, and circuit depth~1. \textcolor{black}{Qiskit's transpiler at \texttt{optimization\_level=3} may drop
below $m$ gates: rotation
angles $\theta_j = 2\arctan(r^{2^j})$ vanish super-exponentially
in $j$ once $|r|^{2^j}$ falls below the transpiler's angle
threshold, at which point the corresponding gates are absorbed
as no-ops.  For example, at $r=0.95$ and $m=12$ the transpiled
circuit contains 9 single-qubit gates (Table~\ref{tab:gatecounts}),
not the raw 12 emitted by the synthesizer, because the top three
qubits' rotations are numerically indistinguishable from the
identity.}
The product-state structure of geometric sequences is a direct
consequence of the binary representation of the index~\cite{grover2002creating}.
The full circuit is $m$ independent single-qubit rotations with no
entangling gates and no ancilla qubits.
Complex $r$ is handled by adding a single phase gate per qubit; real
positive $r$ skips this layer and recovers the original
zero-entangling-gate, depth-$1$ circuit unchanged.  A complex
prefactor $c$ contributes a global phase.
\begin{lstlisting}[style=python]
# k_s = 0: product state, depth 1
circuit, info = encode(GEOMETRIC(r=0.5), N=8)
print(info)
\end{lstlisting}

\begin{lstlisting}[basicstyle=\ttfamily\small\color{black},]
----- output --------------
PyEncode  v3.0.0
  Pattern     : GEOMETRIC
  N           : 8  (m = 3 qubits)
  Gate count  : 3
  Complexity  : O(m)
  Validated   : yes
  Gates 1q/2q : 3 / 0
  Depth       : 1
  Success prob: 1.0
  Vector      : numpy array, shape (8,)
  Parameters  : {'r': 0.5, 
            'k_s': 0,
            'k_e': 8, 'c': 1.0}
  Circuit code: 377 chars 
\end{lstlisting}

Figure~\ref{fig:ex_geometric} shows the $k_s=0$ vector and
its depth-1 circuit.

\loadfigs{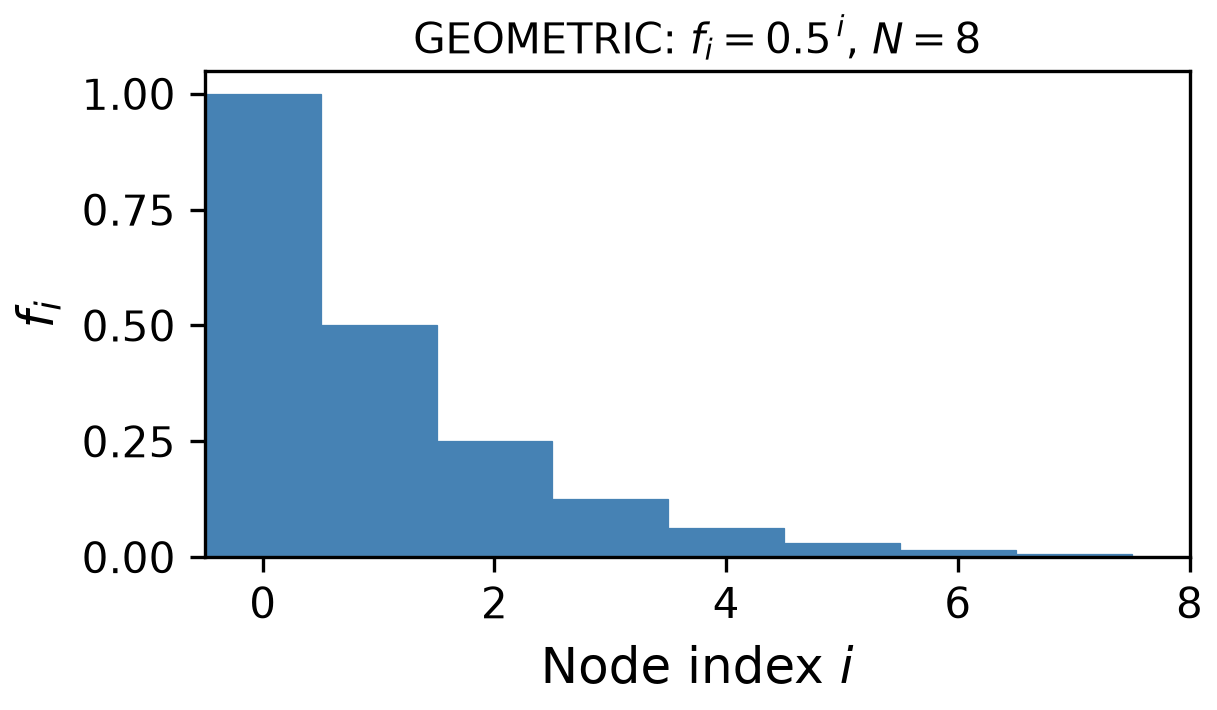}{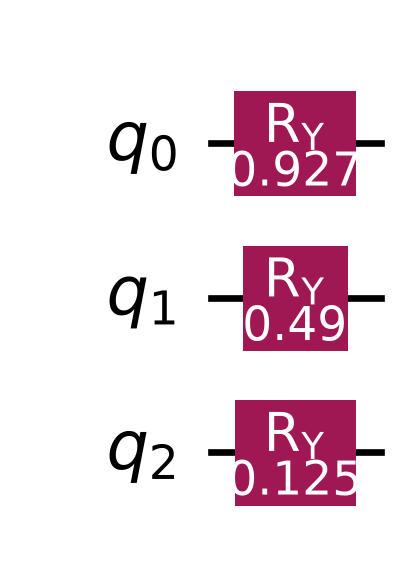}
  {Geometric: $r=0.5$, $N=8$}
  {unused}
  {0.4}{ex_geometric}

\textbf{Example (complex $r$): discrete plane wave.}

Setting $r = e^{i\omega}$ with $|r|=1$ produces the discrete plane
wave $f_i = c\,e^{i\omega i}$, the natural output of a Hadamard
transform on a Fourier-momentum eigenstate and a building block for
quantum signal processing.  PyEncode encodes it as the same depth-1
product state as the real-amplitude case:
\begin{lstlisting}[style=python]
import cmath
omega = 0.7
circuit, info = encode(GEOMETRIC(r=cmath.exp(1j*omega)), N=64)
# info.gate_count -> 6   (m gates, 0 CX, depth 1)
# info.complexity -> "O(m)"
\end{lstlisting}
Figure~\ref{fig:ex_geometric_planewave} shows the real and imaginary
parts of the prepared state and the corresponding depth-1 circuit.

\loadfigs{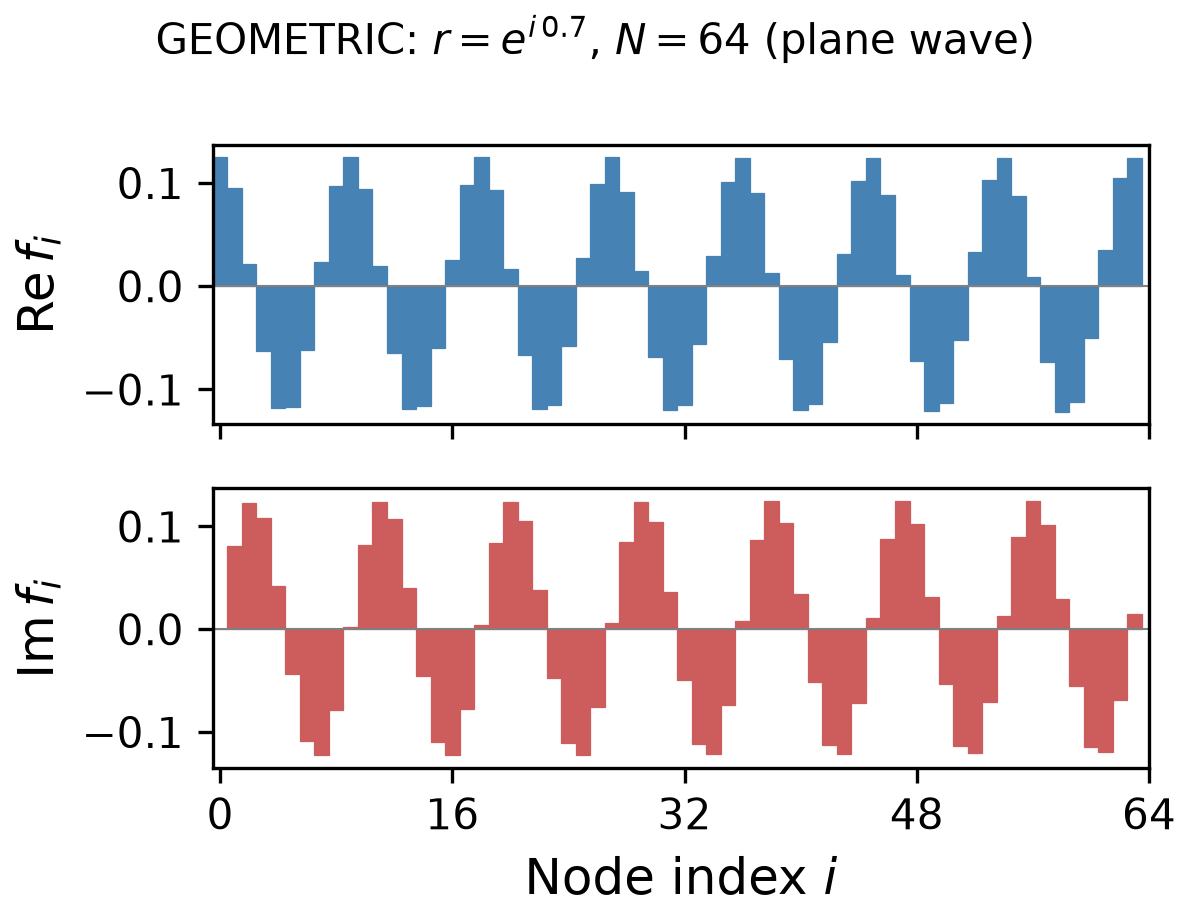}{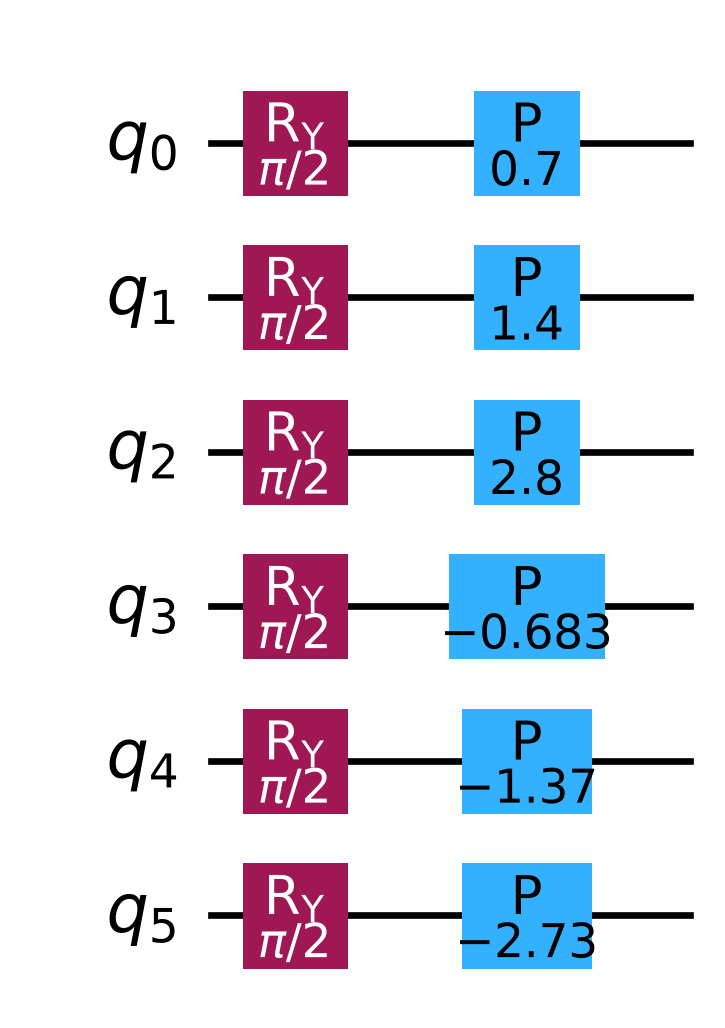}
  {Plane wave: $r=e^{i\,0.7}$, $N=64$ --- Re and Im parts (depth 1)}
  {unused}
  {0.75}{ex_geometric_planewave}

\textbf{Example $k_s>0$.}

For $k_s>0$ the target state has support only on
$[k_s, N)$, and the product-state decomposition above no longer applies
directly (the nonzero window does not coincide with the full register).
A special case remains cheap: when the window width $w = N - k_s$ is a
power of two \emph{and} $k_s \bmod w = 0$, the construction on the lower
$\log_2 w$ qubits is still the product state above, augmented by
$X$-gates on the upper qubits to fix the window's location.  This
covers $k_s \in \{N/2,\, 3N/4,\, 7N/8,\, 15N/16, \ldots\}$ (i.e.\
$k_s = N(1 - 2^{-p})$ for $p = 1,2,\ldots$) and retains the
$\mathcal{O}(m)$ cost.

The general case uses a \emph{dyadic
decomposition}~\cite{bentley1980} of the half-open interval $[k_s, N)$. Total cost: $\mathcal{O}(m^2)$, no ancilla, success probability one
(the block supports are disjoint by construction).
\begin{lstlisting}[style=python]
# Arbitrary k_s: dyadic assembly, unit success probability
circuit, info = encode(GEOMETRIC(r=0.8, k_s=5), N=16)
# Support [5,16) = [5,6) U [6,8) U [8,16): three aligned blocks (L=3)
# info.gate_count           -> 24
# info.complexity           -> "O(m^2)"
# info.success_probability  -> 1.0
\end{lstlisting}
Figure~\ref{fig:ex_geometric_arbitrary} shows the offset decay and its
dyadic-assembly circuit; the latter is visibly denser than the
depth-1 $k_s=0$ case.
  
\loadfigsv{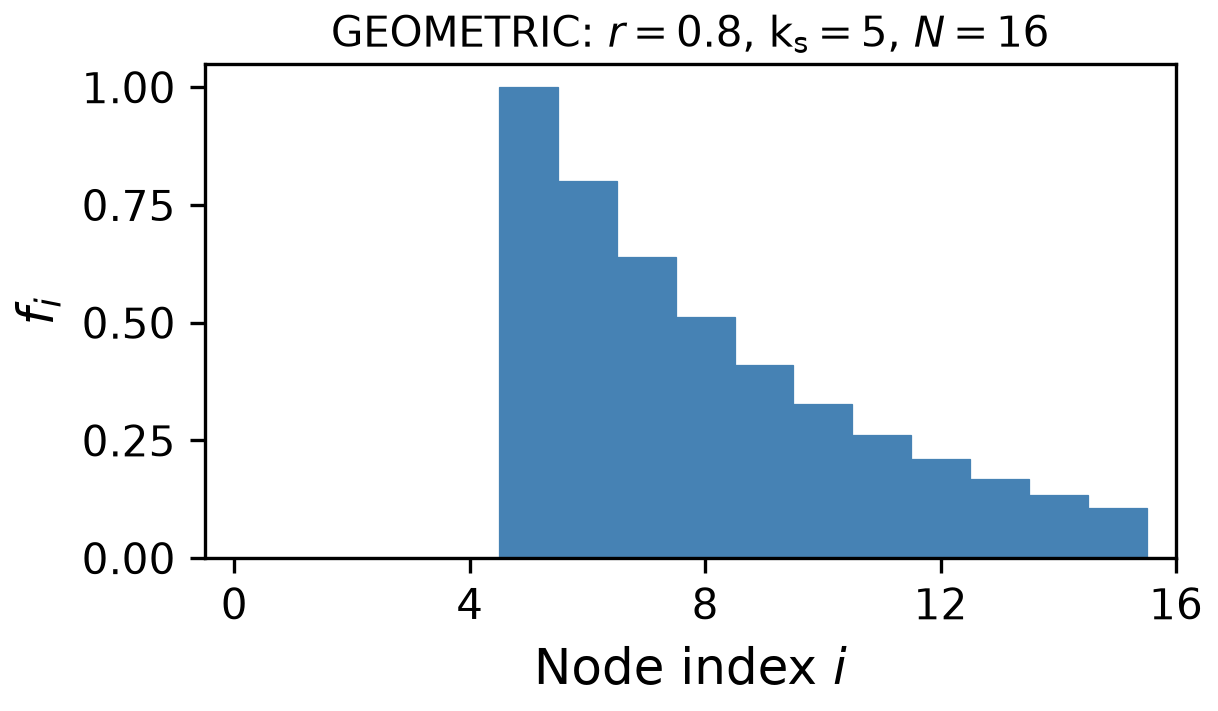}{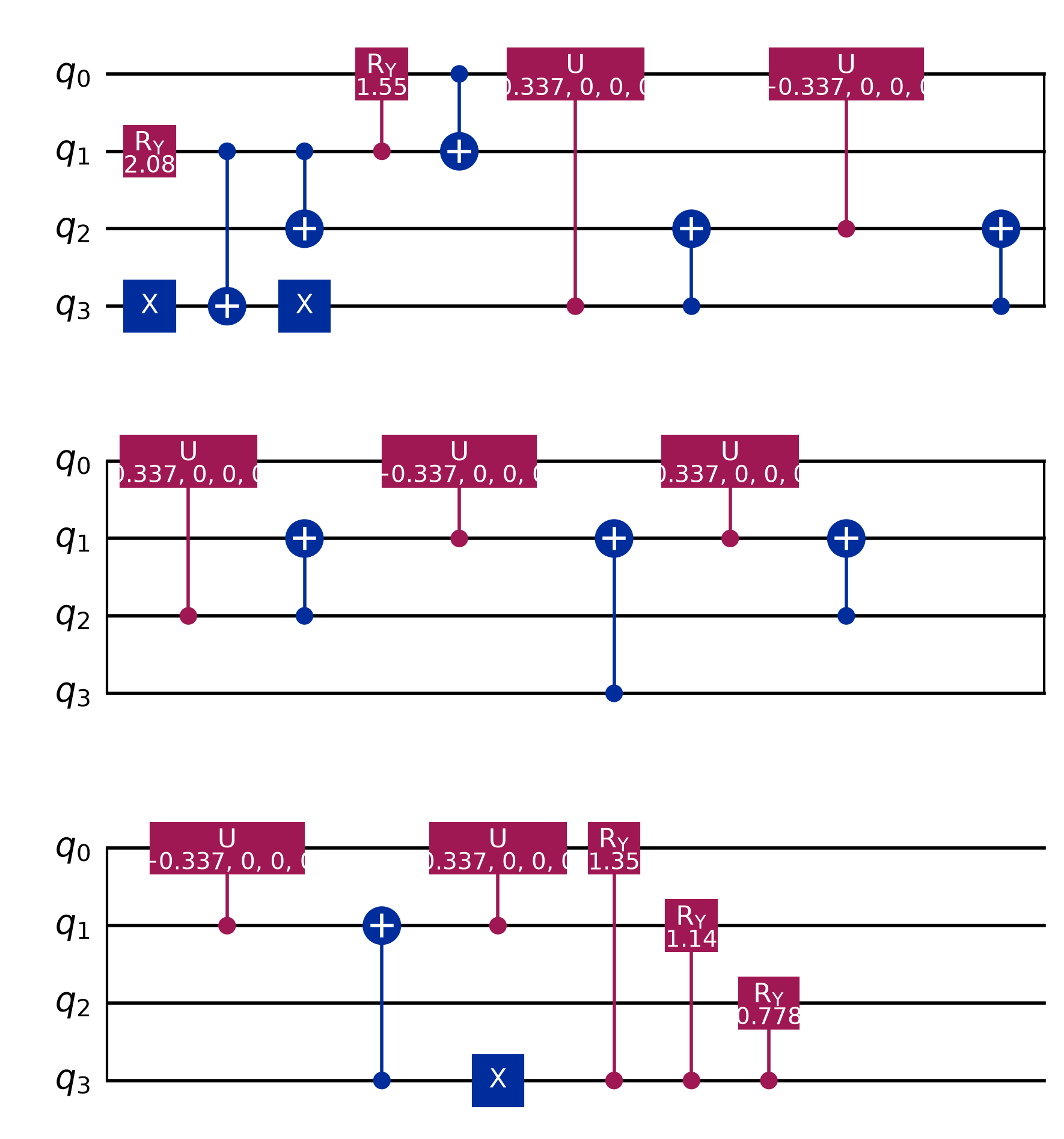}
{Geometric: $r=0.8$, $k_s=5$, $N=16$ --- three dyadic blocks}
{0.5}{0.85}
{ex_geometric_arbitrary}

\subsection{Hamming}
\label{sec:hamming}
A product state whose amplitude depends only on the number of
$1$-bits (Hamming weight) in the index:
\begin{equation}
  f_i = c\,r^{\,\mathrm{wt}(i)}, \qquad r \in \mathbb{C}\setminus\{0\},\; r\neq 1.
\end{equation}
Constructor: \texttt{HAMMING(r, c)}.
Circuit complexity: exactly $m$ single-qubit $R_y$ gates, \emph{zero}
two-qubit gates, and depth~1.

The Hamming construction is the constant-ratio specialization of the
geometric product-state decomposition (Section~\ref{sec:geometric}).
Setting the per-qubit ratio in the \textsc{Geometric} construction
($r^{2^j}$) to a constant $r$ across all qubits yields exactly the
\textsc{Hamming} state. PyEncode surfaces \textsc{Hamming} as a named pattern for notational clarity and
to make it directly composable with \textsc{Sum}, \textsc{Partition},
and \textsc{Tensor}.
For complex $r$ each qubit picks up an additional phase gate, retaining
depth $1$; for real positive $r$ no phase gate is emitted.

\textbf{Example.}

\begin{lstlisting}[style=python]
circuit, info = encode(HAMMING(r=0.5), N=16)
# info.gate_count   -> 4   (m gates, 0 CX, depth 1)
\end{lstlisting}

\loadfigs{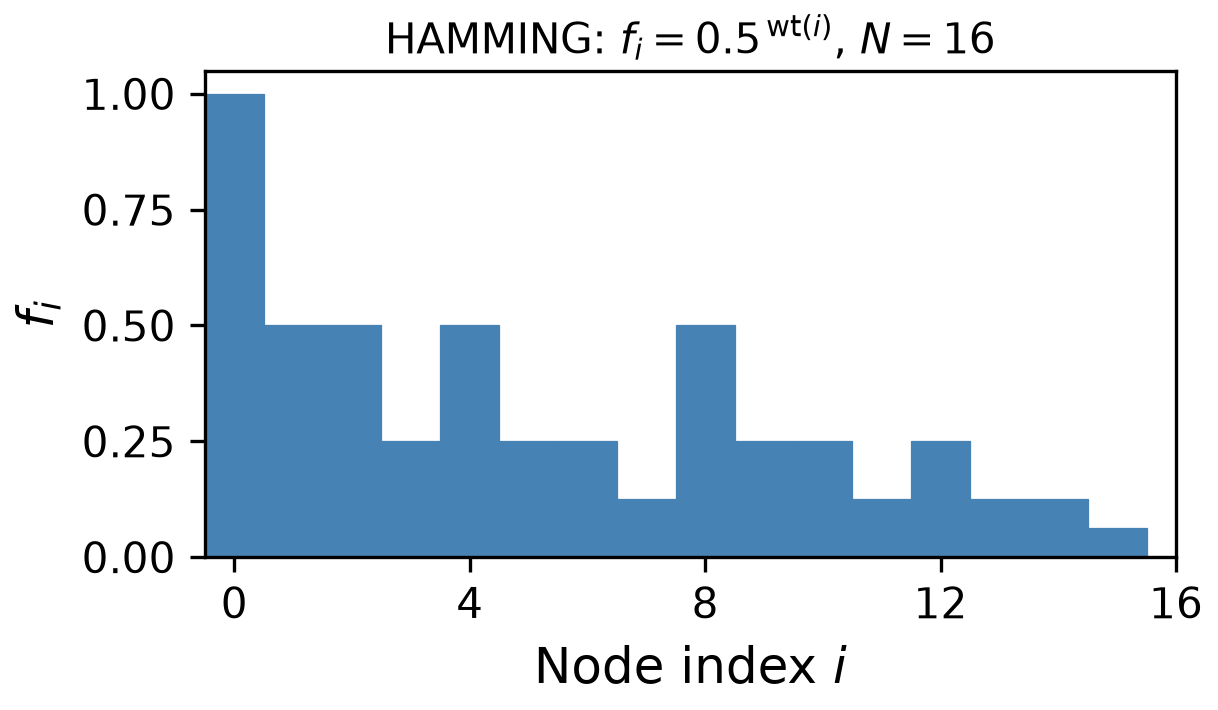}{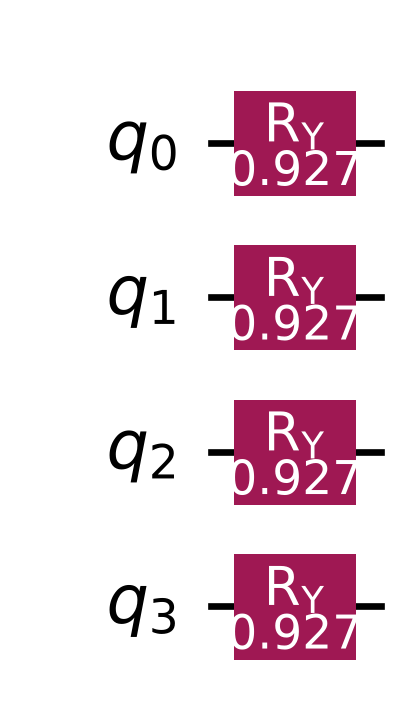}
  {Hamming: $f_i = 0.5^{\mathrm{wt}(i)}$, $N=16$}
  {unused}
  {0.45}{ex_hamming}

\subsection{Staircase}
\label{sec:staircase}
A sparse geometric sequence supported only on the unary indices
$i = 2^k - 1 \in \{0, 1, 3, 7, 15, \ldots, N{-}1\}$:
\begin{equation}
  f_{2^k - 1} = c\,r^{\,k}, \qquad k = 0, 1, \ldots, m,
\end{equation}
and $f_i = 0$ for all other indices. Constructor:
\texttt{STAIRCASE(r, c)}. Circuit complexity: $\mathcal{O}(m)$
with $m$ total gates (one $R_y$ and $m{-}1$ controlled $R_y$) and depth
$\mathcal{O}(m)$.

The target state $|\psi\rangle = \sum_{k=0}^{m}\alpha_k |2^k - 1\rangle$
is prepared by a staircase of controlled rotations. Starting from
$|0\rangle^{\otimes m}$, qubit $0$ is rotated by $R_y(\theta_0)$
to create the first branch; subsequent qubits are rotated by
$\mathrm{CR}_y(\theta_k)$ controlled on qubit $k{-}1$. The angles
$\theta_k$ are fixed by the residual norms:
\begin{equation}
  \theta_k = 2\arctan\!\left(
    \frac{\sqrt{\sum_{j\geq k+1}\alpha_j^2}}{\alpha_k}
  \right),
\end{equation}
producing exactly the geometric ratio $r$ at each step.  The cascade
structure is a specialization of the general controlled-rotation state
preparation tree of Möttönen et al.~\cite{mottonen2005transformation}
to a single-branch sparse support.  Similar constructions underlie
the standard W-state preparation~\cite{cruz2019efficient} (uniform
weights $\alpha_k = 1/\sqrt{m}$) and the tensor-tree hierarchies of
Hackbusch~\cite{hackbusch1999}; PyEncode surfaces the named
\texttt{STAIRCASE} pattern with $R_y$ angles set analytically from
the geometric prefactor, avoiding a separate parameter-tree traversal.
Staircase profiles appear as coarse-to-fine wavelet hierarchies in
multigrid solvers and as geometrically decaying refinement levels in
adaptive FEM.  For complex $r$ the cascade angles are computed
from $|r|$ and a final phase layer (one phase gate per qubit) recovers
the correct argument on the staircase support; the original cascade is
recovered exactly when $r$ is real positive.

\textbf{Example.}
\begin{lstlisting}[style=python]
circuit, info = encode(STAIRCASE(r=0.5), N=16)
# f_0=1, f_1=0.5, f_3=0.25, f_7=0.125, f_15=0.0625, else 0
# info.gate_count  -> 4
\end{lstlisting}

\loadfigs{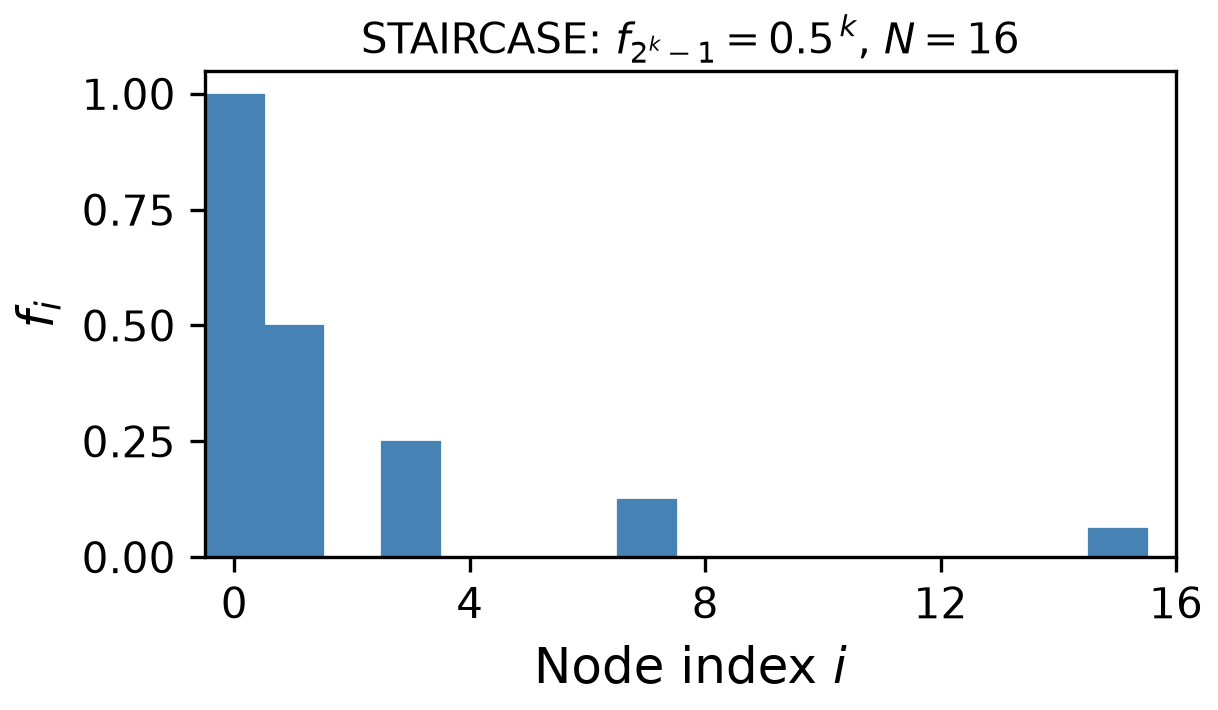}{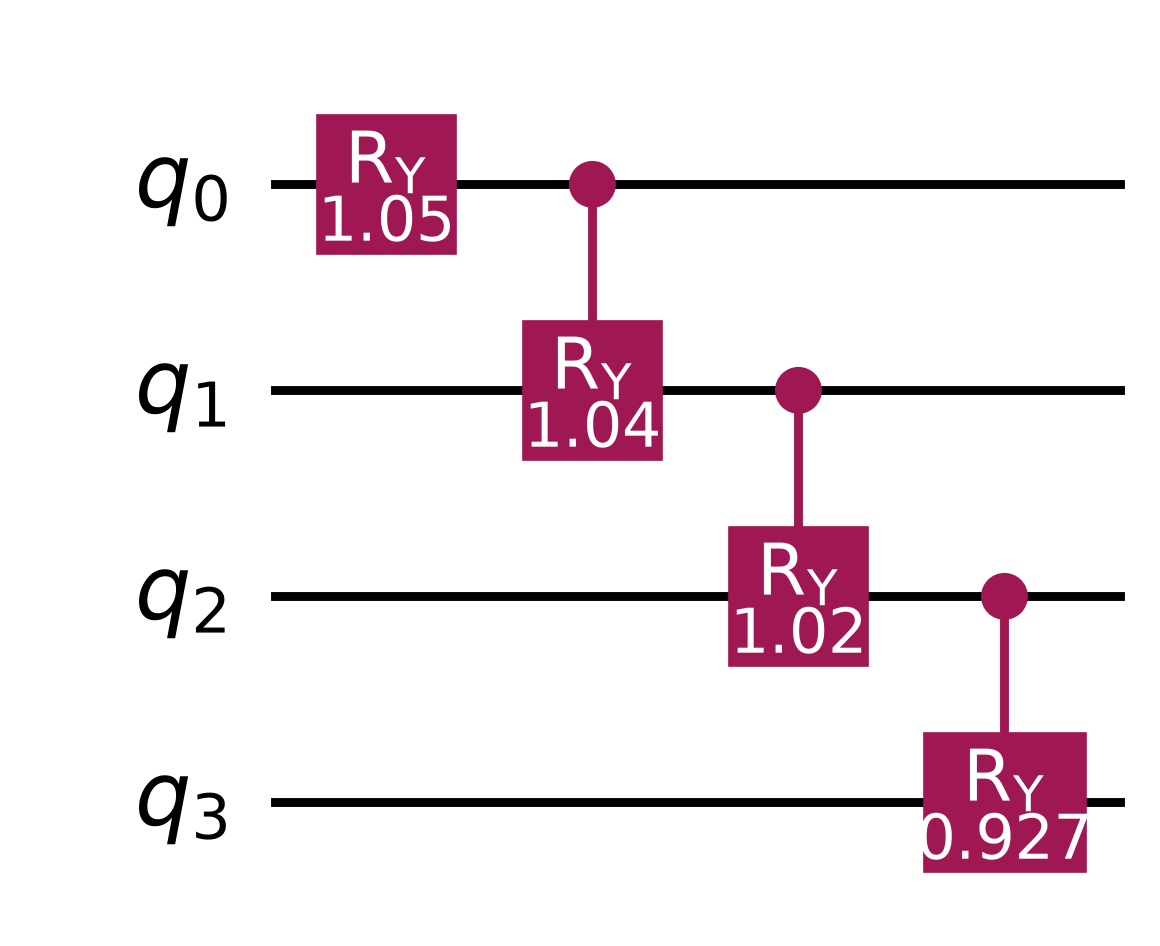}
  {Staircase: $f_{2^k-1} = 0.5^k$, $N=16$}
  {unused}
  {1}{ex_staircase}

\subsection{Dicke}
\label{sec:dicke}
The Dicke state $|D^m_k\rangle$ is the uniform superposition over all
$m$-qubit computational-basis states of Hamming weight exactly $k$:
\begin{equation}
  |D^m_k\rangle \;=\; \binom{m}{k}^{-1/2}\!\!\sum_{\substack{S\subseteq\{0,\ldots,m-1\}\\|S|=k}}\! |e_S\rangle,
  \qquad 0 \leq k \leq m,
\end{equation}
so $f_i = c\,\mathbf{1}[\mathrm{wt}(i) = k]$ --- constant on the
weight-$k$ sphere, zero off it.  Unlike \textsc{Hamming}
(Section~\ref{sec:hamming}), which is a product state with geometric
decay across weight classes, \textsc{Dicke} is genuinely entangled and
supported on a single weight class.  Constructor: \texttt{DICKE(k, c)}.
Circuit complexity: $\mathcal{O}(k(m-k))$ two-qubit gates and
$\mathcal{O}(m)$ depth, ancilla-free with unit success probability. PyEncode implements the deterministic cascade of Bärtschi and
Eidenbenz~\cite{baertschi2019dicke}. 

\textbf{Symmetry optimization.}

Bitwise complementation on the computational basis gives the identity
\begin{equation}
  |D^m_k\rangle \;=\; X^{\otimes m}\,|D^m_{m-k}\rangle,
  \label{eq:dicke_symmetry}
\end{equation}
which PyEncode exploits for $k > m/2$: it synthesizes the lighter
Dicke state $|D^m_{m-k}\rangle$ using $k' = m - k$ inside the cascade,
then appends $X^{\otimes m}$.  The Qiskit transpiler at
\texttt{optimization\_level=3} absorbs this final $X$-layer into
adjacent rotations, so $k$ and $m-k$ produce \emph{identical}
transpiled CX counts and circuit depth.

\textbf{Special cases.}

$k = 0$ returns the empty circuit preparing $|0\rangle^{\otimes m}$;
$k = m$ uses $m$ $X$ gates to prepare $|1\rangle^{\otimes m}$.  The
worst case is $k = m/2$, where $k(m-k) = m^2/4$, placing
\textsc{Dicke} between the $\mathcal{O}(m)$ and $\mathcal{O}(m^2)$
tiers depending on $k$.

\textbf{Example.}
\begin{lstlisting}[style=python]
circuit, info = encode(DICKE(k=2), N=16)
# info.complexity    -> "O(k*(m-k))"
# info.gate_count_2q -> 24   (m=4, k'=2)
\end{lstlisting}

\loadfigsv{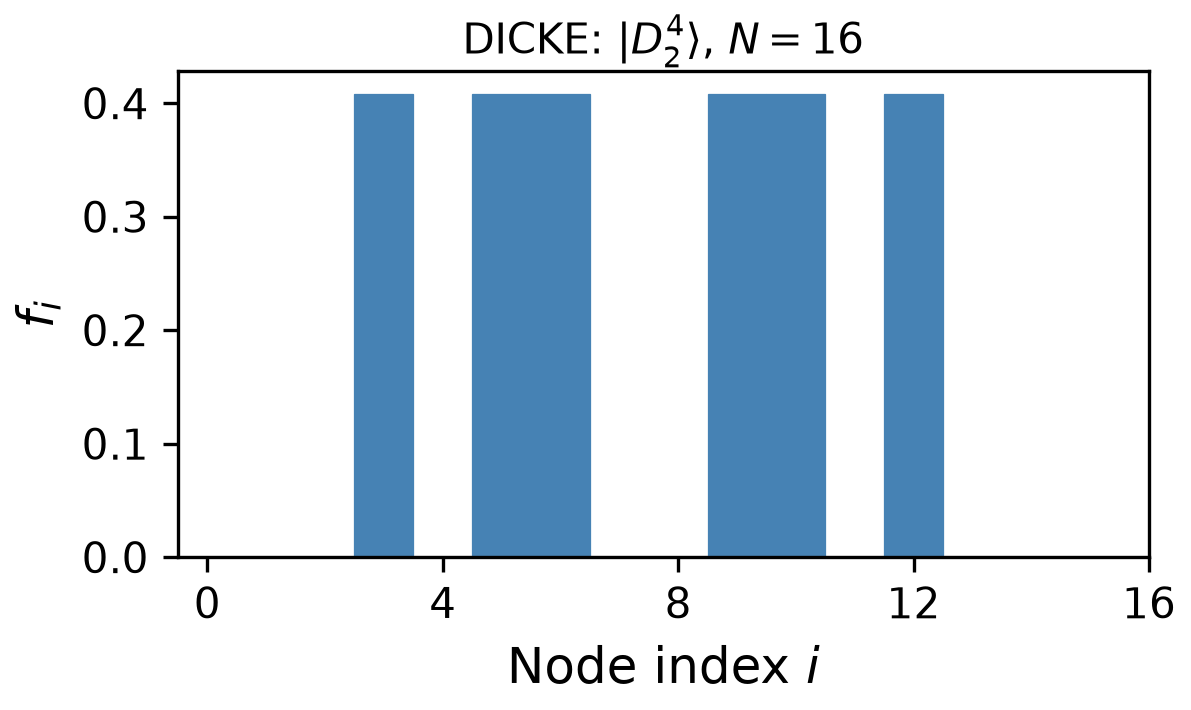}{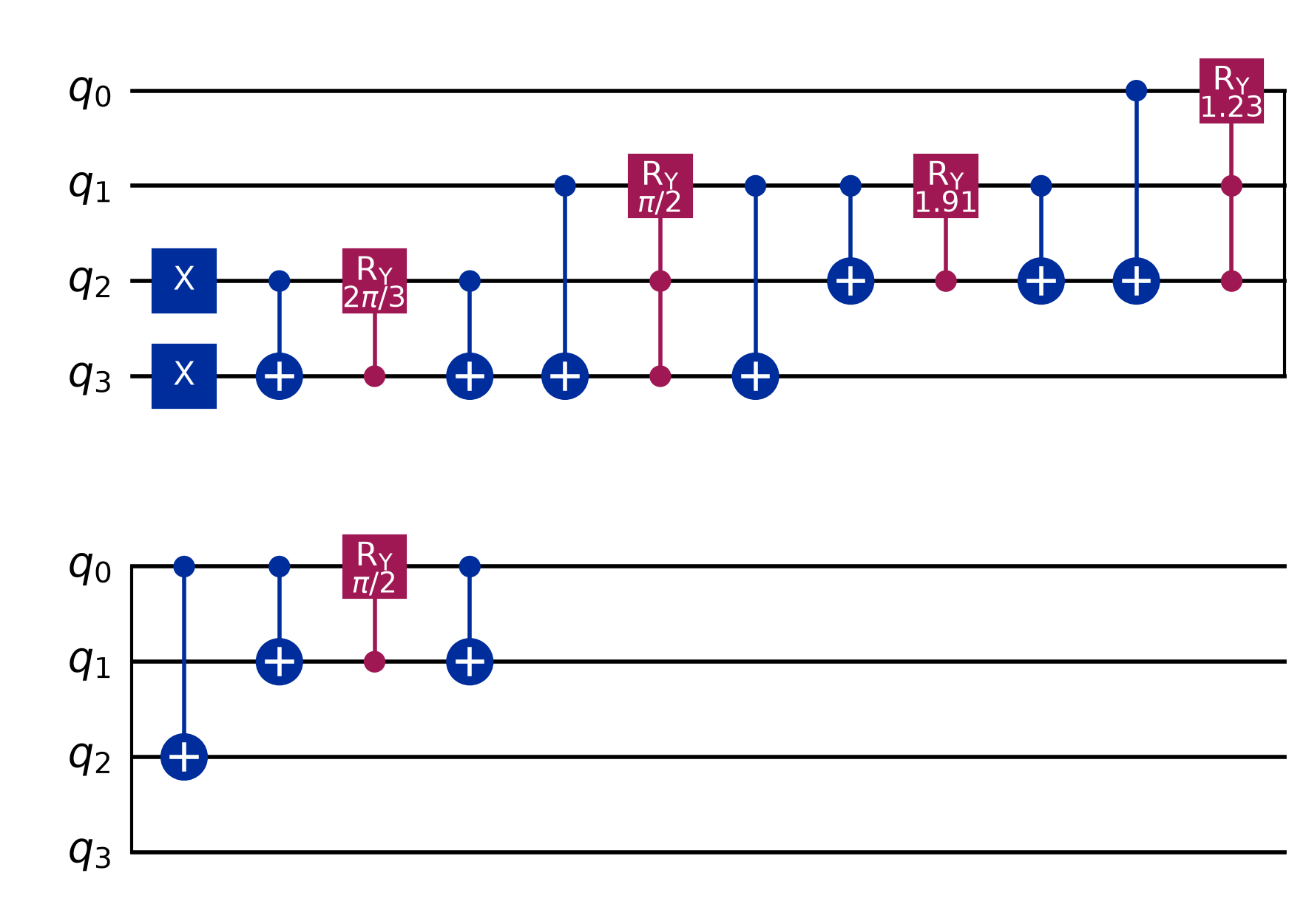}
  {Dicke: $|D^4_2\rangle$ on $N=16$
   (uniform over the $\binom{4}{2}=6$ weight-$2$ indices)}
  {0.5}
  {0.75}{ex_dicke}

\subsection{Polynomial}
\label{sec:polynomial}
A general polynomial function of the grid variable $x = i/(N-1)$
on the unit interval:
\begin{equation}
  f_i = \sum_{j=0}^{d} c_j\,\left(\frac{i}{N-1}\right)^j,
  \qquad \text{degree }d.
\end{equation}
Constructor: \texttt{POLYNOMIAL(coeffs=[$c_0,\ldots,c_d$])}.
Circuit complexity: $\mathcal{O}(m^{d+1})$ gates, providing
exact encoding for ramp ($d=1$), parabolic ($d=2$), cubic ($d=3$),
and higher-degree profiles.

\textbf{Construction.}

The circuit exploits the \emph{Walsh sparsity} of polynomial functions,
a classical result due to Welch et al.~\cite{welch2014} and central
to the polynomial-encoding framework of Gonzalez-Conde et
al.~\cite{gonzalezconde2024}:
\begin{quote}
\emph{If $f$ is a degree-$d$ polynomial in $i$, its Walsh-Hadamard
transform has support only on indices of Hamming weight at most $d$.}
\end{quote}
The number of nonzero Walsh coefficients is therefore
$s = \sum_{k=0}^{d} \binom{m}{k} = \mathcal{O}(m^d)$, independent of $N$.
The synthesizer performs:
\begin{enumerate}[leftmargin=*, label=\arabic*., itemsep=0pt, topsep=2pt]
  \item Classical evaluation of $f$ on the grid and computation of the
        Walsh-Hadamard transform $\mathbf{x} = \mathbf{W}\mathbf{f}/\sqrt{N}$,
        retaining only the $\mathcal{O}(m^d)$ coefficients at
        Hamming-weight indices $\leq d$.
  \item Preparation of the sparse Walsh-coefficient register
        $|\psi_x\rangle = \sum_k x_k\,|k\rangle$ with real or complex
        amplitudes using the Gleinig--Hoefler sparse loader at
        $\mathcal{O}(s\,m)$ gate cost~\cite{gleinig2021}.  No separate
        phase-correction pass is required.
  \item A single layer of Hadamards $H^{\otimes m}$, which is
        self-inverse to the Walsh transform and maps
        $|\psi_x\rangle \mapsto \sum_i f_i |i\rangle$.
\end{enumerate}
The total circuit cost is $\mathcal{O}(m \cdot s) = \mathcal{O}(m^{d+1})$
gates, and the result is exact (no truncation or approximation). Measured transpiled cost for
the linear ramp at $m=12$ is 78 gates (56~$U$ + 22~CX) versus
Qiskit's 8{,}178 --- a 105$\times$ reduction; for the Poiseuille
profile ($d=2$) at $m=12$, 1{,}599 gates versus 8{,}094, widening to
14$\times$ by $m=14$.

\textbf{Example.}
\begin{lstlisting}[style=python]
# Ramp f(x) = x on x in [0,1]
circuit, info = encode(POLYNOMIAL(coeffs=[0.0, 1.0]), N=64)

# Poiseuille parabolic profile f(x) = 4x(1-x)
circuit, info = encode(POLYNOMIAL(coeffs=[0.0, 4.0, -4.0]), N=64)
\end{lstlisting}

\loadfigs{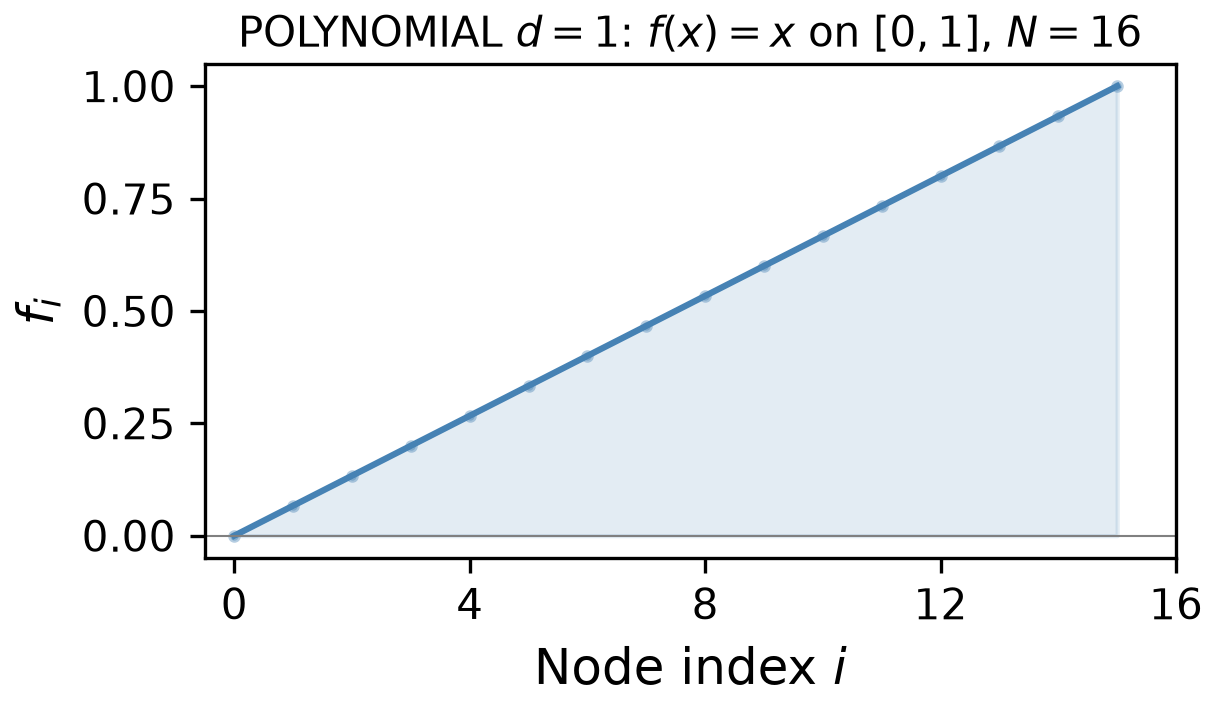}{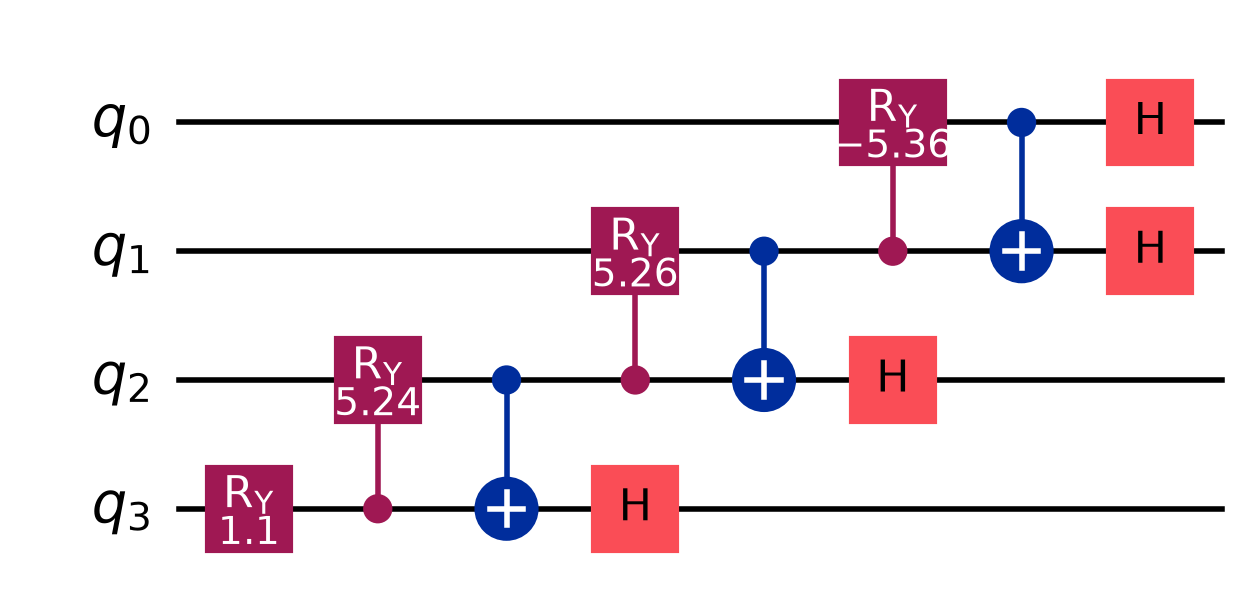}
  {Polynomial $d{=}1$: ramp $f(x) = x$, $N=16$}
  {unused}
  {1}{ex_poly_ramp}

\loadfigsv{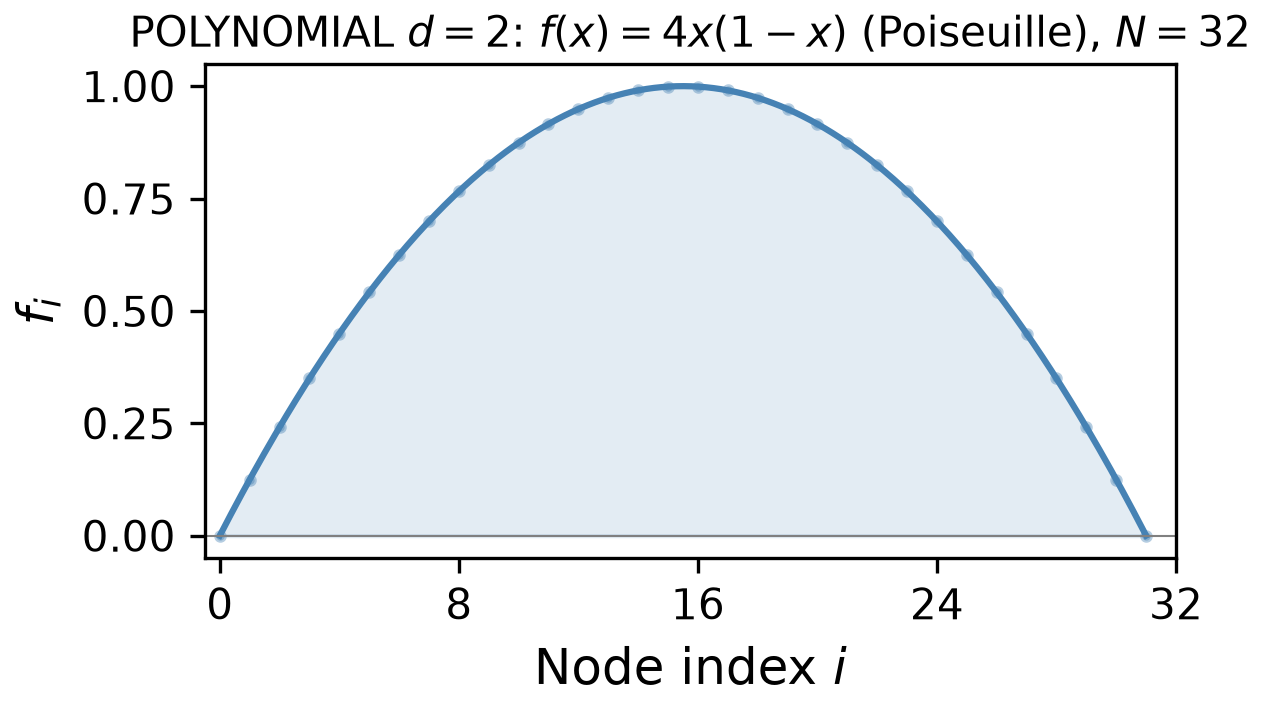}{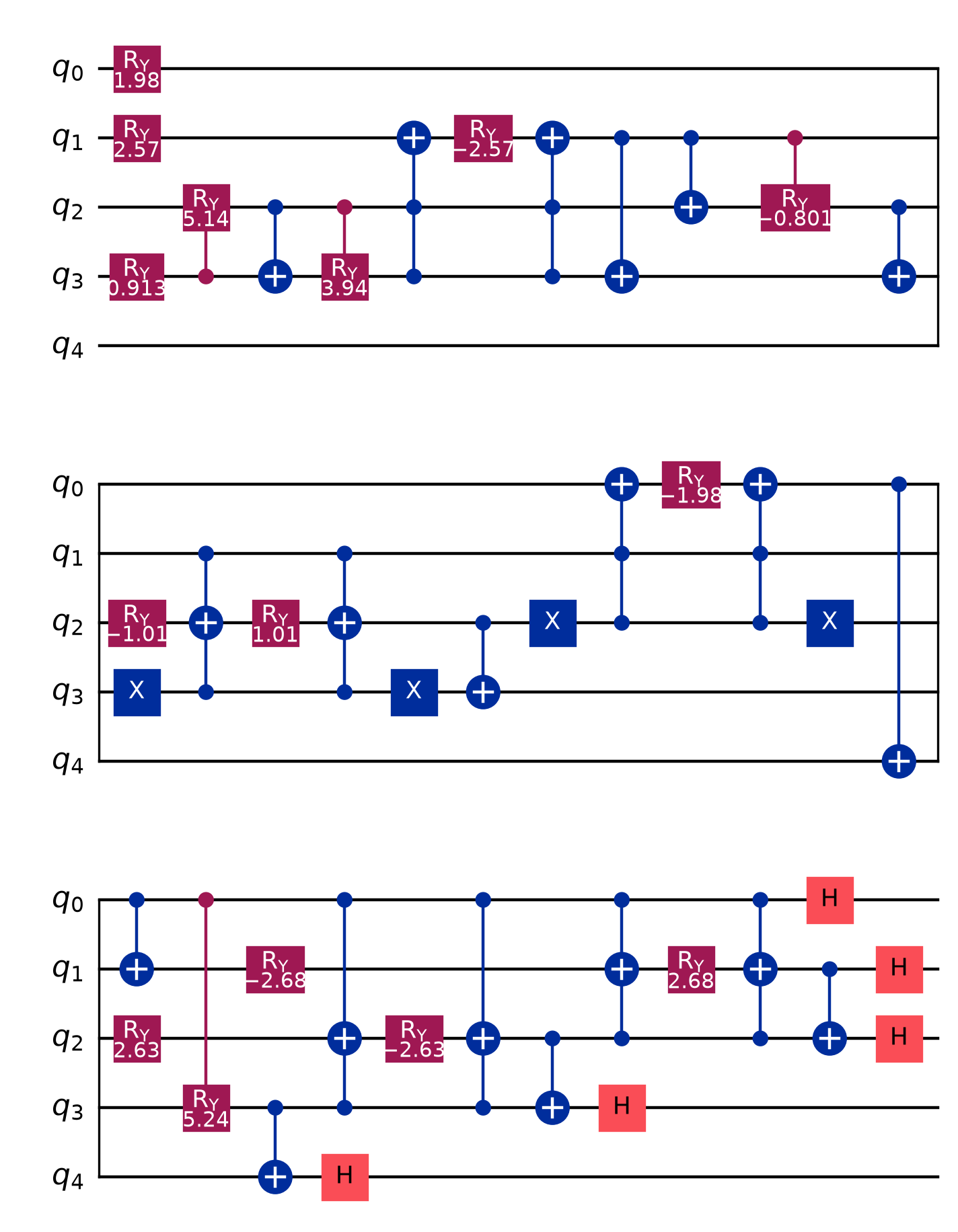}
{Polynomial $d{=}2$: Poiseuille $f(x) = 4x(1{-}x)$, $N=32$}
{0.5}{0.8}
{ex_poly_poiseuille}

\subsection{Sum (weighted superposition)}
\label{sec:sum}
A weighted superposition of $r$ component states:
\begin{equation}
  |\psi\rangle \propto \sum_{j=1}^{r} w_j\,|\hat{f}^{(j)}\rangle,
\end{equation}
where each $|\hat{f}^{(j)}\rangle$ is a normalized state prepared by
any PyEncode pattern.
Constructor: \texttt{SUM([(w1, pattern1), (w2, pattern2), \ldots])}.
The weights $w_j \in \mathbb{C} \setminus \{0\}$ may be arbitrary
nonzero complex numbers; the magnitudes $|w_j|$ are loaded by the
ancilla PREP register and the phases $\arg(w_j)$ are threaded through
the SELECT step.  Component patterns themselves admit real or complex
amplitudes (e.g.\ \texttt{SPARSE} entries with mixed signs or
\texttt{POLYNOMIAL} with complex coefficients), and any global phase
on a component circuit is composed correctly with the weight phase.

The \texttt{SUM} constructor implements the Linear Combination of
Unitaries (LCU) technique of Childs \& Wiebe~\cite{childs2018,
babbush2018}.

\subsubsection{Circuit construction}

The circuit implements LCU~\cite{childs2018, babbush2018}:
\begin{enumerate}[leftmargin=*, label=\arabic*., itemsep=0pt, topsep=2pt]
  \item \textbf{PREP}: a binary $R_y$-tree on
        $n_\text{anc} = \lceil\log_2 r\rceil$ ancilla qubits prepares
        the magnitude superposition
        $\sum_j \beta_j |j\rangle_\text{anc}$ with
        $\beta_j = \sqrt{|w_j|\|f^{(j)}\|}/Z$ and
        $Z^2 = \sum_j |w_j|\|f^{(j)}\|$ ($\beta_j$ are real and
        non-negative).
  \item \textbf{SELECT}: each component circuit $U_j$ is applied
        controlled on ancilla $|j\rangle$, with $\arg(w_j)$ baked
        into the inner global phase of $U_j$ so that the controlled
        operation realises $e^{i\arg(w_j)}\,U_j$.
  \item \textbf{PREP}$^\dagger$: invert the $R_y$-tree to uncompute
        the ancilla.
  \item \textbf{Post-select} ancilla on $|0\rangle^{\otimes n_\text{anc}}$.
\end{enumerate}

Conditional on the post-selection succeeding, the data register is in
the target state $\sum_j w_j |\hat{f}^{(j)}\rangle$ exactly --- no
truncation or approximation.  When the post-selection fails, the data
register is in an unrelated mixture; in that sense the circuit is
\emph{exact only on the post-selected $|0\rangle_\text{anc}$ subspace}
and probabilistic otherwise.  For applications that require
deterministic output, amplitude amplification on the ancilla flag
recovers the target state with $\mathcal{O}(1/\sqrt{p})$ overhead, or
the cheaper \texttt{PARTITION} constructor (Section~\ref{sec:partition})
should be used whenever the components have disjoint support.
For real positive weights, $\arg(w_j) = 0$ and step 2 reduces to the
classical LCU SELECT, recovering the original construction unchanged.

\subsubsection{Success probability}

The post-selection succeeds with probability
\begin{equation}
  p \;=\; \biggl\|\sum_j \beta_j^2\,e^{i\arg(w_j)}\,
                          |\hat{f}^{(j)}\rangle\biggr\|^2 \;\in\; [0, 1],
  \label{eq:sum_p}
\end{equation}
which depends both on the magnitudes $\beta_j$ (via PREP) and on the
inner products $\langle\hat{f}^{(i)}|\hat{f}^{(j)}\rangle$ between
component states.  For real positive weights the phase factors equal
one and~\eqref{eq:sum_p} reduces to the classical LCU expression
$\sum_{i,j}\beta_i^2\beta_j^2\langle\hat{f}^{(i)}|\hat{f}^{(j)}\rangle$.
$p = 1$ if and only if every weighted component
$e^{i\arg(w_j)}|\hat{f}^{(j)}\rangle$ is identical up to a real
non-negative scalar.

\subsubsection{Disjoint versus overlapping support}

PyEncode detects analytically whether the component vectors have
disjoint support:
\begin{itemize}[leftmargin=*, itemsep=0pt, topsep=2pt]
  \item \textbf{STEP} and \textbf{SQUARE}: support is an interval;
        disjointness is checked by interval non-overlap in
        $\mathcal{O}(r^2)$ time.
  \item \textbf{SPARSE}: support is an index set; disjointness is
        checked by set intersection.
  \item \textbf{WALSH} and \textbf{FOURIER}: support is always the
        full register $[0, N)$ --- never disjoint with anything.
\end{itemize}

For disjoint-support components, all cross terms in~\eqref{eq:sum_p}
vanish and the diagonal phase factors equal one, so $p = \sum_j
\beta_j^4$ \emph{independently} of the weight phases.
For $r$ equal-weight equal-norm components, $\beta_j = 1/\sqrt{r}$ and
$p = r\cdot(1/r)^2 = 1/r$.
For overlapping components the cross terms in~\eqref{eq:sum_p} are
generally non-zero, so $p$ depends on the relative geometry of the
component states; a \texttt{UserWarning} is issued in this case since
the post-selection overhead is non-trivial and amplitude amplification
may be warranted.
When the components are known to have disjoint support, the
\texttt{PARTITION} constructor (Section~\ref{sec:partition}) is
strictly preferable: it produces the same state ancilla-free with
$p=1$ at lower gate count.

\textbf{Example.}

Two disjoint \texttt{SQUARE} intervals with different amplitudes.
PyEncode detects disjoint support analytically and reports
$p$ via~\eqref{eq:sum_p}:
\begin{lstlisting}[style=python]
circuit, info = encode(
    SUM([(1.0, SQUARE(k_s=0, k_e=8, c=1)),
         (3.0, SQUARE(k_s=8, k_e=16, c=1))]),N=16)
# f = [1,...,1, 3,...,3] / ||f||
print(info)
\end{lstlisting}

\begin{lstlisting}[basicstyle=\ttfamily\small\color{black},]
----- output --------------
PyEncode  v3.0.0
  Pattern     : SUM
  N           : 16  (m = 4 qubits)
  Gate count  : 47
  Complexity  : O(2 m)
  Validated   : yes
  Success prob: 0.6250  (post-selection required)
  Vector      : numpy array, shape (16,)
  Parameters  : {'components':
  ['SQUARE', 'SQUARE'],
  'weights': [1.0, 3.0], 'disjoint': True}
  Circuit code: 2414 chars
\end{lstlisting}

For overlapping components a \texttt{UserWarning} is issued;
post-selection is always valid and $p$ is reported in
\texttt{info.success\_probability}.

Figure~\ref{fig:ex_lcu_disjoint} shows the disjoint two-interval
vector and SUM circuit. Figure~\ref{fig:ex_lcu_overlap} shows an
overlapping example (uniform $+$ sinusoidal), where $p < 1$.

\loadfigsv{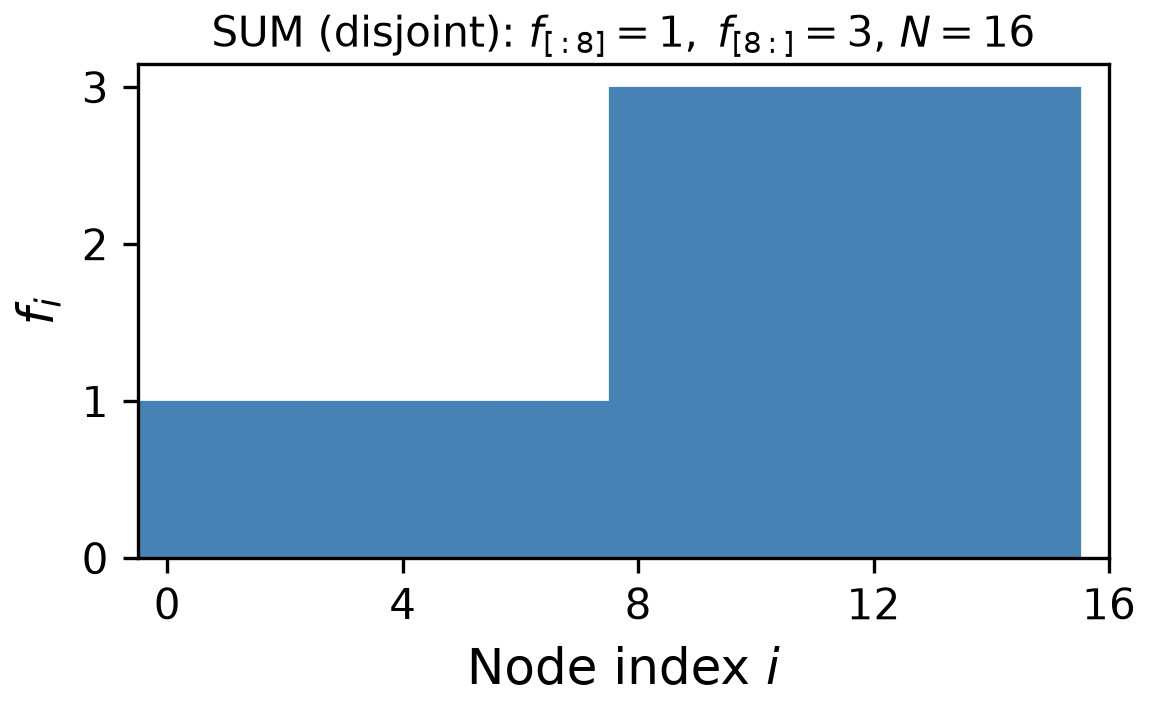}{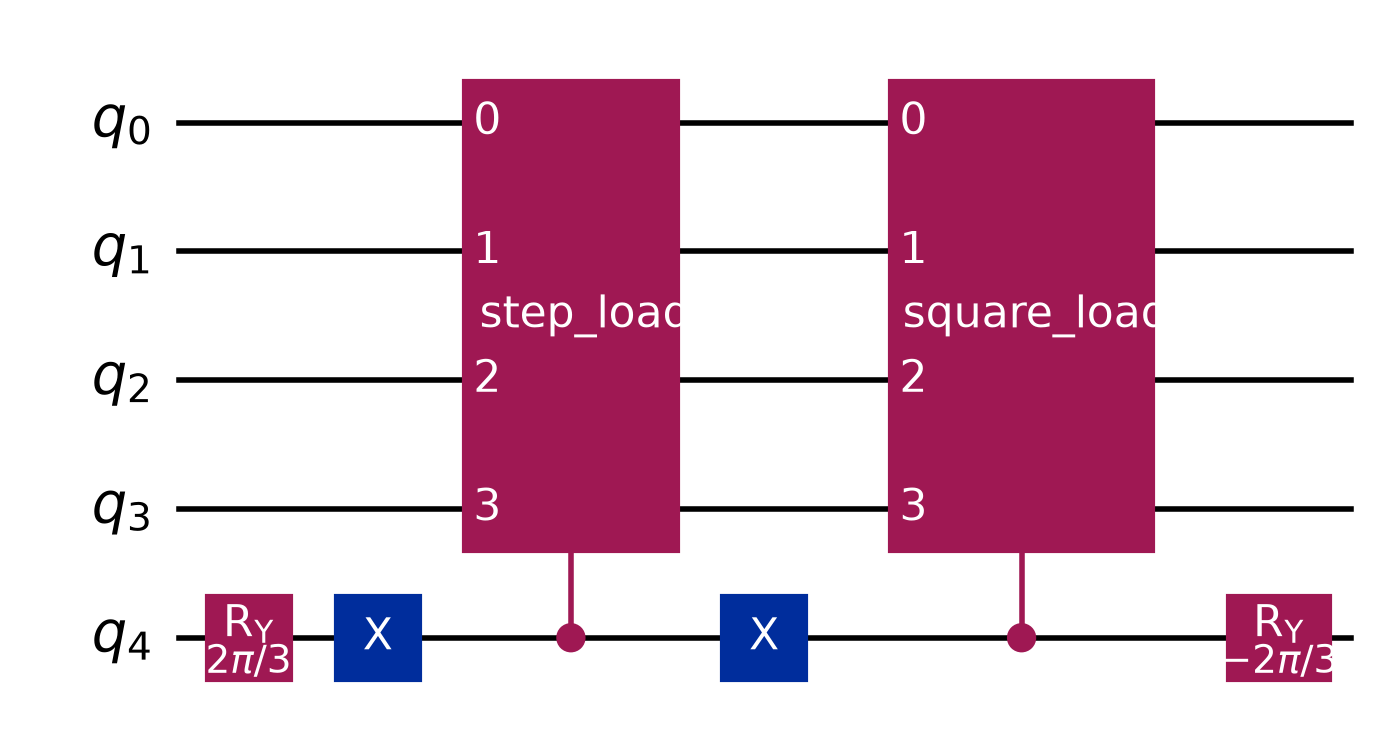}
  {SUM (disjoint): $f_{[:8]}=1,\;f_{[8:]}=3$, $N=16$}
  {0.5}
  {0.8}{ex_lcu_disjoint}

\textbf{Example: complex weights.}

The same construction handles complex weights without API change.
Negative-real and pure-imaginary weights are simply special cases
(magnitude tree unchanged; phase threaded through SELECT):
\begin{lstlisting}[style=python]
# Negative real weight
qc, info = encode(
    SUM([(-1.0, SQUARE(k_s=0, k_e=4, c=1.0)),
         ( 2.0, SQUARE(k_s=4, k_e=8, c=1.0))]),
    N=8)

# Complex weights, disjoint components
qc, info = encode(
    SUM([(1.0+1j, SQUARE(k_s=0, k_e=4, c=1.0)),
         (1.0-1j, SQUARE(k_s=4, k_e=8, c=1.0))]),
    N=8)
# info.success_probability -> 0.5
\end{lstlisting}
For real positive weights the gate count is unchanged from the
original construction; complex weights add at most one phase gate per
component, which the transpiler typically absorbs into an adjacent
rotation at \texttt{optimization\_level=3}.

\loadfigsv{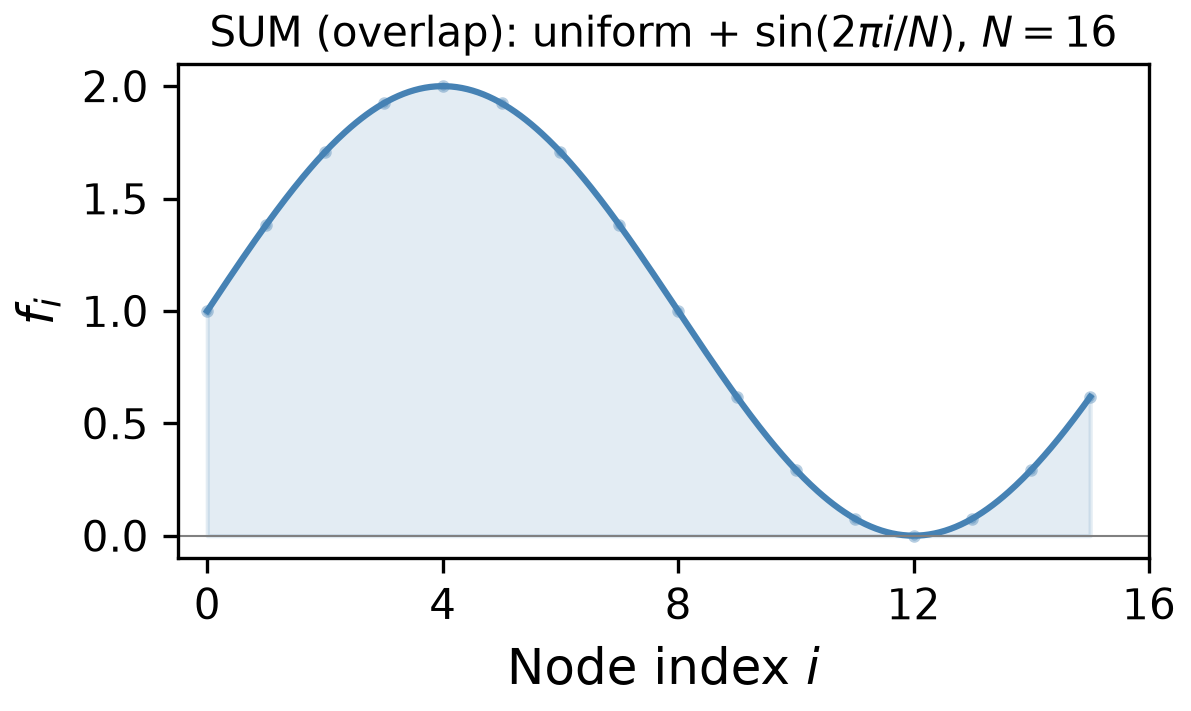}{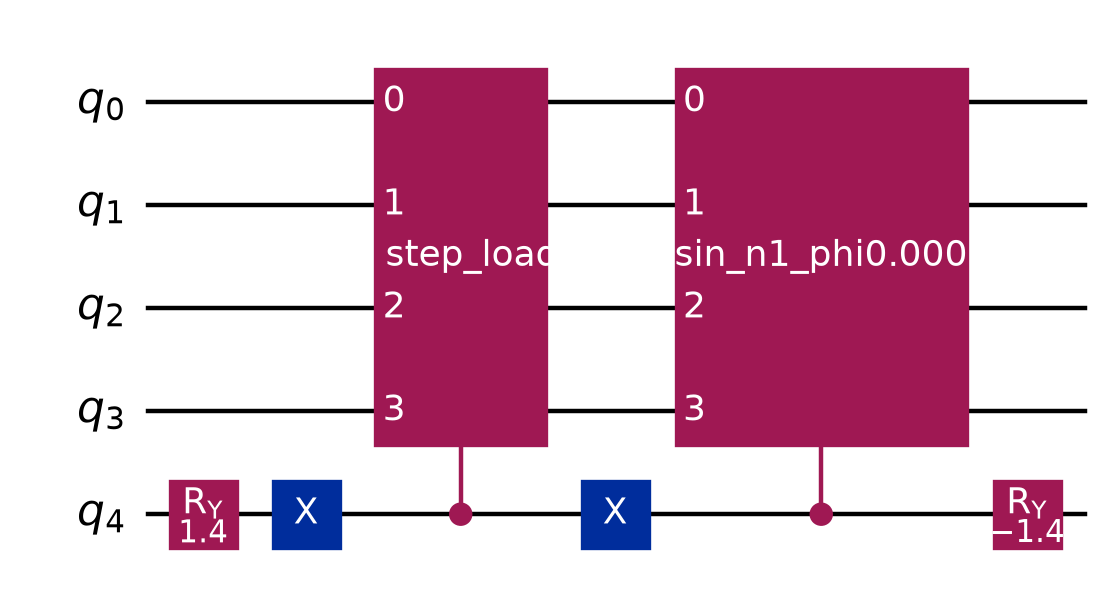}
  {SUM (overlap): uniform $+$ $\sin(2\pi i/N)$, $N=16$}
  {0.5}
  {0.75}{ex_lcu_overlap}

\subsection{Partition (disjoint-support)}
\label{sec:partition}
When the component states have pairwise-disjoint support, the sum
$\sum_j |f^{(j)}\rangle$ can be prepared without ancilla and
with success probability one, at lower gate count than the

\texttt{SUM} construction.  This is the \texttt{PARTITION} constructor:
\begin{equation}
  |\psi\rangle \propto \sum_{j=1}^{K} |f^{(j)}\rangle,
  \qquad \mathrm{supp}(f^{(i)}) \cap \mathrm{supp}(f^{(j)}) = \emptyset.
\end{equation}
Constructor: \texttt{PARTITION([pattern1, pattern2, \ldots])},
where each component is a bounded-support pattern (\texttt{SPARSE},
\texttt{STEP}, \texttt{SQUARE}, or \texttt{GEOMETRIC}).  Dense-support
patterns (\texttt{FOURIER}, \texttt{WALSH}, \texttt{HAMMING},
\texttt{STAIRCASE}, \texttt{POLYNOMIAL}) are rejected at construction
time, as their supports can never be disjoint.
Let $K$ denote the number of components and 
$L$ the total number of dyadic anchor blocks produced by their decomposition, with $L \le K(m+1)$.
PyEncode verifies pairwise disjoint-ness of components  in $\mathcal{O}(K^2)$ naively; on overlap,
\texttt{PARTITION} raises \texttt{ValueError} and recommends
\texttt{SUM}.

\subsubsection{Circuit construction}

Each component support is broken into a small number of
power-of-two-aligned blocks via the dyadic decomposition of
Bentley and Saxe~\cite{bentley1980}; an interval of width $w$
yields at most $\mathcal{O}(\log w)$ such blocks, and a singleton
yields one block of size one.  Assembly then proceeds in two
stages:
\begin{enumerate}[leftmargin=*, label=\arabic*., itemsep=0pt, topsep=2pt]
  \item \textbf{Anchor load.} A sparse state preparation~\cite{gleinig2021} 
        loads one amplitude per block at the
        block's lowest index, weighted by the block's total
        contribution to the target.
  \item \textbf{Block spread.} Each multi-index block is then
        spread into a uniform (for \texttt{STEP}, \texttt{SQUARE})
        or geometric (for \texttt{GEOMETRIC}) superposition over
        its support via controlled rotations on the qubits below
        the anchor. Singleton blocks (\texttt{SPARSE}
        components) are untouched.
\end{enumerate}
Because the block supports are exactly disjoint by construction,
no ancilla is needed, and post-selection succeeds with probability
one. Total gate count is $\mathcal{O}(L \cdot m)$, bounded by
$\mathcal{O}(m^2)$ in the worst case.

\textbf{Example.}

The motivating case: a sparse prefix on $\{2,5,7\}$ followed by a
geometric tail from index $11$, on $N = 256$.
\begin{lstlisting}[style=python]
circuit, info = encode(
    PARTITION([
        SPARSE([(2, 0.3), (5, 0.5), (7, 0.7)]),
        GEOMETRIC(r=0.8, k_s=11),
    ]), N=256)
# info.gate_count          -> ~715
# info.success_probability -> 1.0
# info.complexity          -> "O(L*m)"
\end{lstlisting}
The same state prepared via \texttt{SUM} of the two components would
cost approximately $9{,}300$ gates with an ancilla qubit and
$p\approx0.54$ (post-selection or amplitude amplification required).  When the application guarantees
disjoint-support components, \texttt{PARTITION} is the right choice.

\subsection{Tensor}
\label{sec:tensor}
A separable state over two or more disjoint subregisters:
\begin{equation}
  |\psi\rangle \;=\; \bigotimes_{j=1}^{r}\, |\hat{f}^{(j)}\rangle,
\end{equation}
where each $|\hat{f}^{(j)}\rangle$ is a normalized state prepared by
any PyEncode pattern on $m_j$ qubits, and the total register width is
$m = \sum_j m_j$.
Constructor: \texttt{TENSOR([(pattern$_1$, $N_1$), \ldots, (pattern$_r$, $N_r$)])},
with $N_j = 2^{m_j}$.

\textbf{Construction.}

Since the subregisters are disjoint, the component unitaries
$U_1, \ldots, U_r$ commute and the composite circuit is simply their
Kronecker product.
The total gate count equals the sum of component counts, and all
component circuits can execute \emph{in parallel}, so the circuit
depth is $\max_j \mathrm{depth}(U_j)$.
No ancilla, no post-selection, unit success probability.

Tensor composition is the natural encoder for separable
multi-dimensional fields, for example, Poisson sources of the form
$f(x,y) = g(x)h(y)$. It formalizes the \texttt{circ1.tensor(circ2)}
idiom already used in Qiskit, but surfaces it as a 
pattern with unified validation and complexity reporting.

\textbf{Example.}

A separable 2D source $\sin(2\pi n_x i/N)\sin(2\pi n_y j/N)$ encoded
on $2m$ qubits:
\begin{lstlisting}[style=python]
circuit, info = encode(
    TENSOR([(FOURIER(modes=[(2, 1.0, 0)]), 32),
            (FOURIER(modes=[(3, 1.0, 0)]), 32)]),
    N=32*32)
# info.gate_count = 2 * (single-axis cost); depth = single-axis depth
\end{lstlisting}

\loadfigsv{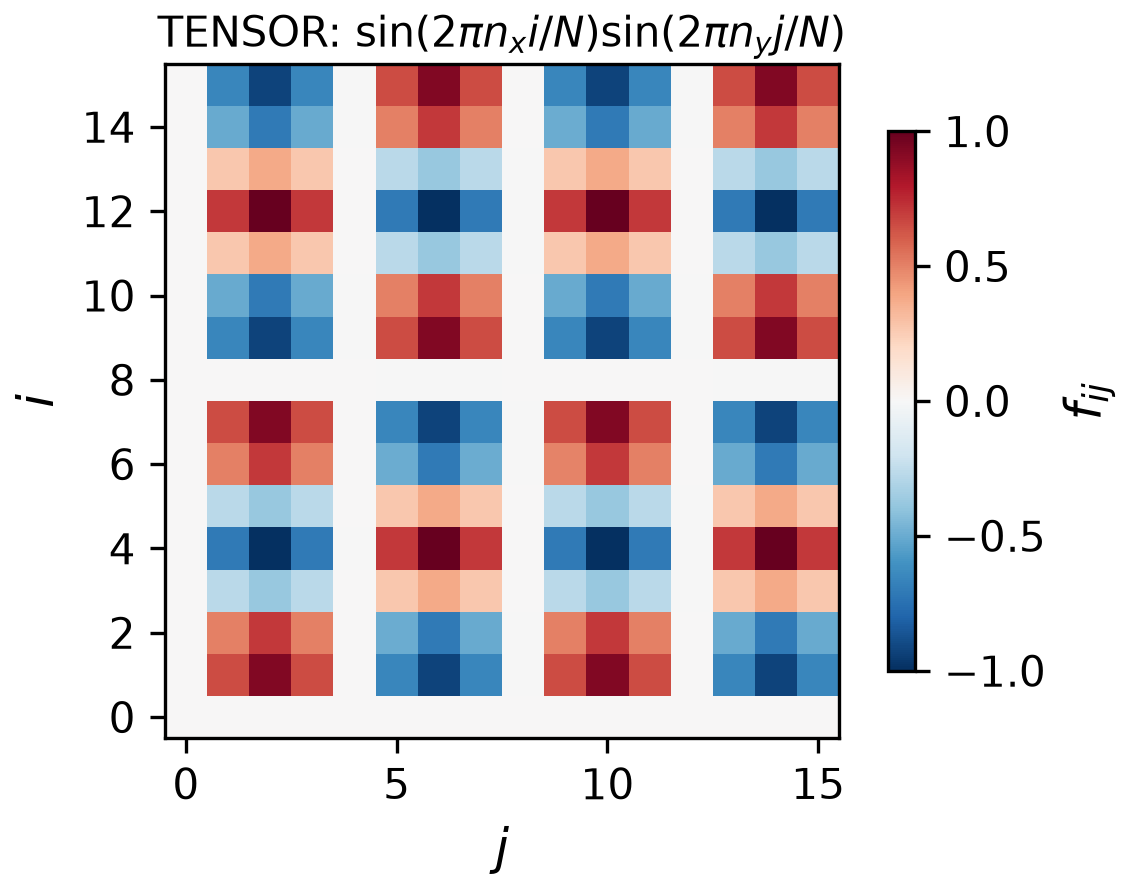}{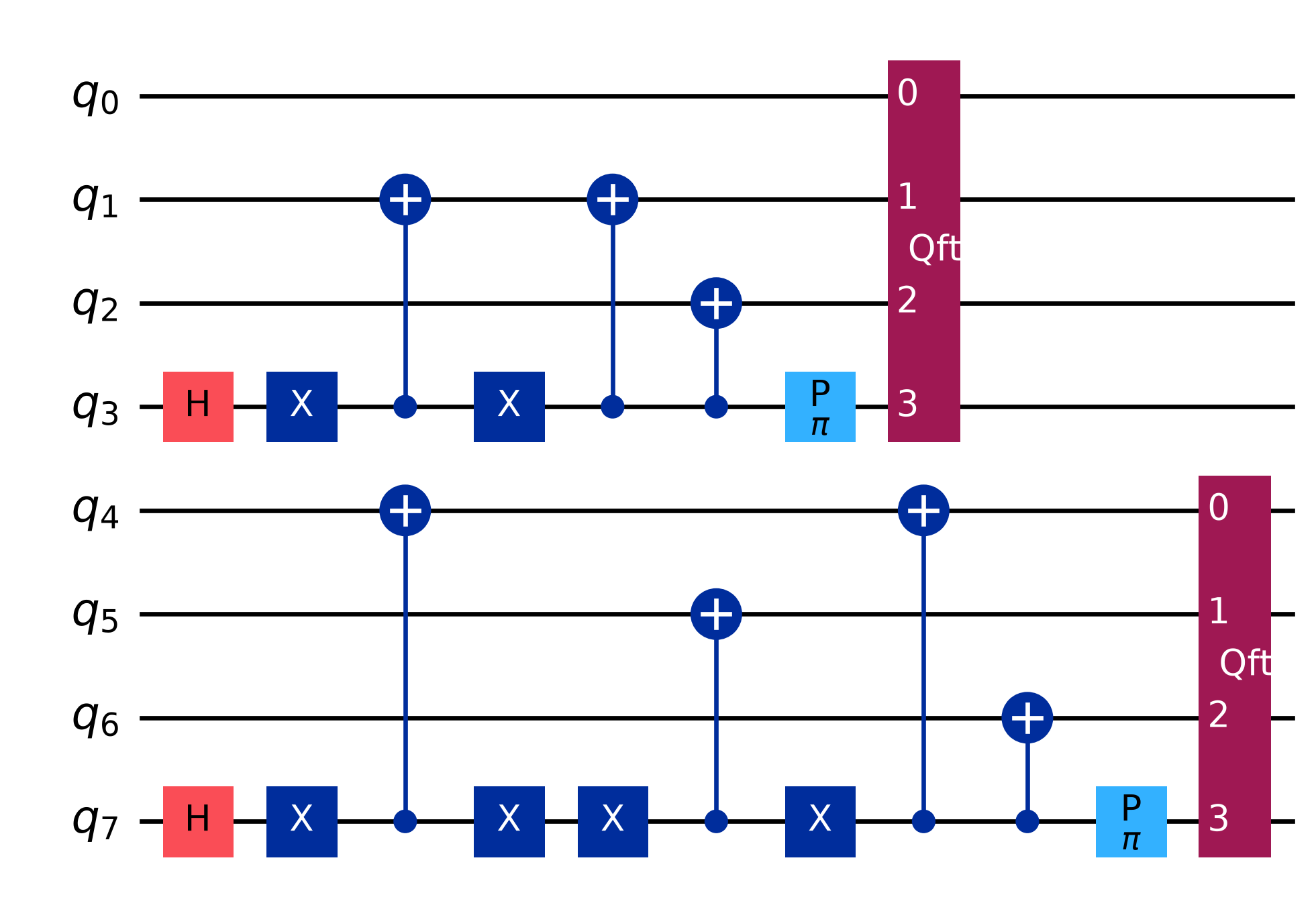}
  {Tensor: $\sin(2\pi n_x i/N)\sin(2\pi n_y j/N)$, $N=16\!\times\!16$}
  {0.5}
  {0.85}{ex_tensor}


\subsection{Return value of \texttt{encode}}
\label{sec:returnValue}
\texttt{encode} returns \texttt{(circuit, info)}, where \texttt{circuit}
is a Qiskit \texttt{QuantumCircuit} and \texttt{info} is an
\texttt{EncodingInfo} dataclass with the following fields:
\begin{itemize}[leftmargin=*, itemsep=1pt, topsep=2pt]
    \item \texttt{pattern\_name} --- name of the recognized pattern
        (e.g.\ \texttt{"SPARSE"}, \texttt{"GEOMETRIC"}).
  \item \texttt{N}, \texttt{m} --- vector length and number of qubits.
  \item \texttt{params} --- supplied vector parameters
        (e.g.\ \texttt{\{"r": 0.95, "c": 1.0\}}).
  \item \texttt{gate\_count} --- total gates in the returned circuit
        (pre-transpilation).
  \item \texttt{gate\_count\_1q}, \texttt{gate\_count\_2q} --- $U$ and
        CX gate counts after transpilation to $\{\textsc{cx}, U\}$.
  \item \texttt{circuit\_depth} --- circuit depth after transpilation;
        determines the minimum execution time when gates on disjoint
        qubits run in parallel.
  \item \texttt{complexity} --- asymptotic gate complexity
        (e.g.\ \texttt{"O(m)"} or \texttt{"O(m\^{}2)"}).
  \item \texttt{success\_probability} --- always 1.0 for single-pattern
        constructors and for \texttt{PARTITION}; $p \in (0,1]$ for
        \texttt{SUM}.
  \item \texttt{circuit\_code} --- a human-readable Qiskit snippet that
        reproduces the circuit independently of PyEncode.
  \item \texttt{validated} --- \texttt{True} if statevector validation
        was performed (see Section~\ref{sec:validation}).
  \item \texttt{vector} --- the classically constructed amplitude vector
        $\mathbf{f}$, populated only when \texttt{validate=True};
        requires $\mathcal{O}(2^m)$ memory.
\end{itemize}

\subsection{Validation}
\label{sec:validation}

By default, no classical vector is constructed during synthesis since
there is no vector to validate. The optional statevector check
(\texttt{validate=True}) constructs $\mathbf{f}$ from the supplied
parameters, runs the circuit on Qiskit's statevector simulator, and
checks agreement up to a single global phase, which is the correct
equivalence for physical quantum states:
\begin{equation}
  \min_{\varphi\in\mathbb{R}}\;
  \bigl\|\hat{\mathbf{f}}-e^{i\varphi}\hat{\mathbf{f}}_\mathrm{sim}\bigr\|_2
  < \varepsilon,
  \label{eq:validation_invariant}
\end{equation}
which is equivalent to fidelity
$|\langle\hat{\mathbf{f}}|\hat{\mathbf{f}}_\mathrm{sim}\rangle|^2
> 1 - \varepsilon^2/2$.  The optimal phase
$e^{i\varphi^\star}=\langle\hat{\mathbf{f}}|\hat{\mathbf{f}}_\mathrm{sim}\rangle
/|\langle\hat{\mathbf{f}}|\hat{\mathbf{f}}_\mathrm{sim}\rangle|$
is closed-form; an unaligned amplitude-magnitude comparison would be
strictly weaker and would admit wrong relative signs between basis
states.  This is the only validation path available and is disabled
by default due to its $\bigO{2^m}$ memory cost.
For large $m$, partial verification via measurement sampling or a
SWAP-test overlap estimate can provide confidence at polynomial
cost; these are not currently implemented.
When enabled, the constructed vector is also returned as \texttt{info.vector}
for inspection and debugging.
\begin{lstlisting}[style=python]
circuit, info = encode(
    FOURIER(modes=[(1, 1.0, 0)]), N=16,
    validate=True, tol=1e-6)
# info.validated-> True
# info.vector  -> numpy array of length N
\end{lstlisting}

\subsection{Cost prediction without synthesis}
\label{sec:predict}

In workflows where many candidate encodings must be evaluated before
committing to a circuit, the cost of a single \texttt{encode()} call
--- $\mathcal{O}(1)$\,s at $m \geq 16$ due to the Qiskit transpile
pass --- can dominate an outer loop that sweeps thousands of
candidates.

\texttt{predict\_gates(pattern, N)} estimates the transpiled gate
counts without any circuit construction, using closed-form formulas
derived from each pattern's analytical structure:
\begin{lstlisting}[style=python]
from pyencode import predict_gates, POLYNOMIAL
p = predict_gates(POLYNOMIAL(coeffs=[0.0, 1.0]), N=4096)
# p = {'pattern_name': 'POLYNOMIAL', 'N': 4096, 'm': 12,
#      'gate_count_1q': 56, 'gate_count_2q': 22,
#      'circuit_depth': 45, 'complexity': 'O(m)', 'exact': True}
\end{lstlisting}

Predictions match \texttt{encode()}'s transpiled counts to the gate
(verified in the test suite) for \texttt{HAMMING}, \texttt{WALSH},
\texttt{STAIRCASE}, \texttt{STEP}, \texttt{SPARSE} ($s=1$),
\texttt{FOURIER} ($T=1$), \texttt{POLYNOMIAL} ($d=1$), and power-of-2-aligned
\texttt{SQUARE}. For patterns whose transpiled count depends on index
bit patterns or multi-mode transpiler optimizations (\texttt{SPARSE}
$s\geq 2$, \texttt{SQUARE} general, \texttt{POLYNOMIAL} $d \geq 2$,
\texttt{FOURIER} ($T \geq 2$), \texttt{TENSOR}, \texttt{SUM},
\texttt{PARTITION}, and \texttt{GEOMETRIC} with nonzero \texttt{k\_s}),
the returned value is either an empirical fit within a few percent
or an upper bound; an \texttt{exact} field in the returned dictionary
flags which regime applies. Prediction is $500$--$8000\times$ faster than full
synthesis and remains sub-millisecond at all $m$ up to $m=16$.

\subsection{\textcolor{black}{Reverse lookup}}
\label{sec:match}

The \texttt{encode} entry point runs forward, from a typed declaration to a
circuit. A user who instead holds a materialized numerical vector, but does
not know which family it belongs to, can run the inverse lookup:
\begin{mdframed}[backgroundcolor=highlightblue,
                 linecolor=codeblue, linewidth=0.8pt,
                 innertopmargin=2pt, innerbottommargin=2pt,
                 innerleftmargin=6pt, innerrightmargin=6pt]
\small
\texttt{match\_vector(v, top\_k=3)}
\end{mdframed}

For each exact pattern family, \texttt{match\_vector} fits that family's free
parameters to \texttt{v}, scores the fit, and returns the closest matches
ranked by error. Each returned match carries a ready-to-\texttt{encode}
constructor with the fitted parameters, so the typical workflow is to inspect
the ranking and then synthesize the top candidate directly.

Because quantum state preparation is insensitive to a global scale and phase
--- the state is normalized and the leading amplitude $c$ of every pattern is
a free parameter --- the fit metric is scale- and phase-invariant. For a
candidate vector $\mathbf{w}$, the optimal complex scale $\alpha$ minimizing
$\lVert \mathbf{v} - \alpha\mathbf{w}\rVert$ is projected out, and the reported
error is
\begin{equation}
\text{rel\_error}
= \frac{\lVert \mathbf{v} - \alpha\mathbf{w}\rVert}{\lVert \mathbf{v}\rVert}
= \sqrt{1 - \frac{|\langle \mathbf{w},\mathbf{v}\rangle|^2}
{\lVert \mathbf{w}\rVert^2\,\lVert \mathbf{v}\rVert^2}}\ \in [0,1],
\end{equation}
where $0$ denotes a perfect structural match; the complementary
$\text{fidelity} = 1 - \text{rel\_error}^2$ is the squared overlap with the
target.

\begin{lstlisting}[style=python]
from pyencode import match_vector, encode
import numpy as np
v = np.array([1, 1, 1, 1, 0, 0, 0, 0], dtype=float)   # STEP(k_e=4)
matches = match_vector(v)
print_matches(matches)
# ----- output --------------
# rank  pattern      rel_error   fidelity  gates  params
#    1  STEP         0.000e+00   1.000000       2  k_e=4, c=0.5
#    2  SQUARE       0.000e+00   1.000000       7  k_s=0, k_e=4, c=0.5
#    3  POLYNOMIAL   1.5e-01     0.977        ...  ...
circuit, info = encode(matches[0].pattern, N=len(v))
\end{lstlisting}

Two properties of the ranking are worth noting. First, the \texttt{SPARSE}
family reproduces any vector exactly from its nonzero entries, so it is a
guaranteed zero-error fallback rather than a structural discovery; the reverse
lookup is therefore best understood as finding the \emph{cheapest structured}
family, not a unique classification. Second, when several families reproduce
the vector exactly, ties are broken by predicted gate count
(Section~\ref{sec:predict}), so the cheapest exact encoding ranks first.

\section{Gate Count Comparison}
\label{sec:gateCount}

\begin{figure}[h]
\centering
\includegraphics[width=\linewidth]{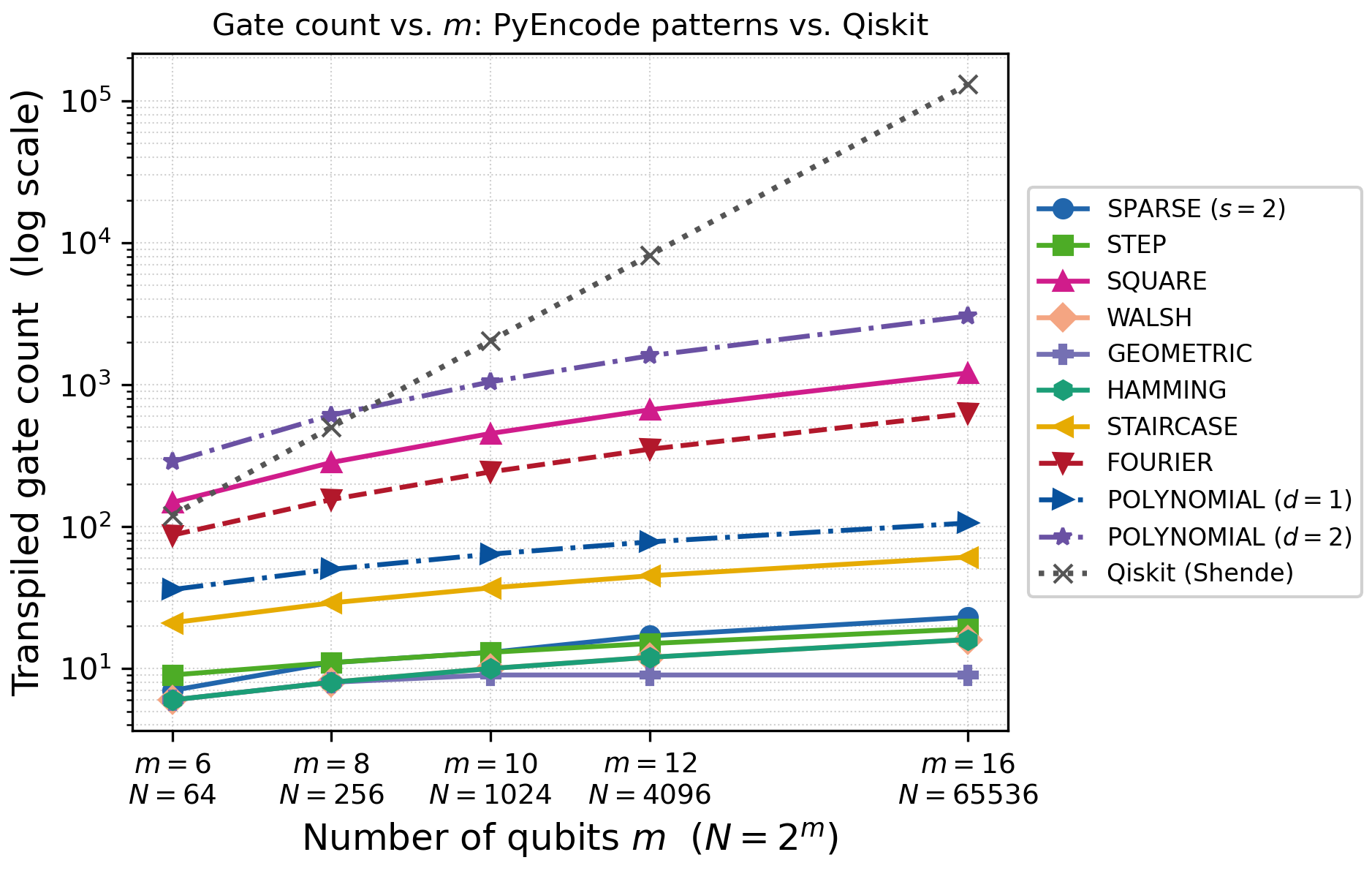}
\caption{Transpiled gate count vs.\ number of qubits $m$ ($N = 2^m$),
for $m \in \{6, 8, 10, 12, 16\}$. Both PyEncode and Qiskit circuits are
transpiled to the $\{\textsc{cx}, U\}$ basis
(\texttt{optimization\_level=3}, Qiskit 2.3.1). Three asymptotic tiers are visible: the
$\mathcal{O}(m)$ patterns (\texttt{SPARSE}, \texttt{STEP}, \texttt{WALSH},
\texttt{GEOMETRIC}, \texttt{HAMMING}) occupy the bottom band at under 25
gates; \texttt{STAIRCASE} and \texttt{POLYNOMIAL} $d=1$ are also
$\mathcal{O}(m)$ with modestly larger constants (20--110 gates);
the $\mathcal{O}(m^2)$ patterns \texttt{FOURIER}, \texttt{SQUARE}, and
\texttt{POLYNOMIAL} $d=2$ form an intermediate band reaching
$\sim 3{,}000$ gates at $m=16$.  Qiskit \texttt{StatePreparation} on a
random vector scales as $\mathcal{O}(2^m)$ and reaches $131{,}053$
gates at $m=16$.
\texttt{SPARSE} uses $s=2$ entries with non-aligned indices;
\texttt{SQUARE} uses a non-aligned interval $[N/4{+}1,\,3N/4{+}1)$,
which activates the general $\mathcal{O}(m^2)$ Draper-adder path.
Circuit depth follows the same asymptotic scaling; see
Table~\ref{tab:gatecounts} for per-pattern depth at $m=12$.}
\label{fig:gate_count_vs_m}
\end{figure}

Table~\ref{tab:gatecounts} summarizes gate counts and circuit depth
at $N = 4096$ ($m = 12$ qubits). All circuits are transpiled to
$\{\textsc{cx}, U\}$ (\texttt{optimization\_level=3}, Qiskit 2.3.1).
Qiskit \texttt{StatePreparation} uses \texttt{reps=3}.
Two-qubit (CX) gates are hardware-critical as they are significantly
noisier than single-qubit gates on near-term devices; circuit depth
determines the minimum execution time when gates on disjoint qubits
run in parallel.
The $\mathcal{O}(m)$ patterns (Sparse, Step, Walsh, Geometric, Hamming)
achieve depth~1 at $m=12$, meaning all gates execute in a single
parallel layer.
Figure~\ref{fig:gate_count_vs_m} shows transpiled gate count as a
function of $m$ for both PyEncode and Qiskit using the same
$\{\textsc{cx}, U\}$ basis.
All ten exact pattern families outperform Qiskit from $m \geq 10$,
and Qiskit's exponential growth separates by orders of magnitude at
$m = 16$. The $\mathcal{O}(m)$ and $\mathcal{O}(m^{d+1})$ patterns form
three distinct tiers: constant-scaling patterns near 10 gates, linear
patterns in the 20--100 range, and quadratic patterns
(\texttt{FOURIER}, \texttt{SQUARE}, \texttt{POLYNOMIAL} $d=2$,
\texttt{DICKE} at $k \approx m/2$) in the
300--3{,}000 range at $m = 16$.
Gate count is independent of the number of Fourier modes $T$
(see Table~\ref{tab:gatecounts}, rows for $T=1$ and $T=2$).

\begin{table*}[ht]
\centering
\caption{Transpiled gate counts and circuit depth at $N=4096$ ($m=12$ qubits).
All circuits transpiled to $\{\textsc{cx}, U\}$
(\texttt{optimization\_level=3}, Qiskit 2.3.1).
Qiskit uses general-purpose $\mathcal{O}(2^m)$ state
preparation~\cite{shende2006}.
$^\dagger$\texttt{SQUARE} uses a Draper QFT-based constant adder~\cite{draper2000}:
$\mathcal{O}(m)$ for $k_s = 0$ or power-of-2-aligned blocks; $\mathcal{O}(m^2)$
in general.
$^\ast$Qiskit for \texttt{WALSH} and \texttt{HAMMING} reflect
optimizer detection of the two-level-constant and Hamming-symmetric
structures; PyEncode delivers this $\mathcal{O}(m)$ cost by analytical
construction for all parameter settings, whereas Qiskit's performance on less symmetric inputs degrades to the general case
(cf.\ the \texttt{GEOMETRIC} row, where Qiskit requires $4{,}088$ gates).}
\label{tab:gatecounts}
\setlength{\tabcolsep}{5pt}
\renewcommand{\arraystretch}{1.15}
\begin{tabular}{lcccccccc}
\toprule
 & \multicolumn{4}{c}{\textbf{PyEncode}}
 & \multicolumn{3}{c}{\textbf{Qiskit}} \\
\cmidrule(lr){2-5}\cmidrule(lr){6-8}
\textbf{Pattern} & $U$ & CX & Depth & $\mathcal{O}(\cdot)$
                 & $U$ & CX & Depth \\
\midrule
Sparse ($s\!=\!1$, $k\!=\!N/4$)
    & 1 & 0 & 1 & $\mathcal{O}(sm)$   & 2 & 1 & 3 \\
Sparse ($s\!=\!2$)
    & 7 & 11 & 12 & $\mathcal{O}(sm)$   & 4{,}095 & 4{,}083 & 8{,}167 \\
Step ($k_e\!=\!N/2$)
    & 11 & 0 & 1 & $\mathcal{O}(m)$    & 22 & 11 & 13 \\
Square ($[N/4{+}1,\,3N/4{+}1)$, general)
    & 405 & 261 & 162 & $\mathcal{O}(m^2)^\dagger$  & 4{,}095 & 4{,}083 & 8{,}167 \\
Walsh ($k\!=\!6$, $c_+\!=\!1,\,c_-\!=\!4$)
    & 12 & 0 & 1 & $\mathcal{O}(m)$    & 12$^\ast$ & 0 & 1 \\
Geometric ($r\!=\!0.95$)
    & 9 & 0 & 1 & $\mathcal{O}(m)$    & 4{,}088 & 4{,}079 & 8{,}159 \\
Hamming ($r\!=\!0.7$)
    & 12 & 0 & 1 & $\mathcal{O}(m)$    & 12$^\ast$ & 0 & 1 \\
Staircase ($r\!=\!0.5$)
    & 23 & 22 & 34 & $\mathcal{O}(m)$    & 4{,}095 & 4{,}083 & 8{,}167 \\
Dicke ($k\!=\!2$)
    & 146 & 112 & 185 & $\mathcal{O}(k(m{-}k))$ & 4{,}091 & 4{,}083 & 8{,}163 \\
Dicke ($k\!=\!11$)
    & 56 & 22 & 45 & $\mathcal{O}(k(m{-}k))$ & 4{,}091 & 4{,}083 & 8{,}163 \\
Polynomial ($d\!=\!1$, ramp)
    & 56 & 22 & 45 & $\mathcal{O}(m)$   & 4{,}095 & 4{,}083 & 8{,}167 \\
Polynomial ($d\!=\!2$, Poiseuille)
    & 874 & 725 & 1{,}132 & $\mathcal{O}(m^2)$  & 4{,}014 & 4{,}083 & 8{,}086 \\
Fourier ($T\!=\!1$, $n\!=\!1$, $\varphi\!=\!0$)
    & 192 & 159 & 98 & $\mathcal{O}(m^2)$  & 4{,}025 & 4{,}083 & 8{,}097 \\
Fourier ($T\!=\!1$, $n\!=\!3$, $\varphi\!=\!\pi/4$)
    & 192 & 161 & 101 & $\mathcal{O}(m^2)$  & 4{,}010 & 4{,}083 & 8{,}082 \\
Fourier ($T\!=\!2$)
    & 195 & 161 & 94 & $\mathcal{O}(m^2)$  & 4{,}020 & 4{,}083 & 8{,}092 \\
\bottomrule
\end{tabular}
\end{table*}

\section{Approximate Encoding}
\label{sec:mps}

The ten pattern families in Section~\ref{sec:patterns} synthesize
\emph{exact} circuits at $\mathcal{O}(\mathrm{poly}(m))$ cost, but
require the input to fit a declared algebraic structure. Many
smooth amplitude vectors of practical interest fall outside these
families; for example, the discretized Gaussian
$f_i = \exp(-\alpha (i - i_0)^2 / N^2)$. 

For such inputs, PyEncode provides an \emph{approximate} loader
based on a bounded-bond matrix product state
representation (MPS)~\cite{oseledets2011,melnikov2023quantum}, exposed
through a dedicated entry point in a separate submodule:
\begin{mdframed}[backgroundcolor=highlightblue,
                 linecolor=codeblue, linewidth=0.8pt,
                 innertopmargin=2pt, innerbottommargin=2pt,
                 innerleftmargin=6pt, innerrightmargin=6pt]
\small
\texttt{encode\_mps(v, bond\_dim, validate=False, tol=1e-6)}
\end{mdframed}
The supplied amplitude vector $\mathbf{v}$ may be real or complex
and of any length; non-power-of-2 lengths are zero-padded to the
next $N = 2^m$.

\textbf{Construction.}

The vector is reduced to a right-canonical matrix product state
\begin{equation}
  v_{i_1 i_2 \cdots i_m}
  = A^{(i_1)}_1\, A^{(i_2)}_2\, \cdots\, A^{(i_m)}_m,
  \qquad i_j \in \{0, 1\},
\end{equation}
via a right-to-left sequence of singular value decompositions (SVD), each
truncated at bond dimension $\chi$ (\texttt{bond\_dim}).  The site
tensors $A^{(i_j)}_j$ are then assembled into a deterministic
sequential quantum cascade following Sch\"on et
al.~\cite{schon2005} and Ran~\cite{ran2020}: each tensor is
completed to a $(2\chi)\times(2\chi)$ unitary $U_j$ via
SVD null-space completion, and the unitaries are applied
in sequence on the bond register together with one physical qubit
per site.  The leftmost site tensor is renormalized to absorb the
cumulative truncation deficit, so the bond register starts and
ends in $|0\rangle$ \emph{deterministically}: the prepared
physical state has success probability $p = 1$, with no
post-selection required.  At
$\chi \ge 2^{m-1}$ the MPS is exact for arbitrary
$\mathbf{v}$; at smaller $\chi$ the tail singular values are
discarded and
\texttt{info.params["truncation\_error\_sq"]} reports the
cumulative discarded weight, an upper bound on
$1 - |\langle\hat{\mathbf{v}}|\psi_{\mathrm{MPS}}\rangle|^2$.

\textbf{Cost.}

The circuit acts on $n_{\mathrm{bond}} + m$ qubits, where
$n_{\mathrm{bond}} = \lceil \log_2 \chi \rceil$ are bond-register
ancilla qubits.  Each of the $m$ site unitaries acts on
$n_{\mathrm{bond}} + 1$ qubits, giving a total gate cost of
$\mathcal{O}(m\,\chi^2)$ two-qubit gates and depth
$\mathcal{O}(m\,\chi^2)$ in the worst case.  The single user-facing
knob $\chi$ trades approximation error against gate count:
$\chi = 1$ collapses to a depth-$m$ product-state cascade with no
entanglement, while $\chi = 2^{m-1}$ reproduces $\mathbf{v}$
exactly but at a cost that itself exceeds the
$\mathcal{O}(2^m)$ general state-preparation bound. The
practical regime is therefore \emph{small} $\chi$ on inputs with
bounded entanglement entropy, where $\chi = \mathcal{O}(1)$ or
$\mathcal{O}(\mathrm{polylog}\,m)$ already drives the truncation
error below threshold.

\textbf{Standalone usage.}

Because MPS encoding operates on a classical numerical vector
rather than an analytic pattern declaration, it differs from the
ten exact families in two practical respects.  First, it requires
materialization of $\mathbf{v}$ at $\mathcal{O}(N)$ classical
cost.  Second, it does not
currently compose with \texttt{SUM}, \texttt{PARTITION}, or
\texttt{TENSOR}.  Apart from these restrictions, the API matches
\texttt{encode}: the return value is a \texttt{(circuit, info)}
tuple, with transpiled gate counts, depth, and the MPS
diagnostics
$\{\,$\texttt{bond\_dim},
\texttt{n\_bond},
\texttt{truncation\_error\_sq},
\texttt{n\_padded}$\,\}$
exposed via \texttt{info.params}.
A second entry point,
\begin{mdframed}[backgroundcolor=highlightblue,
                 linecolor=codeblue, linewidth=0.8pt,
                 innertopmargin=2pt, innerbottommargin=2pt,
                 innerleftmargin=6pt, innerrightmargin=6pt]
\small
\texttt{encode\_mps\_from\_tensors(tensors)}
\end{mdframed}
accepts pre-built right-canonical site tensors of shape $(\chi_l, 2, \chi_r)$ and
skips PyEncode's SVD sweep; this is the recommended path when the
tensors come from an external source such as a DMRG ground-state
calculation.

\textbf{Example.}

The discretized Gaussian on $N = 256$ ($m = 8$), prepared at bond
dimension $\chi = 8$:
\begin{lstlisting}[style=python]
import numpy as np
from pyencode.mps import encode_mps

N = 256
i = np.arange(N)
alpha = 50.0
v = np.exp(-alpha * ((i - N/2) / N) ** 2)
v /= np.linalg.norm(v)

circuit, info = encode_mps(v, bond_dim=8, validate=True)
# info.complexity                          -> "O(m*chi^2) with chi=8"
# info.params["n_bond"]                    -> 3
# info.params["truncation_error_sq"]       -> < 1e-12
# info.success_probability                 -> 1.0
\end{lstlisting}
At $\chi = 4$, the same construction yields a circuit whose
truncation error exceeds the default validation tolerance
($\|\hat{v}_{\mathrm{prep}} - \hat{v}\|_2 > 10^{-6}$); doubling
the bond dimension to $\chi = 8$ drives the error well below
$10^{-12}$.  This illustrates the practical workflow: increase
$\chi$ until \texttt{info.params["truncation\_error\_sq"]} falls
below the application's accuracy threshold.

\loadfigs{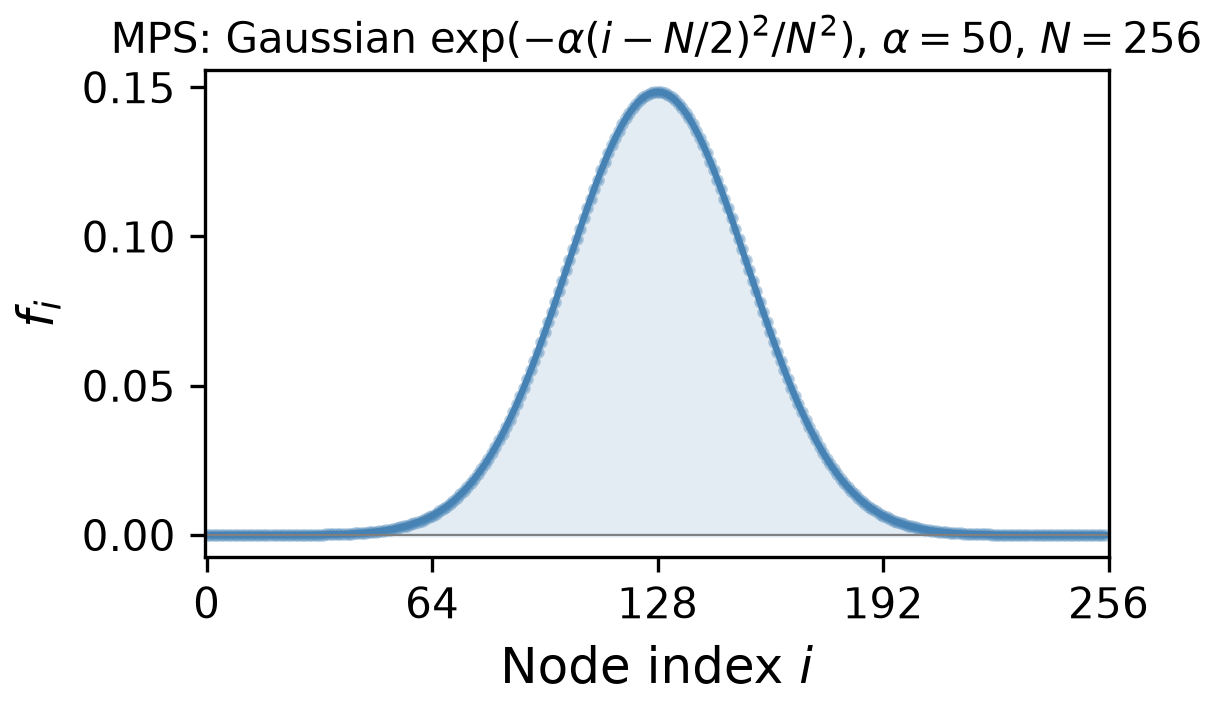}{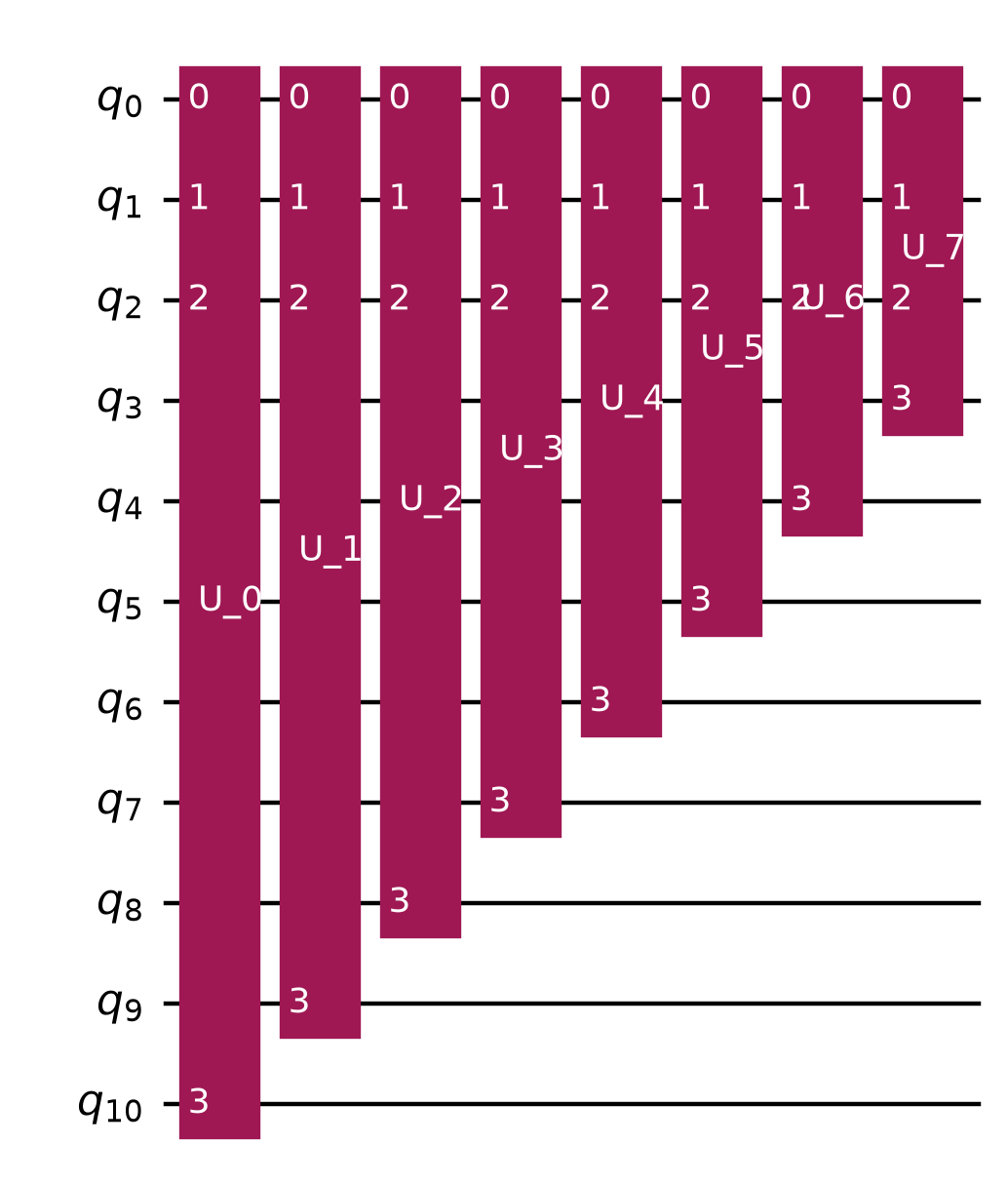}
  {Gaussian via MPS: $\alpha = 50$, $\chi = 8$, $N = 256$.
   Truncation error squared $< 10^{-12}$;  circuit cost
   $\mathcal{O}(m\,\chi^2)$ on $n_{\mathrm{bond}} + m = 11$ qubits,
   versus $\mathcal{O}(2^m)$ for Qiskit
   \texttt{StatePreparation}.}
  {unused}
  {1}{ex_mps_gaussian}

\section{Applications}
\label{sec:applications}

This section demonstrates the PyEncode framework
through three applications drawn from distinct fields.

\subsection{Quantum Chemistry}
\label{sec:app_chemistry}
Fault-tolerant quantum algorithms for chemistry represent the molecular
Hamiltonian as a linear combination of unitaries (LCU),
$H = \sum_j \alpha_j \hat{P}_j$~\cite{cao2019}. This requires a PREP
oracle that prepares
$|\boldsymbol{\alpha}\rangle \propto \sum_j \sqrt{|\alpha_j|}\,|j\rangle$~\cite{childs2018},
whose gate cost directly affects the total $T$-gate
count~\cite{babbush2018}.

\textbf{The extended Fermi--Hubbard coefficient vector.}
The extended Hubbard model adds a nearest-neighbor density-density
interaction to the standard Hubbard
Hamiltonian~\cite{hubbard1963,jordan1928}. After Jordan--Wigner on an $L$-site chain, the Pauli coefficient
vector takes three distinct values: hopping $t$ on the first $L$
terms, on-site $U$ on the next $L$ terms, and nearest-neighbor $V$
on the final $L$ terms ($N = 3L$, padded to the next power of two).
The natural composition is \texttt{PARTITION} of three disjoint
intervals, which prepares the coefficient vector ancilla-free with
success probability one at $\mathcal{O}(L \cdot m)$ gate cost:
\begin{lstlisting}[style=python]
import math
L = 8;  t = 1.0;  U = 4.0;  V = 0.5
N = 1 << (3*L - 1).bit_length()   # next power of two >= 3L
circuit, info = encode(
    PARTITION([
        STEP(k_e=L, c=math.sqrt(t)),
        SQUARE(k_s=L,     k_e=2*L, c=math.sqrt(U)),
        SQUARE(k_s=2*L,   k_e=3*L, c=math.sqrt(V)),
    ]), N=N)
# info.complexity          -> "O(L*m)"
# info.gate_count          -> 138  (64 U + 74 CX, depth 83)
# info.success_probability -> 1.0
\end{lstlisting}
Figure~\ref{fig:sec6_hubbard} shows the three-block coefficient
vector and assembled circuit. The padding region $[3L, N)$ carries
zero amplitude by construction; the disjoint-support guarantee of
\texttt{PARTITION} ensures no amplitude leaks into the unused
indices and no post-selection is required.

\loadfigsv{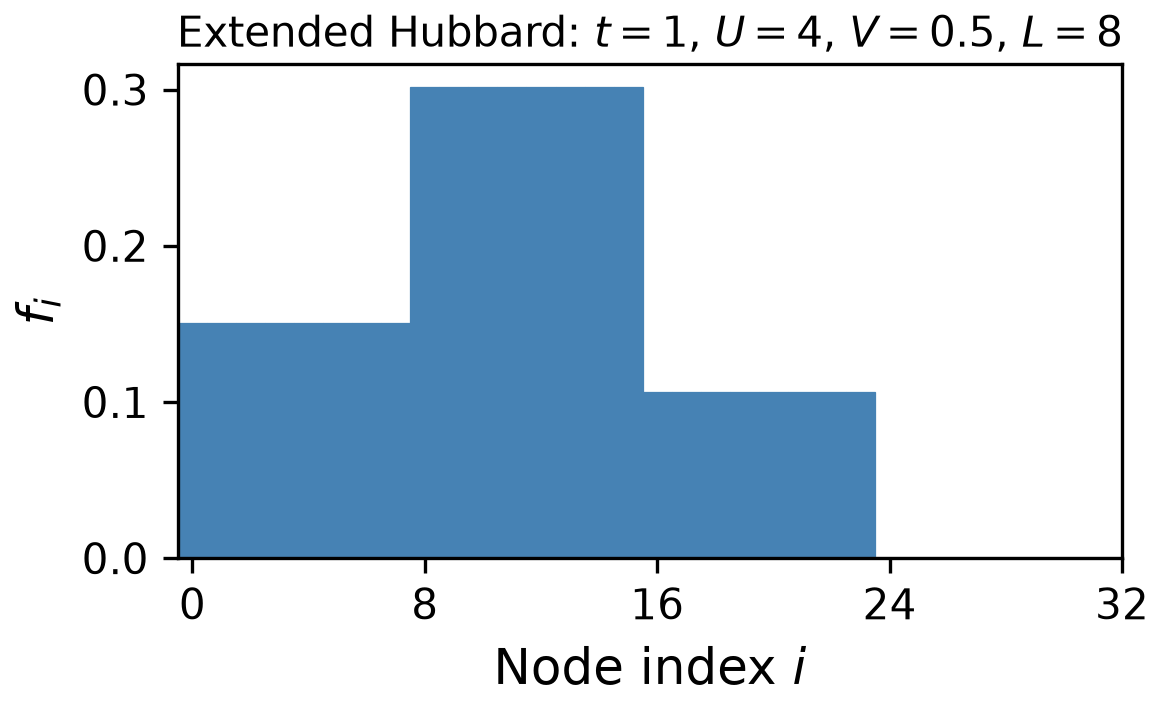}{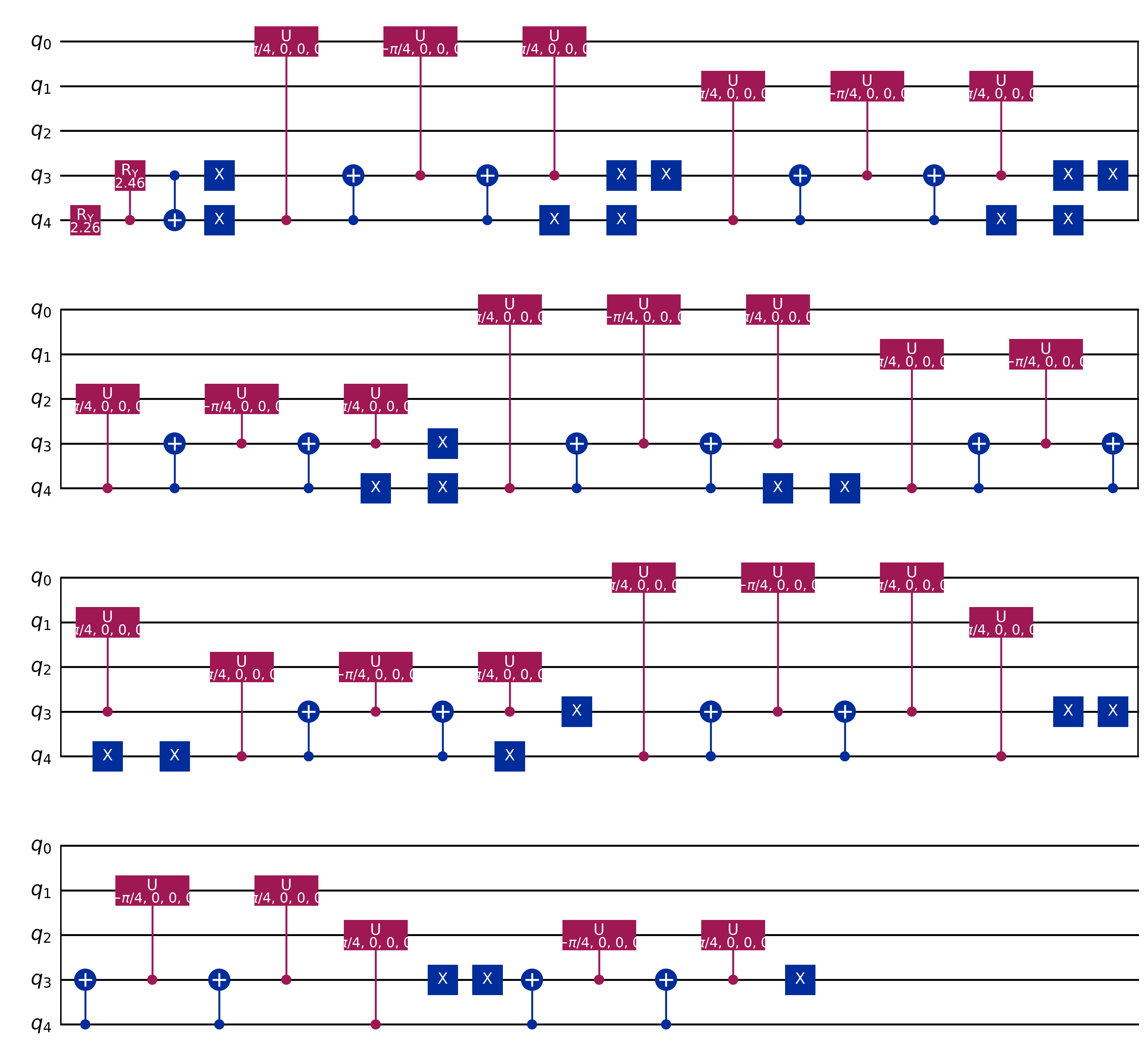}
  {Extended Fermi--Hubbard PREP: $t\!=\!1$, $U\!=\!4$, $V\!=\!0.5$,
   $L\!=\!8$, $N\!=\!32$.  Three constant blocks of length $L$
   followed by $L$ zero-padded indices.}
  {0.5}
  {1}{sec6_hubbard}

\subsection{Computational Mechanics}
\label{sec:app_poisson}

The Poisson equation $-\nabla^2 u = f$ on the unit square with a
separable sinusoidal source,
\begin{equation}
  f(x,y) = \sin(2\pi n x)\sin(2\pi p y),
\end{equation}
arises naturally in elliptic PDE solvers when the source term is
periodic~\cite{strang2007}. After discretization on an $N\times N$
grid, the right-hand side vector $\mathbf{f}$ is a tensor product:
\begin{equation}
  \mathbf{f} = \mathbf{u} \otimes \mathbf{v},
  u_i = \sin(2\pi n i/N), 
  v_j = \sin(2\pi p j/N).
\end{equation}
A tensor product of two normalized states is a product state on the
qubit register, so the $2m$-qubit encoding separates exactly into two
independent $m$-qubit \texttt{FOURIER} circuits composed via the
\texttt{TENSOR} pattern (Section~\ref{sec:tensor}):
\begin{lstlisting}[style=python]
circuit, info = encode(
    TENSOR([(FOURIER(modes=[(2, 1.0, 0)]), 32),
            (FOURIER(modes=[(3, 1.0, 0)]), 32)]),
    N=32*32)
\end{lstlisting}
Figure~\ref{fig:sec6_poisson} shows the separable source term and combined circuit.

\loadfigsv{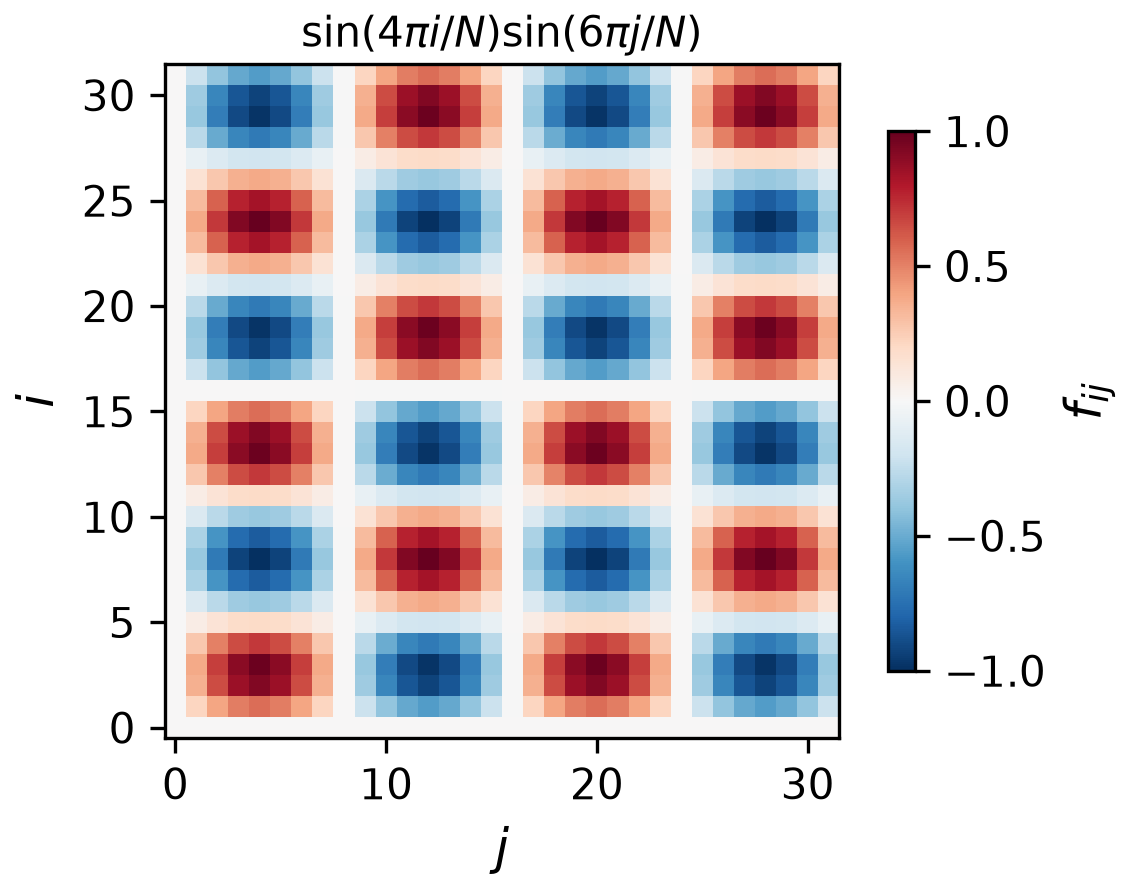}{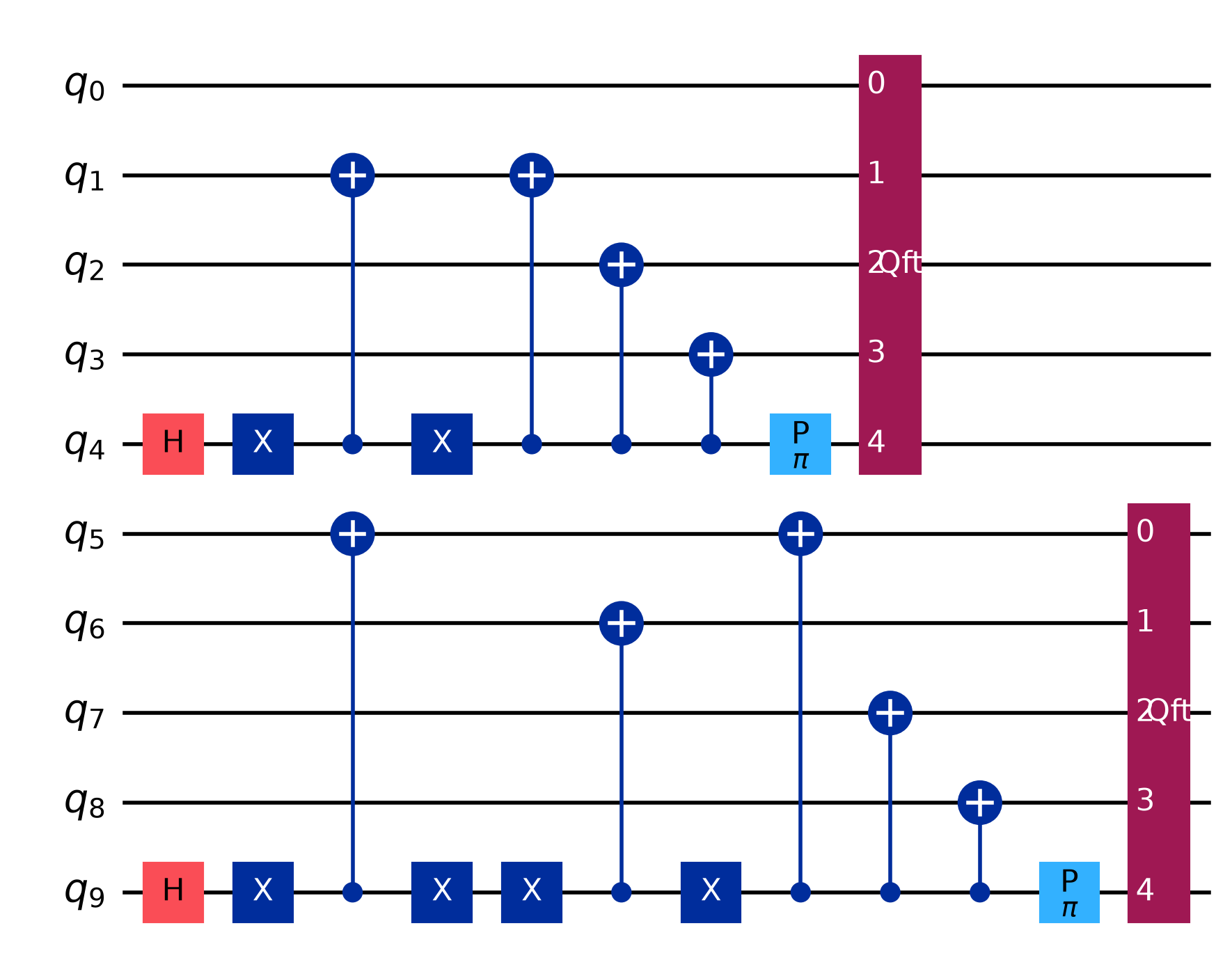}
  {2D Poisson source: $\sin(4\pi i/N)\sin(6\pi j/N)$, $N=32$}
  {0.5}
  {0.75}{sec6_poisson}

The total gate count is $2\times\bigO{m^2}$ for $2m$ qubits encoding
$N^2$ amplitudes, compared to $\bigO{N^2}$ for general state preparation.
At $m=5$ per axis ($N^2 = 1024$ amplitudes, $2m=10$ qubits), the
\texttt{TENSOR}-composed circuit transpiles to 119 gates (62~$U$
+ 57~CX) versus 2{,}036 gates for Qiskit's
\texttt{StatePreparation}\,---\,a 17$\times$ reduction, with the
advantage widening exponentially in $m$.

\subsection{Quantitative Finance}
\label{sec:app_finance}

Quantum amplitude estimation provides a quadratic speedup over
classical Monte Carlo for derivative pricing, conditional on
efficient preparation of the underlying probability
distribution~\cite{montanaro2015,Herbert2022,stamatopoulos2020}.
Under the Black--Scholes model, the asset price $S$ at maturity $T$
follows a log-normal distribution
\begin{equation}
  p(S) = \frac{1}{S\,\sigma\sqrt{2\pi T}}
  \exp\!\left(-\frac{(\ln S - \mu)^2}{2\sigma^2 T}\right),
\end{equation}
with $\mu = \ln S_0 + (r - \sigma^2/2)T$.  After truncation to
some $[S_{\min}, S_{\max}]$ and discretization on $N = 2^m$ grid
points, the target amplitude vector is $f_i \propto \sqrt{p(S_i)}$.

The log-normal density falls outside the exact pattern families.  However,
 \texttt{encode\_mps} (Section~\ref{sec:mps}) applies.

\textbf{Example.}
Black--Scholes parameters $S_0 = 100$, $r = 0.05$, $\sigma = 0.2$,
$T = 1$, on $N = 2^{16}$ grid points spanning $\pm 3\sigma\sqrt{T}$
in log-space:
\begin{lstlisting}[style=python]
import numpy as np
from pyencode.mps import encode_mps

m = 16;  N = 2**m
S0, r, sig, T = 100.0, 0.05, 0.2, 1.0
mu = np.log(S0) + (r - 0.5*sig**2)*T
S  = np.linspace(S0*np.exp(-3*sig*np.sqrt(T)), S0*np.exp( 3*sig*np.sqrt(T)), N)
p  = np.exp(-(np.log(S)-mu)**2 / (2*sig**2*T))/ (S*sig*np.sqrt(2*np.pi*T))
v  = np.sqrt(p);  v /= np.linalg.norm(v)

circuit, info = encode_mps(v, bond_dim=8)
\end{lstlisting}
With $\chi=8$, the circuit acts on $8 + m = 19$ qubits and transpiles
to $1{,}425$ two-qubit gates ($3{,}791$ total) at
\texttt{optimization\_level=3}, with truncation error squared below
$10^{-9}$. Figure~\ref{fig:ex_lognormal} shows the distribution and MPS circuit.

\loadfigs{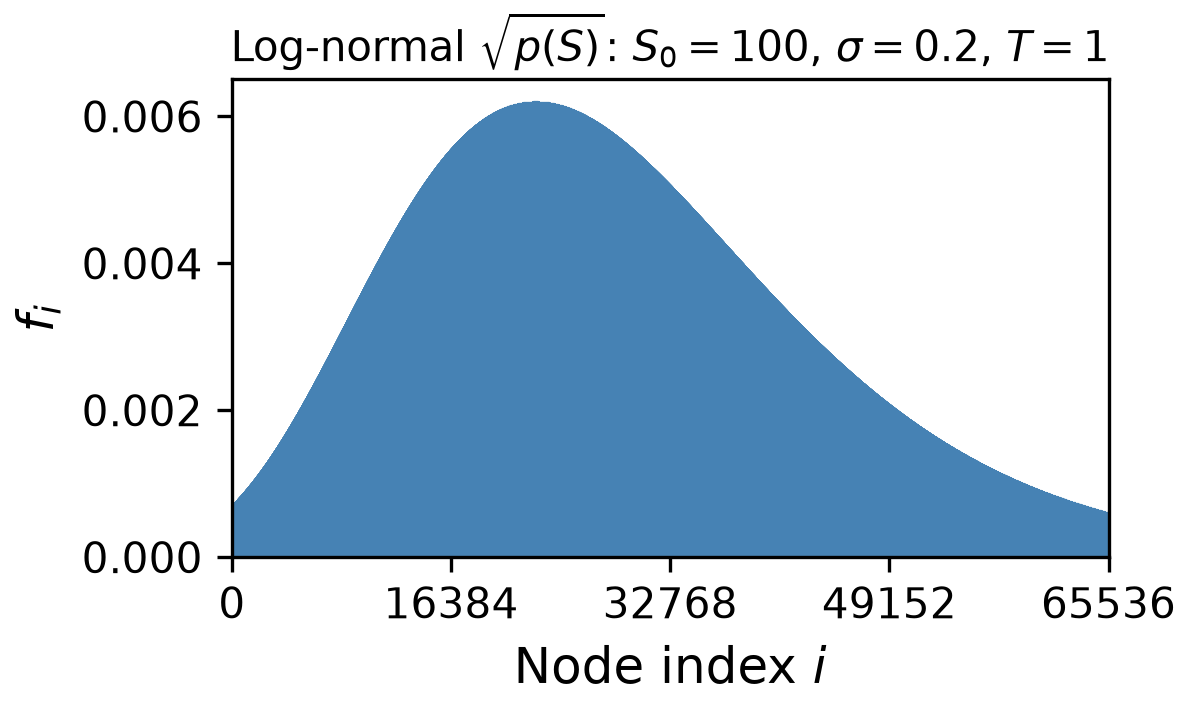}{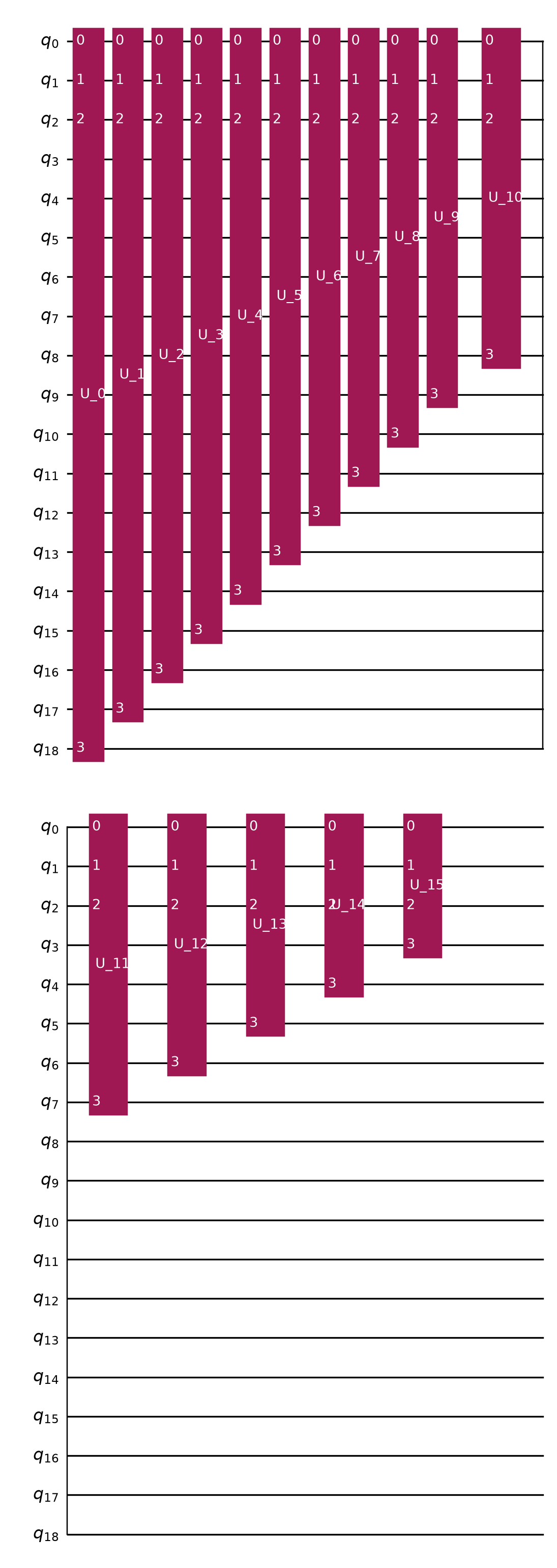}
  {Log-normal density: $S_0=100$, $\sigma=0.2$, $T=1$, $m=16$,
   $\chi=8$.}
  {unused}
  {1}{ex_lognormal}

\section{Conclusions}
\label{sec:conclusions}

PyEncode packages a decade of structured quantum state preparation
theory into a convenient Python API. The library's core idea is
that a declared pattern such as \emph{sparse, step, Fourier, Dicke,} and
so on, carries enough algebraic structure to collapse the general
$\mathcal{O}(2^m)$ synthesis cost to $\mathcal{O}(\mathrm{poly}(m))$; \emph{sum}, \emph{partition}, and
\emph{tensor} preserve these gains under composition. Users specify a
vector by its mathematical form; the library returns a verified Qiskit circuit that never materializes the vector classically.
A companion \texttt{predict\_gates} entry point returns transpiled
gate counts and depth in closed form, enabling design-space
exploration at problem sizes where circuit synthesis would be
prohibitive, and a reverse-lookup utility \texttt{match\_vector} maps a
materialized numerical vector back to the cheapest structured family
that fits it.

Across the ten pattern families, PyEncode outperforms Qiskit's
general-purpose \texttt{StatePreparation} from $m \geq 10$ onward; by
$m = 16$ the separation is two to four orders of magnitude in
two-qubit gate count.

PyEncode also provides a standalone matrix product state loader
\texttt{encode\_mps} for approximate encoding.  Extending \texttt{encode\_mps} to compose with
\texttt{SUM}, \texttt{PARTITION}, and \texttt{TENSOR} is a next step.

Characterizing which classes of vectors --- beyond the ones covered here --- admit
$\mathcal{O}(\mathrm{poly}(m))$ exact circuits remains an open and
potentially rich theoretical question. 

\section*{Acknowledgments}
The first author would like to acknowledge the Vilas Associate Grant from the
University of Wisconsin Graduate School.

\section*{Use of AI Tools}
The authors used generative AI assistants, specifically, Claude (Anthropic) and Gemini (Google) during the preparation of this manuscript. The tools were used for code development and drafting of text. All content was reviewed, verified, and edited by the authors, who take full responsibility for the accuracy and integrity of the work.

\section*{Declarations}
The authors declare no conflict of interest.

\section*{Code Availability}
The code developed in this work is available at
\url{https://github.com/UW-ERSL/PyEncode.git}.
The package can be installed with
\begin{center}
\small\texttt{pip install git+\url{https://github.com/UW-ERSL/PyEncode.git}}
\end{center}
\vspace{6pt}
\noindent
(The \texttt{pyencode} package on PyPI is unrelated to this library.)

\bibliographystyle{unsrt}
\bibliography{references}

\end{document}